\newif\iflong
\newif\ifshort

\shortfalse
\longtrue

\ifshort
\documentclass[acmsmall]{acmart} 
\fi\iflong
\documentclass[acmsmall,nonacm]{acmart} 
\fi

\usepackage[utf8]{inputenc}
\usepackage[T1]{fontenc}
\usepackage{xspace}
\usepackage{xparse}
  \NewDocumentCommand{\li}{v}{\textbf{\footnotesize\texttt{#1}}}

\usepackage[frozencache=true]{minted}
  \ifshort
  \setminted{fontsize=\footnotesize,linenos}
  \fi
  \iflong
  \setminted{fontsize=\footnotesize,linenos}
  \fi
\usepackage{tikz}
  \usetikzlibrary{calc}
  \usetikzlibrary{shadows,graphs}
\usepackage{amsmath}
\usepackage{mathpartir}
  
  \newcommand{\Rule}[1]{\hyperlink{#1}{\TirName {#1}}}

\usepackage{listings}

\usepackage{marvosym} 

\usepackage[skins,theorems]{tcolorbox} 
\tcbset{highlight math style={enhanced,
  colframe=white,colback=white,
  arc=0pt,outer arc = 0pt, 
  top=0pt,bottom=0pt,left=0mm,right=0mm,
  leftrule=0mm,rightrule=0mm,toprule=0mm,bottomrule=0mm,
  boxsep=0.7mm, 
  borderline={0.2mm}{0mm}{white,solid},
  boxrule=0pt}}

\usepackage[b]{esvect} 

\newcommand{\sref}[1]{\ref{sec:#1}}
\newcommand{\fref}[1]{\ref{fig:#1}}

\newcommand\lnref[1]{\ref{line:#1}}
\newcommand\thref[1]{\ref{thm:#1}}
\newcommand\defref[1]{\ref{def:#1}}
\newcommand\appendixref[1]{\ref{appendix:#1}}

\newcommand\lnlabel[1]{\label{line:#1}}
\newcommand\flabel[1]{\label{fig:#1}}
\newcommand\slabel[1]{\label{sec:#1}}
\newcommand\thlabel[1]{\label{thm:#1}}
\newcommand\deflabel[1]{\label{def:#1}}
\newcommand\appendixlabel[1]{\label{appendix:#1}}

\newcommand\lbox[1]{\ensuremath{
  \tcbhighmath[boxrule=0.4pt,borderline={0.2mm}{0mm}{black,solid}]{#1}}}
\newcommand\rbox[1]{\ensuremath{
  \tcbhighmath[boxrule=0.4pt,borderline={0.2mm}{0mm}{black,dashed}]{#1}}}
\newcommand\projmarkedname{\ensuremath{\kw{proj\_marked}}}
\newcommand\projmarked[2]{\ensuremath{\projmarkedname\;{#1}\;{#2}}}
\newcommand\sloansinclname{\ensuremath{\kw{sloans\_incl}}}
\newcommand\sloansincl[2]{\ensuremath{\sloansinclname\;{#1}\;{#2}}}
\newcommand\wfjoinholename{\ensuremath{\kw{wf\_join\_hole}}}
\newcommand\wfjoinhole[2]{\ensuremath{\wfjoinholename\;{#1}\;{#2}}}

\newcommand{\fstar}{F$^\ast$\xspace}

\newcommand\llbcs{\ensuremath{\text{LLBC}^{\#}}\xspace}
\newcommand\llbcp{\ensuremath{\text{LLBC}^{+}}\xspace}
\newcommand\hlplp{\ensuremath{\text{HLPL}^{+}}\xspace}
\newcommand\llbconly{\textnormal{\textbf{(LLBC only)}}}
\newcommand\llbcsonly{\textnormal{\textbf{(\llbcs only)}}}
\newcommand\hlplonly{\textnormal{\textbf{(HLPL only)}}}

\DeclareRobustCommand{\son}[1]{}
\DeclareRobustCommand{\jonathan}[1]{}
\DeclareRobustCommand{\aymeric}[1]{}

\newcommand\kw[1]{\ensuremath{\mathsf{#1}}}
\newcommand\tbrw[2]{\ensuremath{\mathsf{\&}^#1\,#2}}
\newcommand\tmbrw[2]{\ensuremath{\mathsf{\&}^#1\mathsf{mut}\,#2}}
\newcommand\ebrw[1]{\ensuremath{\mathsf{\&}\,#1}}
\newcommand\embrw[1]{\ensuremath{\mathsf{\&mut}\,#1}}
\newcommand\erbrw[1]{\ensuremath{\mathsf{\&reserved}\,#1}}
\newcommand\eassign[2]{\ensuremath{#1 := #2}}

\newcommand\emove[1]{\ensuremath{\kw{move}\,#1}}
\newcommand\ecopy[1]{\ensuremath{\kw{copy}\,#1}}
\newcommand\epanic{\kw{panic}}
\newcommand\ereturn{\kw{return}}
\newcommand\eseq[2]{#1;\,#2}
\newcommand\eite[3]{\kw{if}\,#1\,\kw{then}\,#2\,\kw{else}\,#3}

\newcommand\ematch[2]{\kw{match}\,#1\,\kw{ with }\;#2}
\newcommand\ematchsum[3]{\kw{match}\,#1\,\kw{ with }
  \;|\; \kw{Left}\; \Rightarrow #2
  \;|\; \kw{Right}\; \Rightarrow #3
}

\newcommand\krv{\ensuremath{rv}}
\newcommand\kop{\ensuremath{op}}
\newcommand\eloop[1]{\ensuremath{\kw{loop}\,#1}}
\newcommand\ebreak[1]{\ensuremath{\kw{break}\,#1}}
\newcommand\econtinue[1]{\ensuremath{\kw{continue}\,#1}}
\newcommand\kfalse{\mathsf{false}}
\newcommand\ktrue{\mathsf{true}}
\newcommand\enew[1]{\ensuremath{\kw{new}\,#1}}
\newcommand\efree[1]{\ensuremath{\kw{free}\,#1}}
\newcommand\esizeof[1]{\ensuremath{\kw{sizeof}\,#1}}
\newcommand\eundef{\text{\Biohazard}}
\newcommand\eblockofname{\ensuremath{\kw{blockof}}}
\newcommand\eaddrofname{\ensuremath{\kw{addrof}}}
\newcommand\eblockof[1]{\ensuremath{\kw{blockof}\,{#1}}}
\newcommand\eblockofinv[1]{\ensuremath{\kw{blockof}^{-1}\,{#1}}}
\newcommand\eaddrof[1]{\ensuremath{\kw{addrof}\,{#1}}}
\newcommand\ecstatename{\ensuremath{\mathcal{C}_{\Omega}}}
\newcommand\ecstatefull[3]{\ensuremath{\ecstatename\;{#1}\;{#2}\;{#3}}}
\newcommand\ecstate[1]{\ensuremath{\ecstatename\;{\kw{blockof}}\;{\kw{addrof}}\;{#1}}}
\newcommand\ecvalname{\ensuremath{\mathcal{C}_{v}}}
\newcommand\ecvalfull[3]{\ensuremath{\ecvalname\;{#1}\;{#2}\;{#3}}}
\newcommand\ecval[1]{\ensuremath{\ecvalname\;{\kw{blockof}}\;{\kw{addrof}}\;{#1}}}
\newcommand\readaddress[3]{\eblockofname,\, \eaddrofname \vdash #1(#2) \readarrow #3}
\newcommand\readvaladdress[3]{\vdash #1 + #2 \readarrow #3}
\newcommand\Omegapl{\ensuremath{\Omega^{\text{pl}}}}
\newcommand\Omegahlpl{\ensuremath{\Omega^{\text{hlpl}}}}
\newcommand\Omegallbc{\ensuremath{\Omega^{\text{llbc}}}}
\newcommand\ecompatiblefull[3]{\kw{compatible}\;{#1}\;{#2}\;{#3}}
\newcommand\ecompatible[1]{\kw{compatible}\;\eblockofname\;\eaddrofname\;#1}

\newcommand\emborrow[2]{\ensuremath{\mathsf{borrow}^m\,#1\;#2}}
\newcommand\esborrow[1]{\ensuremath{\mathsf{borrow}^s\,#1}}
\newcommand\esrborrow[1]{\ensuremath{\mathsf{borrow}^{s,r}\,#1}}
\newcommand\erborrow[1]{\ensuremath{\mathsf{borrow}^r\,#1}}
\newcommand\erborrowv{\ensuremath{\mathsf{borrow}^r}}
\newcommand\eborrowv{\ensuremath{\mathsf{borrow}^{s,r,m}}}

\newcommand\emloanv{\ensuremath{\mathsf{loan}^m}}
\newcommand\esloan[2]{\ensuremath{\mathsf{loan}^s\,#1\,#2}}
\newcommand\emloan[1]{\ensuremath{\mathsf{loan}^m\,#1}}
\newcommand\ebox[1]{\ensuremath{\mathsf{Box}\,#1}}
\newcommand\eptr[1]{\ensuremath{\mathsf{ptr}\,#1}}
\newcommand\eloc[2]{\ensuremath{\mathsf{loc}\,#1\,#2}}
\newcommand\eleft[1]{\ensuremath{\mathsf{Left}\,#1}}
\newcommand\eright[1]{\ensuremath{\mathsf{Right}\,#1}}
\newcommand\elist[1]{\ensuremath{\mathsf{List}\,#1}}
\newcommand\enil{\ensuremath{\mathsf{Nil}}}
\newcommand\econs[2]{\ensuremath{\mathsf{Cons}\,#1\,#2}}
\newcommand\elength[1]{\ensuremath{\mathsf{length}\,#1}}
\newcommand\econcat{\,\texttt{++}\,}
\newcommand\esubst[3]{\ensuremath{\left[{#2}\Big/{#1}\right]{#3}}}
\newcommand\onehole[2]{\ensuremath{{#1}[\,{#2}\,]}}
\newcommand\omegaonehole[2]{\ensuremath{\onehole{\Omega_{#1}}{#2}}}
\newcommand\twoholes[3]{\ensuremath{{#1}[\,{#2}\,,\,{#3}\,]}}
\newcommand\omegatwoholes[3]{\ensuremath{\twoholes{\Omega_{#1}}{#2}{#3}}}
\newcommand\threeholes[4]{\ensuremath{{#1}[\,{#2}\,,\,{#3}\,,\,{#4}\,]}}

\newcommand\mapstoemborrow[2]{\mapsto \emborrow{\ell_{#1}}{#2}}
\newcommand\mapstoemloan[1]{\mapsto \emloan{\ell_{#1}}}

\newcommand\vdashPl{\ensuremath{\vdash_{\kw{pl}}}}
\newcommand\vdashLlbc{\ensuremath{\vdash_{\kw{llbc}}}}
\newcommand\vdashLlbcp{\ensuremath{\vdash_{\kw{llbc+}}}}
\newcommand\vdashHlpl{\ensuremath{\vdash_{\kw{hlpl}}}}
\newcommand\vdashHlplp{\ensuremath{\vdash_{\kw{hlpl+}}}}
\newcommand\vdashLlbcs{\ensuremath{\vdash_{\kw{llbc\#}}}}
\newcommand\vdashLone{\ensuremath{\vdash_{\kw{l}}}}
\newcommand\vdashLtwo{\ensuremath{\vdash_{\kw{h}}}}
\newcommand\vdashLplus{\ensuremath{\vdash_{\kw{+}}}}

\newcommand\stmtarrow{\ensuremath{\rightsquigarrow}}
\newcommand\stmtres[2]{\ensuremath{\stmtarrow (#1,\,#2)}}

\newcommand\bool{\kw{bool}}
\newcommand\eor{\;\vee\;}
\newcommand\eand{\;\wedge\;}

\newcommand\readarrow{\ensuremath{\Rightarrow}}
\newcommand\readarrowcap[1]{\overset{#1}{\readarrow}}
\newcommand\eqimmut{\ensuremath{\readarrowcap{\kw{imm}}}}
\newcommand\eqmut{\ensuremath{\readarrowcap{\kw{mut}}}}
\newcommand\eqmove{\ensuremath{\readarrowcap{\kw{mov}}}}
\newcommand\eqimmutmut{\ensuremath{\readarrowcap{\kw{imm,mut}}}}
\newcommand\eqk{\ensuremath{\readarrowcap{\kw{k}}}}

\newcommand\updtoperator{\ensuremath{\leftarrow}}
\newcommand\updtimmut{\ensuremath{\updtoperator}}
\newcommand\updtmut{\ensuremath{\updtoperator}}
\newcommand\updtmove{\ensuremath{\updtoperator}}
\newcommand\updtimmutmut{\ensuremath{\updtoperator}}
\newcommand\updtk{\ensuremath{\updtoperator}}

\newcommand\updtarrow{\ensuremath{\Rightarrow}}
\newcommand\updtarrowcap[1]{\overset{#1}{\updtarrow}}
\newcommand\ueqimmut{\ensuremath{\updtarrowcap{\kw{imm}}}}
\newcommand\ueqmut{\ensuremath{\updtarrowcap{\kw{mut}}}}
\newcommand\ueqmove{\ensuremath{\updtarrowcap{\kw{mov}}}}
\newcommand\ueqimmutmut{\ensuremath{\updtarrowcap{\kw{imm,mut}}}}
\newcommand\ueqk{\ensuremath{\updtarrowcap{\kw{k}}}}

\newcommand\updtstateplace[3]{\ensuremath{#1 \vdash #2 \updtoperator #3}}

\newcommand\writeres[2]{\ensuremath{(#1,\,#2)}}
\newcommand\wimmut[2]{\ensuremath{\ueqimmut\writeres{#1}{#2}}}

\newcommand\wmove[2]{\ensuremath{\ueqmove\writeres{#1}{#2}}}
\newcommand\wimmutmut[2]{\ensuremath{\ueqimmutmut\writeres{#1}{#2}}}
\newcommand\wk[2]{\ensuremath{\ueqk\writeres{#1}{#2}}}

\newcommand\exprarrow{\ensuremath{\Downarrow}}
\newcommand\exprres[2]{\ensuremath{\exprarrow (#1,\,#2)}}

\newcommand\arrowThree{\ensuremath{\Downarrow}}

\newcommand\steparrow[1]{\ensuremath{\underset {#1} \rightsquigarrow}}
\newcommand\stmtstepres[3]{\ensuremath{\steparrow{#1} (#2,\,#3)}}
\newcommand\pushstack{\ensuremath{\vdash\kw{push\_stack}\;}}
\newcommand\popstack{\ensuremath{\vdash\kw{pop\_stack}\;}}
\newcommand\plpushstack{\ensuremath{\vdash\kw{pl\_push\_stack}\;}}
\newcommand\plpopstack{\ensuremath{\vdash\kw{pl\_pop\_stack}\;}}

\newcommand\borrowchecks[1]{\ensuremath{\kw{borrow\_checks}\; {#1}}}

\newcommand\projinput[3]{\ensuremath{\kw{proj\_input}\;{#1}\;{#2}={#3}}}

\newcommand\projoutput[4]{\ensuremath{\kw{proj\_output}\;{#1}\;{#2}=({#3},\;#4)}}

\newcommand\absinput{\ensuremath{A_{\kw{in}}}}

\newcommand\absinit{\ensuremath{A_{\kw{init}}}}
\newcommand\absfinal{\ensuremath{A_{\kw{final}}}}
\newcommand\absreborrow{\ensuremath{A_{\kw{reborrow}}}}
\newcommand\abssig{\ensuremath{A_{\kw{sig}}}}
\newcommand\absinputrho{\absinput(\rho)}

\newcommand\absinitrho{\absinit(\rho)}
\newcommand\absfinalrho{\absfinal(\rho)}
\newcommand\absreborrowrho{\absreborrow(\rho)}
\newcommand\vecabssigrho{\ensuremath{\overrightarrow{\abssigrho}}}
\newcommand\vecabsfinalrho{\ensuremath{\overrightarrow{\absfinalrho}}}
\newcommand\vecabsinitrho{\ensuremath{\overrightarrow{\absinitrho}}}
\newcommand\vecabsinputrho{\ensuremath{\overrightarrow{\absinputrho}}}

\newcommand\vecabsreborrowrho{\ensuremath{\overrightarrow{\absreborrowrho}}}

\newcommand\vecabsrho{\ensuremath{\overrightarrow{A(\rho)}}}
\newcommand\vecabsrhoi[1]{\ensuremath{\overrightarrow{A_{#1}(\rho)}}}
\newcommand\abssigrho{\abssig(\rho)}
\newcommand\tauout{\ensuremath{\tau^{\kw{out}}}}
\newcommand\vout{\ensuremath{v_{\kw{out}}}}
\newcommand\xret{\ensuremath{x_{\kw{ret}}}}

\newcommand\fsig[1]{\ensuremath{
  \kw{fn}\ \langle{#1}\rangle
     \ (\overrightarrow{x_i} : \overrightarrow{\tau_i})
     \ (\overrightarrow{y_j} : \overrightarrow{\tau_j})
     \ (x_\mathsf{ret}: \tau) \ \{ \ s \ \}
}}
\newcommand\fsignoindices[1]{\ensuremath{
  \kw{fn}\ \langle{#1}\rangle
     \ (\overrightarrow{x} : \overrightarrow{\tau})
     \ (\overrightarrow{y} : \overrightarrow{\tau'})
     \ (x_\mathsf{ret}: \tau_\kw{ret}) \ \{ \ s \ \}
}}
\newcommand\fcall{
  f\langle{\vec{\_}},\,\vec{\tau}\rangle(\overrightarrow{op_j})
}
\newcommand\fcallnoindices{
  f\langle{\vec{\_}},\,\vec{\tau}\rangle(\overrightarrow{op})
}
\newcommand\feqsig[1]{
  f\langle{#1},\,\vec{\tau}\rangle=
  \fsig{#1}
}
\newcommand\feqsignoindices[1]{
  f\langle{#1},\,\vec{\tau}\rangle=
  \fsignoindices{#1}
}
\newcommand\Omegas{\ensuremath{\Omega^{\#}}}
\newcommand\Ss{\ensuremath{S^{\#}}}

\newcommand\toabs{\ensuremath{\prec^{\text{to-abs}}}}
\newcommand\merge{\ensuremath{=}}
\newcommand\MergeAbs{\ensuremath{\Join}}

\newcommand\join[2]{\ensuremath{\mathsf{join}\left(#1,\;#2\right)}}
\newcommand\joinres{\ensuremath{\;\vert\;}}

\newcommand\joinenv[5]{\ensuremath{#1,\;#2 \vdash \mathsf{join}_\Omega\; #3\;#4 \rightsquigarrow #5}}
\newcommand\joinvals[2]{\mathsf{join}_v\; #1\;#2}
\newcommand\joinval[6]{\ensuremath{#1,\;#2 \vdash \mathsf{join}_v\; #3\;#4 \Downarrow #5 \joinres #6}}

\newcommand\collapse{\ensuremath{\searrow}}

\newcommand\Abs[2]{\ensuremath{{#1}\,\{\,#2\,\}}}
\newcommand\AbsI[2]{\ensuremath{A_{#1}\,\{\,#2\,\}}}

\newcommand\emem{\ensuremath{\mathsf{mem}}}
\newcommand\eenv{\ensuremath{\mathsf{env}}}

\newcommand\Thm[1]{\hyperlink{#1}{\TirName{#1}}}

\newcommand\Section[1]{§\sref{#1}}
\newcommand\Theorem[1]{Theorem~\thref{#1}}
\newcommand\Figure[1]{Figure~\fref{#1}}
\newcommand\Appendix[1]{Appendix~\appendixref{#1}}

\newcommand\myparagraph[1]{\emph{#1}.\ }

\newlength{\characterlength}
\definecolor{envcolor}{RGB}{61, 122, 122}
\newcommand\thmvspace{\ifshort\vspace{-5mm}\fi}
\newcommand\figvspace{\ifshort\vspace{-3mm}\fi}

\ifshort
\setcopyright{acmlicensed}
\acmDOI{10.1145/3674640}
\acmYear{2024}
\copyrightyear{2024}
\acmSubmissionID{icfp24main-p55-p}
\acmJournal{PACMPL}
\acmVolume{8}
\acmNumber{ICFP}
\acmArticle{251}
\acmMonth{8}
\received{2024-02-28}
\received[accepted]{2024-06-18}
\fi

\begin{document}

\iflong
\title{Sound Borrow-Checking for Rust via Symbolic Semantics (Long Version)}
\fi

\author{Son Ho}
\orcid{0000-0003-3297-9156}
\affiliation{%
  \institution{Inria}
  \city{Paris}
  \country{France}
}
\email{son.ho@inria.fr}

\author{Aymeric Fromherz}
\orcid{0000-0003-2642-543X}
\affiliation{%
  \institution{Inria}
  \city{Paris}
  \country{France}
}
\email{aymeric.fromherz@inria.fr}

\author{Jonathan Protzenko}
\orcid{0000-0001-7347-3050}
\affiliation{%
  \institution{Microsoft Azure Research}
  \city{Redmond}
  \country{USA}
}
\email{protz@microsoft.com}





\begin{abstract}
The Rust programming language continues to rise in popularity, and as such,
warrants the close attention of the programming languages community. In this
work, we present a new foundational contribution towards the theoretical
understanding of Rust's semantics. We prove that LLBC, a high-level, borrow-centric model
previously proposed for Rust's semantics and execution, is sound with regards to a low-level
pointer-based language \emph{à la} CompCert. Specifically, we prove the
following: that LLBC is a correct view over a traditional model of execution;
that LLBC's symbolic semantics are a correct abstraction of LLBC programs;
and that LLBC's symbolic semantics act as a borrow-checker for LLBC, i.e. that
symbolically-checked LLBC programs do not get stuck when executed on a
heap-and-addresses model of execution.

To prove these results, we introduce a new proof style that considerably
simplifies our proofs of simulation, which relies on a notion of hybrid
states. Equipped with this reasoning framework, we show that a new addition to
LLBC's symbolic semantics, namely a join operation, preserves the abstraction
and borrow-checking properties. This in turn allows us to add support for
loops to the Aeneas framework; we show, using a series of examples and case
studies, that this unlocks new expressive power for Aeneas.
\end{abstract}



\maketitle

\newcommand\lambdar{\ensuremath{\lambda_{\text{Rust}}}}

\section{Introduction}

The Rust programming language continues to rise in popularity. Rust has by now
become the darling of developers, voted the most admired language for the 8th
year in a row~\cite{stackoverflow2023rust}; a favorite of governments,
who promote safe languages in the context of cybersecurity~\cite{nsa2022safelang};
an industry bet, wherein large corporations are
actively migrating to Rust~\cite{microsoft2022rust}; and an active topic in the
programming languages research community.

Two research directions have emerged over the
past few years. First, many tools now set out to verify Rust
programs and prove properties about their behavior, e.g., show that a Rust
program matches its specification. But to confidently reason about Rust
programs, one needs a solid semantic foundation to build upon. This is the
second research direction, namely, understanding the semantics of Rust itself,
clarifying the language specification, and showing the soundness of Rust's type
system.

For Rust verification, we find tools such as Creusot~\cite{denis2022creusot},
Prusti~\cite{prusti}, Verus~\cite{lattuada2023verus}, or hax~\cite{hax}. All
of those leverage one key insight: the strong ownership
discipline enforced by the Rust type system makes verification easier. For Creusot
and Verus, this observation turns into a first-order logical encoding of Rust
programs that can then be discharged to an SMT solver. For Prusti, the Rust
discipline guides the application of separation logic rules in the underlying
Viper framework~\cite{mueller16viper}. And for hax, restricting programs to a pure value-passing
subset of Rust allows writing an almost identity-like translation to
backends such as \fstar~\cite{swamy2016fstar} or ProVerif~\cite{blanchet2001proverif}.

For Rust semantics, RustBelt~\cite{jung2017rustbelt} aims to prove the
soundness of Rust's type system
using a minimalistic model called $\lambdar$, whose operational semantics is
defined by compilation to a core language. The MIRI project~\cite{miri} aims to
provide a reference operational semantics of Rust, even in the presence of
unsafe code. Stacked~\cite{jung2019stacked}
and Tree Borrows~\cite{treeborrows2023} aim to clarify the
aliasing contract between the programmer and the Rust compiler.

At the intersection of these two axes is RustHornBelt~\cite{RustHornBelt},
which aims to prove that
the RustHorn~\cite{matsushita2020rusthorn} logical
encoding of Rust programs (as used in Creusot) is sound
with regards to the semantics of $\lambdar$, and consequently, that properties
proven thanks to the logical encoding hold for the original program.



Also straddling both axes is Aeneas~\cite{aeneas}, a Rust verification project that
tackles the question of Rust source semantics too. For semantics, Aeneas
introduces the low-level borrow calculus (LLBC), a core language inspired by
Rust's Mid-Level IR (MIR) and that directly models borrows, loans, and places as
found in source Rust programs, as well as delicate patterns such as reborrows or
two-phase borrows. Importantly, the LLBC semantics relies on a map from
variables to borrow-centric values, instead of a traditional heap and addresses. For
verification, Aeneas relies on two steps: first, an abstraction of LLBC called
the symbolic semantics, which operates in a modular fashion by summarizing the
ownership behavior of functions based on their types; second, a translation that
consumes the execution traces of the symbolic semantics to construct
a pure, executable program that is functionally equivalent to the source Rust.
Properties can then be proven upon the pure, translated code, and carry over to
the original Rust code.

Aeneas, as currently presented, exhibits two issues.
%
First, the correctness of the Aeneas endeavor is predicated on the LLBC model
being a sound foundation. LLBC has several unusual features, such as attaching
values to pointers rather than to the underlying memory location, or not relying on
an explicit heap; right now, it requires a leap of faith to
believe that this is an acceptable way to model Rust programs. The value of LLBC
is that it explains, checks and provides a semantic foundation for reasoning
about many Rust features; but the drawback is that one has to trust that this
semantics makes sense. We remark already that this question is orthogonal to the
RustBelt line of work. RustBelt attempts to establish the soundness of Rust's
type system with regards to $\lambdar$, whose classic, unsurprising semantics
does not warrant scrutiny.
Clarifying the link between LLBC and a standard heap-and-addresses model is not
just a matter of theory: once the Rust compiler emits LLVM bitcode, the heap and
addresses become real; if Aeneas is proving properties about an execution model
that cannot be connected to such semantics, then the validity of the whole
Aeneas endeavor is threatened.
The second shortcoming of Aeneas concerns its symbolic semantics,
which trades concrete program executions for abstract ones that forget about
precise values, and that rely on function signatures for modularity. This symbolic
semantics is an essential step to materialize the functional translation. Yet,
in spite of carefully-crafted rules that emphasize readability and simplicity,
there is no formal argument that it correctly approximates LLBC. Specifically,
the Aeneas paper claims that the interpreter for the symbolic semantics acts as
a borrow-checker for LLBC programs, but has no formal justification for this
claim. The symbolic semantics also suffer from limitations related to
control-flow: continuations of if-then-elses are duplicated in each branch, and
there is no support for loops. We posit that, lacking a formal basis in which to
reason about the correctness of the symbolic semantics, designing a
general-purpose ``join'' operation that can reconcile two symbolic states
was out of reach.

In this work, we set out to address both shortcomings of Aeneas, and introduce
new proof techniques to do so.
First, we establish that LLBC is, indeed, a reasonable model of execution, and
that in spite of some mildly exotic features, it really does connect to a
traditional heap-and-addresses model of execution.
Second, we show that the symbolic semantics of LLBC correctly approximates its
concrete semantics, and thus that successful executions in the symbolic
semantics guarantee the soundness of concrete executions -- in short, that the
symbolic interpreter acts as a borrow checker.

Because LLBC can, in some situations, analyze code more precisely than the Rust
borrow-checker, we are therefore able to prove the soundness of more
programs than the Rust compiler itself. We thus hope that this work sheds some
light on potential improvements to the current implementation of
the borrow-checker, backed by an actual formal argument.
%
%
%
Our contributions are as follows:

We introduce PL, for ``pointer language'', which uses a traditional,
    explicit heap inspired by CompCert's C memory model, and show that it
    refines LLBC (\Section{refine}). This allows us to establish that a low-level model (where
    values live at a given address) refines the Rust model given by LLBC (where
    borrows hold the value they are borrowing). These two semantics
    have opposite perspectives on what it \emph{means} to have a pointer; to the
    best of our knowledge, such a connection between two \emph{executable}
    semantics is novel.

We prove that LLBC itself refines the symbolic version of LLBC,
    henceforth known as \llbcs (\Section{llbc-sharp}).
    Combined with the previous result, this allows us to precisely state why
    \llbcs is a borrow-checker for LLBC: if an \llbcs execution
    succeeds, then any corresponding low-level PL execution is safe for all
    inputs.
    We obtain this result by a combination of forward simulations, along with
    the determinism of the target (PL) language; we also reason about what it
    means for a low-level (PL) initial state to satisfy a function signature
    with ownership constraints in \llbcs.

To conduct these proofs of refinement, we introduce a novel proof
    technique that relies on the fact that our languages operate over the same
    grammar of expressions, but give it different \emph{meanings}, or
    \emph{views} (\Section{proofmeth}). For instance: both our heap-and-address (PL) and
    borrows-and-loans (LLBC) views operate over the same program syntax, but
    have different types for their state and reduction rules.
    Rather than go full-throttle with a standard compilation-style proof of
    simulation, we reason modularly over local or pointwise
    transformations that rewrite one piece of state to another -- proofs over
    those elementary transformations are much easier. We then show that two
    states that are related by the transitive closure of these elementary
    transformations continue to relate throughout their execution, in the
    \emph{union} of the semantics, ultimately giving a proof of refinement.

Equipped with a framework in which to reason about the soundness of
    \llbcs with regards to LLBC, we define and prove correct a new operation
    that was previously missing from \llbcs: the join operation, which can
    reconcile two branches of control-flow (\Section{joins}). Inspired by joins in
    abstract interpretation (and specifically, joins in shape analysis), this
    new operation gives us a symbolic interpreter (i.e., a borrow-checker)
    that can handle loops (\Section{loops}), and does not exhibit pathological
    complexity on conditionals.
    Furthermore, our support naturally extends to the functional translation,
    meaning that Aeneas now supports verification of loops.
    We leave a discussion of the correctness of the Aeneas
    functional translation to future work.

We evaluate the effectiveness of our join operation (\Section{evaluation}), specifically when
    used to compute shape fixpoints for loops, over a series of small examples
    and case studies. We find that, in spite of being (naturally) incomplete,
    our join operation handles all of the examples we could muster. Our
    interpretation is that because Rust imposes so many invariants, a join
    procedure can leverage this structure and fare better than, say,
    a general-purpose join operation for analyzing C programs.


\begin{figure}[h]
  \centering
  \includegraphics[width=\columnwidth]{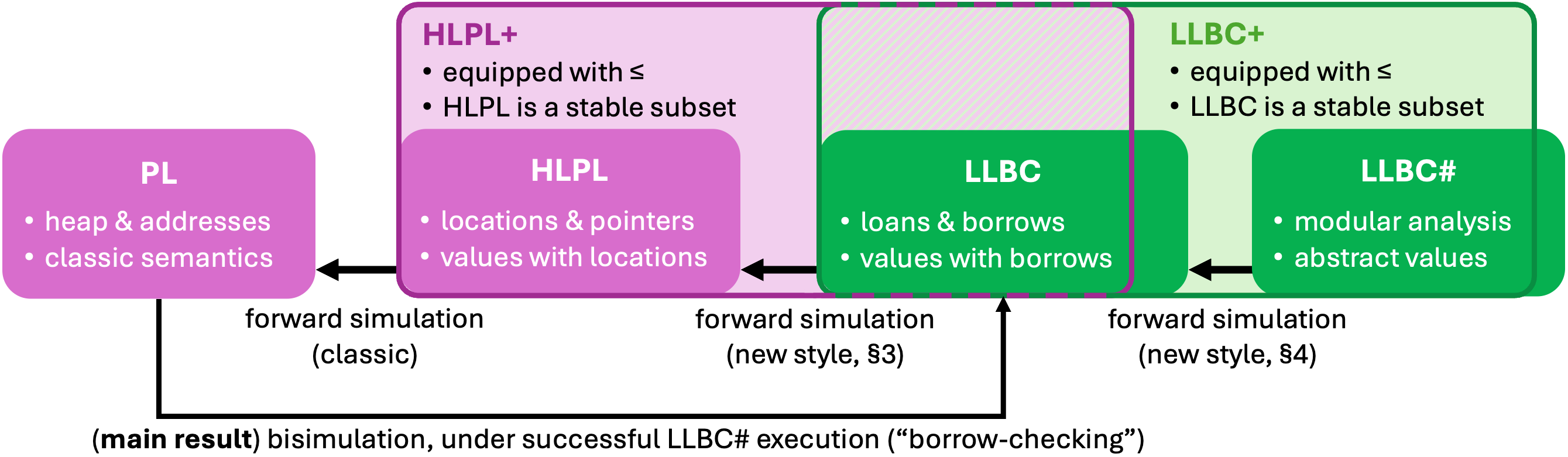}
  \caption{The architecture of our proof}
  \flabel{architecture}
  \figvspace
\end{figure}

\NewDocumentCommand{\LlbcSyntax}{s}{
\smaller
\arraycolsep=1pt %
\centering
\[
\IfBooleanF{#1}{\begin{array}{ll}}
\begin{array}[t]{llll}
  \tau & ::= & & \text{type} \\
    &&
    \kw{bool}\mid\kw{uint32}\IfBooleanT{#1}{\mid\kw{int32}}\mid\ldots\; & \text{literal types} \\
    && \tmbrw\rho\tau & \text{mutable borrow} \\
    && \tbrw\rho\tau & \text{immutable (shared) borrow}\\
\IfBooleanF{#1}{&&\ldots\\[1ex]}
\IfBooleanT{#1}{
    %
    && \alpha, \beta, \ldots & \text{type variables} \\
    && \tau_0 + \tau_1 & \text{sum}\\
    && (\vec \tau) & \text{pair ($\kw{len}(\vec \tau) = 2$) or unit ($\kw{len}(\vec \tau) = 0$)}\\
    && \mu X.\, \tau & \text{equirecursive type}
  \\[1ex]

  T & ::= & & \text{type constructor application} \\
  && t& \text{user-defined data type} \\
  && \kw{Box}& \text{boxed type}
  \\[1ex]
}

  \rho & & & \text{region (or lifetime)}
  \\[1ex]

  s & ::= & & \text{statement} \\
  && \emptyset & \text{empty statement (nil)} \\
  && \eseq s s & \text{sequence (cons)} \\
  && \eassign p \krv & \text{assignment} \\
\IfBooleanF{#1}{&& \epanic & \text{unrecoverable error} \\}
\IfBooleanF{#1}{&&\ldots\\[1ex]}
\IfBooleanT{#1}{
  && \eite \kop s s & \text{conditional}\\
  && \ematch p {\overrightarrow{ C \to s}} & \text{ data type case analysis} \\
  && \efree{p} & \text{free} \\
  && \ereturn & \text{function exit} \\
  && \epanic & \text{unrecoverable error} \\
  && \eloop{s} & \text{loop} \\
  && \ebreak{i} & \text{break to an outer loop} \\
  && \econtinue{i} & \text{continue to an outer loop} \\
  && p := f(\vec\kop) & \text{function call}
  \\[1ex]
}

  x & & & \text{variable}
  \\[1ex]

  p & ::= & P[x] & \text{place}
  \\[1ex]

\IfBooleanF{#1}{\end{array} & \begin{array}[t]{llll}}

  rv & ::= & & \text{assignable rvalues} \\
  && \kop & \text{operand} \\
  && \embrw p & \text{mutable borrow} \\
  && \ebrw p & \text{immutable (shared) borrow}\\
\IfBooleanF{#1}{&&\ldots\\[1ex]}
\IfBooleanT{#1}{
  && !\kop \mid \kop + \kop \mid \kop - \kop \mid \ldots & \text{operators}\\
  && \enew{op} & \text{new}
  \\[1ex]
}

  \kop & ::= & & \text{operand} \\
  && \emove p & \text{ownership transfer} \\
  && \ecopy p & \text{scalar copy} \\
  && \ktrue \mid \kfalse \IfBooleanT{#1}{\mid n_{\mathsf{i32}}} \mid n_{\mathsf{u32}} \mid \ldots\; &
  \text{literal constants} \\
\IfBooleanF{#1}{&&\ldots\\[1ex]}
\IfBooleanT{#1}{
  && \eleft{\kop} \mid \eright{\kop} & \text{sum constructor}\\
  && (\vec{op}) & \text{pair ($\kw{len}(\vec{op}) = 2$) or unit ($\kw{len}(\vec{op}) = 0$)}
  \\[1ex]
}

  P & ::= & & \text{path} \\
  && [.] & \text{base case} \\
  && * P & \text{deref} \\
  && P.f & \text{field selection}
  \\[1ex]

\IfBooleanT{#1}{
  D & ::= & & \text{top-level declaration} \\
  &&
  \mathsf{fn}\,f\,\langle\vec\rho,\, \vec\tau\rangle
    \ (\vec x_\mathsf{arg}: \vec\tau)
    \ (\vec x_\mathsf{local}: \vec \tau)
    \ (x_\mathsf{ret}: \tau)
    \ \{ \ s \ \} \quad
    & \text{function declaration}
}
\IfBooleanF{#1}{\end{array}}
\end{array}
\]
}

\NewDocumentCommand{\LlbcRvaluesEvalRules}{s}{
  \inferrule[E-MutBorrow \llbconly]{
    \hypertarget{E-MutBorrow} 
    \vdash \Omega(p) \eqmut v \\\\
    \IfBooleanF{#1}{\bot, \kw{loan}^{s, m} \not\in v}
    \IfBooleanT{#1}{\bot, \kw{loan}^{s, m}, \kw{borrow}^r \not\in v}\\
    \ell \text{ fresh} \\\\
    \updtstateplace{\Omega}{p}{\emloan\ell} \ueqmut \Omega'
  }{
    \Omega \vdash \embrw p \exprres{\emborrow \ell v}{\Omega'}
  }

  \inferrule[E-SharedBorrow \llbconly]{
    \hypertarget{E-SharedBorrow} 
    \vdash \Omega(p) \eqimmut v\\
    \IfBooleanF{#1}{\bot, \emloan{}{} \notin v}
    \IfBooleanT{#1}{\bot, \emloan{}{}, \erborrow{} \notin v}\\\\
    \updtstateplace{\Omega}{p}{v'} \ueqimmut \Omega'\\\\
    v' = \begin{cases}
      \esloan {\ell} v'' & \text { if } v = \esloan{\ell} v'' \cr
      \esloan {\ell} v & \ell \text { fresh otherwise}
    \end{cases}
  }{
    \Omega \vdash \ebrw{p} \exprres{\esborrow\ell}{\Omega'}
  }
\IfBooleanT{#1}{

  \inferrule[E-ReservedBorrow \llbconly]{
    \hypertarget{E-ReservedBorrow} 
    \vdash \Omega(p) \eqmut v \\
    \bot, \emloan{}{}, \erborrow{} \notin v\\\\
    \updtstateplace{\Omega}{p}{v'} \ueqmut \Omega'\\\\
    v' = \begin{cases}
      \esloan {\ell} v'' & \text { if } v = \esloan{\ell} v'' \cr
      \esloan {\ell} v & \ell \text { fresh otherwise}
    \end{cases}
  }{
    \Omega \vdash \erbrw{p} \exprres{\erborrow\ell}{\Omega'}
  }
}

  \inferrule[E-Move]{
    \vdash \Omega(p) \eqmove v \\\\
    \IfBooleanF{#1}{\bot, \kw{loan}^{s,m} \not\in v}
    \IfBooleanT{#1}{\bot, \kw{loan}^{s,m}, \kw{borrow}^r \not\in v}\\\\
    \updtstateplace{\Omega}{p}{\bot} \ueqmove \Omega'
  }{
    \Omega \vdash \emove p \exprres{v}{\Omega'}
  }
\IfBooleanT{#1}{

  \inferrule[E-Copy]{
    \vdash \Omega(p) \eqimmut v \\
    \bot, \emloan\relax, \kw{borrow}^{m,r} \not\in v \\\\
    \vdash \ecopy v = v'
  }{
    \Omega \vdash \ecopy p \exprres{v'}{\Omega}
  }


  \inferrule[E-Pair]{
  \Omega_0 \vdash op_0 \exprres{v_0}{\Omega_1}\\\\
  \Omega_1 \vdash op_1 \exprres{v_1}{\Omega_2}
  }{
  \Omega_0 \vdash (op_0, op_1) \exprres{(v_0, v_1)}{\Omega_2}
  }

  \inferrule[E-Sum-Left]{
  \Omega \vdash op \exprres{v}{\Omega'}
  }{
  \Omega \vdash \kw{Left}\;op \exprres{\eleft{v}}{\Omega'}
  }

  \inferrule[E-Sum-Right]{
  \Omega \vdash op \exprarrow v \dashv \Omega'
  }{
  \Omega \vdash \kw{Right}\;op \exprres{\eright{v}}{\Omega'}
  }
  }
}

\NewDocumentCommand{\LlbcCopyRules}{s}{
\IfBooleanT{#1}{
    \inferrule[Copy-SharedBorrow]{
    }{
    \vdash \ecopy \esborrow\ell = \esborrow\ell
    }

    \inferrule[Copy-SharedLoan]{
    \vdash \ecopy v = v'
    }{
    \vdash \ecopy \esloan{\ell}{v} = v'
    }

    \inferrule[Copy-Scalar]{
    v \text{ literal value (boolean, integer, etc.)}
    }{
    \vdash \ecopy v = v
    }

    \inferrule[Copy-Sum]{
    C = \kw{Left} \eor C = \kw{Right}\\\\
    \vdash \ecopy v = v'
    }{
    \vdash \ecopy C\;v = C\;v'
    }

    \inferrule[Copy-Pair]{
    \vdash \ecopy v_0 = v'_0\\\\
    \vdash \ecopy v_1 = v'_1
    }{
    \vdash \ecopy (v_0,\,v_1) = (v'_0,\,v'_1)
    }
  }
}

\NewDocumentCommand{\LlbcStatementsEvalRules}{s}{
  \inferrule[E-Reorg]{
    \Omega_0 \hookrightarrow \Omega_1\\
    \Omega_1 \vdash s \stmtres{r}{\Omega_2}
  }{
    \Omega_0 \vdash s \stmtres{r}{\Omega_2}
  }
\IfBooleanT{#1}{

  \inferrule[E-Return]{
  }{
    \Omega \vdash \ereturn \stmtres{\ereturn}{\Omega}
  }
}

  \inferrule[E-Panic]{
  }{
    \Omega \vdash \epanic \stmtres{\epanic}{\Omega}
  }

  \inferrule[E-Seq-Unit]{
    \Omega_0 \vdash s_0 \stmtres{()}{\Omega_1}\\\\
    \Omega_1 \vdash s_1 \stmtres{r}{\Omega_2}
  }{
    \Omega_0 \vdash s_0; s_1 \stmtres{r}{\Omega_2}
  }
\IfBooleanT{#1}{

  \inferrule[E-Break]{
  }{
    \Omega \vdash \ebreak i \stmtres{\ebreak i}{\Omega}
  }

  \inferrule[E-Continue]{
  }{
    \Omega \vdash \econtinue i \stmtres{\econtinue i}{\Omega}
  }

  \inferrule[E-Seq-Propagate]{
    r \in \{\ereturn, \epanic, \econtinue{i}, \ebreak{i}\}\\\\
    \Omega \vdash s_0 \stmtres{r}{\Omega'}
  }{
    \Omega \vdash s_0; s_1 \stmtres{r}{\Omega'}
  }

  \inferrule[E-IfThenElse-T]{
    \Omega \vdash \kop \exprres{v}{\Omega'}\\\\
    v = \kw{true} \vee v = \esloan{\ell}{\kw{true}}\\\\
    \Omega' \vdash s_1 \stmtres{r}{\Omega''}
  }{
    \Omega \vdash \eite{\kop}{s_1}{s_2} \stmtres{r}{\Omega''}
  }

  \inferrule[E-IfThenElse-F]{
    \Omega \vdash \kop \exprres{v}{\Omega'}\\\\
    v = \kw{false} \vee v = \esloan{\ell}{\kw{false}}\\\\
    \Omega' \vdash s_2\stmtres{r}{\Omega''}
  }{
    \Omega \vdash \eite{\kop}{s_1}{s_2} \stmtres{r}{\Omega''}
  }

  \inferrule[E-Match]{
    \vdash \Omega(p) \eqimmut v\\
    v = C\,v' \vee v = \esloan{\ell}{(C\,v')}\\\\
    (C = \kw{Left} \wedge s = s_1) \vee (C = \kw{Right} \wedge s = s_2)\\
    \Omega \vdash s \stmtres{r}{\Omega'}
  }{
    \Omega \vdash (\ematchsum{p}{s_1}{s_2}) \stmtres{r}{\Omega'}
  }
}

  \inferrule[E-Assign \llbconly]{
    \hypertarget{E-Assign} 
    \Omega \vdash rv \exprres{v}{\Omega'} \\
    \vdash \Omega'(p) \eqmut v_p \\
    v_p \text{ has no outer } \kw{loan}^{s,m} \\\\
    \updtstateplace{\Omega'}{p}{v} \ueqmut \Omega'' \\
    \Omega''' = \Omega'',\; \_ \rightarrow v_p
  }{
  \Omega \vdash \eassign p {rv} \stmtres{()}{\Omega'''}
  }
\IfBooleanT{#1}{

  \inferrule[E-Box-New \llbconly]{
    \hypertarget{E-Box-New} 
    \Omega \vdash op \exprres{v}{\Omega'}
  }{
    \Omega \vdash \enew op \stmtres{\ebox{v}}{\Omega'}
  }

  \inferrule[E-Box-Free]{
    \vdash \Omega(p) \eqmove \ebox v \\\\
    \Omega \vdash \eassign{*p}{\bot} \exprres{()}{\Omega'}\\\\
    \updtstateplace{\Omega}{p}{\bot} \ueqmove \Omega''
  }{
    \Omega \vdash \efree{p} \stmtres{()}{\Omega''}
  }

  \inferrule[E-PushStack]{
    \Omega' = \{\; \Omega\; \kw{with}\;
    \kw{env} = [\overrightarrow{x \rightarrow v}] \econcat \Omega.\kw{env},\;
    \kw{stack} = [\overrightarrow{x}] :: \Omega.\kw{stack}\;
    \}
  }{
    \pushstack [\overrightarrow{x \rightarrow v}]\; \Omega\; =\; \Omega'
  }

  \inferrule[E-PopStack]{
    \Omega_0.\kw{stack} = ([x_{ret}] \econcat [\overrightarrow{x_i}]) :: \kw{stack'}\\
    \forall\; i, \Omega_i \vdash x_i := \bot \exprres{()}{\Omega_{i+1}}\\
    \Omega_m.env = [x_{ret} \rightarrow v_{ret}] \econcat [\overrightarrow{x_i \rightarrow v_i}] \econcat \kw{env'}\\
    \bot, \kw{loan}, \kw{borrow}^r\not\in v_{ret} \\
    \Omega_{end} = \{\; \Omega_m\; \kw{with}\;
    \kw{stack}\; =\; \kw{stack'},\;
    \kw{env}\;=\;\kw{env'} \;\}
  }{
    \popstack \Omega\; =\; (v_{ret},\; \Omega_{end})
  }

  \inferrule[E-Call]{
    \overrightarrow{\rho}\text{ fresh}\\
    \feqsig{\vec{\_}}\\
    \forall\; j \in [0;\, m[, \Omega_j \vdash op_j \exprres{v_j}{\Omega_{j+1}}\\
    \pushstack
      \left([x_{\text{ret}} \rightarrow \bot] \econcat
      [\overrightarrow{x_j \rightarrow v_j}] \econcat
      [\overrightarrow{y_k \rightarrow \bot}]\right) \Omega_m = \Omega_{begin}\\
    \Omega_{\text{begin}} \vdash body \stmtres{r}{\Omega_{\text{end}}}\\
    (r',\, \Omega_1) = \begin{cases}
      (\epanic,\, \Omega_{\text{end}}) & \text { if } r = \epanic \cr
      ((),\; \Omega''_{\text{end}}) &
        \text { if } r = \ereturn \;\wedge
        \popstack \Omega_{\text{end}}\;=\; (v_{\text{ret}},\; \Omega'_{\text{end}})
        \;\wedge
        \Omega'_{\text{end}} \vdash \eassign p {v_{\text{ret}}} \stmtres{()}{\Omega''_{\text{end}}}
    \end{cases}
  }{
    \Omega_0 \vdash \eassign{p}{\fcall} \stmtres{r'}{\Omega_1}
  }
  }
}

\NewDocumentCommand{\LlbcLoopsEvalRules}{s}{
  \inferrule[E-Loop-Break-Inner]{
    \Omega \vdash s \stmtres{\ebreak{0}}{\Omega}
  }{
    \Omega \vdash \eloop{s} \stmtres{()}{\Omega}
  }

  \inferrule[E-Loop-Break-Outer]{
    \Omega \vdash s \stmtres{\ebreak{i + 1}}{\Omega}
  }{
    \Omega \vdash \eloop{s} \stmtres{\ebreak{i}}{\Omega}
  }

  \inferrule[E-Loop-Continue-Inner]{
    \Omega \vdash s \stmtres{\econtinue{0}}{\Omega'}\\\\
    \Omega' \vdash \eloop{s} \stmtres{r}{\Omega''}
  }{
    \Omega \vdash \eloop{s} \stmtres{r}{\Omega''}
  }

  \inferrule[E-Step-Loop-Continue-Inner]{
    \Omega \vdash s \stmtstepres{n}{\econtinue{0}}{\Omega'}\\\\
    \Omega' \vdash \eloop{s} \stmtstepres{n}{r}{\Omega''}
  }{
    \Omega \vdash \eloop{s} \stmtstepres{n+1}{r}{\Omega''}
  }

  \inferrule[E-Loop-Continue-Outer]{
    \Omega \vdash s \stmtres{\econtinue{i + 1}}{\Omega}
  }{
    \Omega \vdash \eloop{s} \stmtres{\econtinue{i}}{\Omega}
  }

  \inferrule[E-Loop-Panic]{
    \Omega \vdash s \stmtres{\epanic}{\Omega}
  }{
    \Omega \vdash \eloop{s} \stmtres{\epanic}{\Omega}
  }

  \inferrule[E-Loop-Return]{
    \Omega \vdash s \stmtres{\ereturn}{\Omega}
  }{
    \Omega \vdash \eloop{s} \stmtres{\ereturn}{\Omega}
  }
}

\NewDocumentCommand{\LlbcReorgRules}{s}{
    \inferrule[Reorg-End-MutBorrow]{
      \IfBooleanT{#1}{
      \text{hole of } \Omega[\emloan{\ell},\, .]
        \text{ not inside a borrowed value or an abs.}}
      \IfBooleanF{#1}{
      \text{hole of } \Omega[\emloan{\ell},\, .]
        \text{ not inside a borrow}}\\\\
      \IfBooleanT{#1}{\text{$\kw{loan}^{s,m}$, $\erborrowv$} \not\in v}
      \IfBooleanF{#1}{\text{$\kw{loan}^{s,m}$} \not\in v}
    }{
      \Omega[\emloan{\ell},\, \emborrow{\ell}{v}] \hookrightarrow \Omega[v,\bot]
    }
\IfBooleanT{#1}{

    \inferrule[Reorg-End-SharedReservedBorrow]{
      \text{hole of } \Omega[.]
        \text{ is not inside a borrowed value or an abs.}\\\\
      \esloan{\ell}{v'} \in \Omega[v]\\
      v = \esrborrow{\ell}
    }{
      \Omega[v] \hookrightarrow \Omega[\bot]
    }

    \inferrule[Reorg-End-SharedLoan]{
      \esrborrow{\ell} \not\in \Omega[\esloan{\ell}{v}]
    }{
      \Omega[\esloan{\ell}{v}] \hookrightarrow \Omega[v]
    }

    \inferrule[Reorg-Activate-Reserved]{
      \kw{loan},\kw{borrow}^r \notin v\\\\
      \text{hole of }\Omega[., \erborrow{\ell}]\text{ not inside a shared value or a region abstraction}\\\\
      \ell \notin \Omega[.,.], v
    }{
      \Omega[\esloan{\ell}{v}, \erborrow{\ell}] \hookrightarrow\\\\
      \Omega[\emloan{\ell}, \emborrow{\ell}{v}]
    }

    \inferrule[Reorg-Seq]{
      \Omega_0 \hookrightarrow \Omega_1\\
      \Omega_2 \hookrightarrow \Omega_2
    }{
      \Omega_0 \hookrightarrow \Omega_2
    }

    \inferrule[Reorg-None]{
    }{
      \Omega \hookrightarrow \Omega
    }
  }
}

\NewDocumentCommand{\LlbcReadRules}{s}{
    \inferrule[Read]{
    p = P[x]\\
    \Omega(x) = v\\\\
    \Omega \vdash P(v) \eqk v'
    }{
    \vdash \Omega(p) \eqk v'
    }

    \inferrule[R-Base]{
    }{
    \Omega \vdash [.](v) \eqk v
    }
\IfBooleanT{#1}{

    \inferrule[R-ProjSum-Left]{
    \Omega \vdash P(v) \eqk v'
    }{
    \Omega \vdash P((\kw{Left}\;v).0) \eqk v'
    }

    \inferrule[R-ProjSum-Right]{
    \Omega \vdash P(v) \eqk v'
    }{
    \Omega \vdash P((\kw{Right}\;v).0) \eqk v'
    }

    \inferrule[R-ProjPair-Left]{
    \Omega \vdash P(v) \eqk v'
    }{
    \Omega \vdash P((v, v_1).0) \eqk v'
    }

    \inferrule[R-ProjPair-Right]{
    \Omega \vdash P(v) \eqk v'
    }{
    \Omega \vdash P((v_0, v).1) \eqk v'
    }

    \inferrule[R-SharedLoan]{
    P \neq [.]\\
    \Omega \vdash P(v) \eqimmut v'
    }{
    \Omega \vdash P(\esloan{\ell}{v}) \eqimmut v'
    }

    \inferrule[R-Deref-Box]{
    \Omega \vdash P(v) \eqk v'
    }{
    \Omega \vdash P(*(\ebox{v}) \eqk v'
    }
}

    \inferrule[R-Deref-SharedBorrow]{
    \esloan{\ell}{v} \in \Omega\\\\
    \Omega \vdash P(v) \eqimmut v'
    }{
    \Omega \vdash P(*(\esborrow{\ell})) \eqimmut v'
    }

\IfBooleanT{#1}{
    \inferrule[R-Deref-MutBorrow]{
    \Omega \vdash P(v) \eqimmutmut v'
    }{
    \Omega \vdash P(*(\emborrow{\ell}{v})) \eqimmutmut v'
    }
  }
}

\NewDocumentCommand{\LlbcWriteRules}{s}{
    \inferrule[Write]{
    p = P[x]\\
    \Omega(x) = v\\\\
    \Omega \vdash P(v) \updtk w \wk{v'}{\Omega}\\\\
    \Omega'' = (\Omega'(x) := v')
    }{
    \vdash \Omega(p) \updtk w \eqk \Omega''
    }

    \inferrule[W-Base]{
    }{
      \IfBooleanT{#1}{\Omega \vdash [.](v : \tau) \updtk (w : \tau) \wk{w}{\Omega}}
      \IfBooleanF{#1}{\Omega \vdash [.](v) \updtk w \wk{w}{\Omega}}
    }
\IfBooleanT{#1}{

    \inferrule[W-ProjSum-Left]{
    \Omega \vdash P(v) \updtk w \wk{v'}{\Omega'}
    }{
    \Omega \vdash P((\eleft{v}).0) \updtk w\\\\
      \wk{\eleft{v'}}{\Omega'}
    }

    \inferrule[W-ProjSum-Right]{
    \Omega \vdash P(v) \updtk w \ueqk v' \dashv \Omega'
    }{
    \Omega \vdash P((\eright{v}).0) \updtk w\\\\
      \wk{\eright{v'}}{\Omega'}
    }

    \inferrule[W-ProjPair-Left]{
    \Omega \vdash P(v) \updtk w \wk{v'}{\Omega'}
    }{
    \Omega \vdash P((v, v_1).0) \updtk w\\\\
      \wk{(v', v_1)}{\Omega'}
    }

    \inferrule[W-ProjPair-Right]{
    \Omega \vdash P(v) \updtk w \wk{v'}{\Omega'}
    }{
    \Omega \vdash P((v_0, v).1) \updtk w\\\\
      \wk{(v_0, v')}{\Omega'}
    }

    \inferrule[W-SharedLoan]{
    P \neq [.]\\
    \Omega \vdash P(v) \updtimmut w \wimmut{v'}{\Omega'}
    }{
    \Omega \vdash P(\esloan{\ell}{v}) \updtimmut w \wimmut{\esloan{\ell}{v'}}{\Omega'}
    }

    \inferrule[W-Deref-Box]{
    \Omega \vdash P(v) \updtk w \wk{v'}{\Omega'}
    }{
    \Omega \vdash P(*(\ebox{v})) \updtk w
      \wk{\ebox{v'}}{\Omega'}
    }

    \inferrule[W-Deref-SharedBorrow]{
    \esloan{\ell}{v} \in \Omega\\
    \Omega \vdash P(\esloan{\ell}{v}) \updtimmut w
      \wimmut{v'}{\Omega'[\esloan{\ell}{v''}]}\\
    \Omega'' = \Omega'[\esloan{\ell}{v'}]
    }{
    \Omega \vdash P(*(\esborrow{\ell})) \updtimmut w
      \wimmut{\esborrow{\ell}}{\Omega''}
    }

    \inferrule[W-Deref-MutBorrow]{
    \Omega \vdash P(v) \updtimmutmut w \wimmutmut{v'}{\Omega'}
    }{
    \Omega \vdash P(*(\emborrow{\ell}{v})) \updtimmutmut w
      \wimmutmut{\emborrow{\ell}{v'}}{\Omega'}
    }
  }
}

\section{A Generic Approach to Proving Language Simulations}
\slabel{proofmeth}

As is standard in this kind of work, we go through several intermediate
languages in order to relate our high-level (\llbcs) and low-level (PL)
languages (\Figure{architecture}). This allows for modular reasoning, where each step focuses on a
particular semantic concept.

To conduct these various refinement steps, we design a
new, reusable proof methodology that allows
efficiently establishing simulations between languages via the use of local and
pointwise reasoning. We use this generic methodology repeatedly throughout
the rest of the paper, so as to make our various reasoning steps much more effective;
we now present the idea in the abstract, and apply it to our use-cases in
subsequent sections (\Section{refine}, \Section{llbc-sharp}).

Our approach consists of a
generic proof template for establishing simulations between two
languages $L_l$ and $L_h$ that share a grammar of statements,
but whose semantics and notions of states differ.
To simplify the presentation, we focus on forward simulations, but the methodology can be
easily adapted to work dually for backward simulations. We briefly touch on this
at the end of this section, and later discuss why we
focus on forward simulations in this work, in \Section{fw-vs-bw}.

Formally, we write $\Omega_l$, $\Omega_h$ for low-level and high-level
states respectively; $\le$, for a given relation that \emph{refines} a high-level state
into a low-level state, i.e., $\Omega_l \le \Omega_h$; and
$\Omega_l \vdash_l e \arrowThree_l (e', \Omega_l')$ (respectively, $h$), for the
reduction of expressions into another expression and resulting state.
We consider variations of such a relation throughout the paper, to relate
the reduction of statements to \emph{outcomes}~\cite{c-hol, java-verif} used to model non-local control-flow (e.g.,
\li+return+, \li+break+), or the reduction of expressions to values, in which
case we consider a relation $\le$ which refines pairs of high-level states and values to
low-level states and values.
We assume $\arrowThree$ defines a big-step reduction over which one can perform inductive
reasoning; we leave the problem of adapting this proof technique to small-step semantics
for future work.
We may omit some of the indices when clear from the context.

We aim to establish (variations of) the following property:
%
%
%
\begin{definition}[Forward Simulation]
For all $\Omega_l$, $\Omega_h$, $\Omega_h'$ states, $e$, $e'$ expressions, we have:
\begin{align*}
\iflong
&  \Omega_l \le \Omega_h \Rightarrow\\
&  
    \Omega_h \vdashLtwo\, e \arrowThree_h (e', \Omega_h') \Rightarrow\\
&  \exists\; \Omega'_l,\,
    \Omega_l \vdashLone\, e \arrowThree_l (e', \Omega_l') \wedge \Omega'_l \le \Omega'_h
\fi
\end{align*}
\thlabel{fwdsim}
\thmvspace
\end{definition}
Proving this property in a direct fashion can be tedious.
When states $\Omega_l$ and $\Omega_h$ heavily differ, for instance because they operate at different levels of abstraction,
the state relation commonly consists of complex global invariants, which are
tricky to both correctly define and reason about, and oftentimes require
maintaining auxiliary data structures, such as maps between high and low, for
the purposes of the proof.
To circumvent this issue, our approach relies on two key components. First,
instead of a global relation between states $\Omega_l$ and $\Omega_h$, we
define a set of small, local rules, the transitive closure of which
constitutes $\le$~(\Section{transformations}). These can then be reasoned
upon individually.
%
%
%
Second, we add the ability to reason about states that contain a mixture of high
and low, which we dub ``hybrid''~(\Section{unionlang}). These now occur
because our rewritings operate incrementally,
and thus may give rise to states that belong neither to $L_l$ nor $L_h$. Since
the proof of forward simulation involves an induction on
$\arrowThree$, we now must reason about the reduction of terms in hybrid states.
Note that, if the empty states in $L_l$ and $L_h$ are related, we retrieve
standard simulation properties, namely that the execution of a closed program
in $L_l$ is related to its execution in $L_h$. This will be the case for the
state relations in \Section{refine} and \Section{llbc-sharp}.


\subsection{Local State Transformations}
\slabel{transformations}

%
To illustrate the idea of state relations based on local transformations,
we take the simplistic example of a state that
contains two integer variables $x$ and $y$; in $\Omega_l$, $x$ and $y$ contain concrete
values; in $\Omega_h$, $x$ and $y$ contain symbolic (or abstract) values. This is
a much-simplified version of the full proof we later develop in
§\sref{llbc-sharp}; the setting of abstract/concrete values makes it easy to
illustrate the concept.
In this example, our goal is to
prove that $L_h$ implements a sound symbolic execution for $L_l$. Instead of
defining a global relation using universal quantification on all variables, we
instead define a local relation that, for a given variable, swaps its 
concrete value $n$ and its corresponding abstract version $\sigma$ in $L_h$. We concisely write
$\Omega[n] \le \Omega[\sigma]$, leveraging a state-with-holes notation for
the state: $\Omega[.]$ is a state with one hole, while $\Omega[v]$ is the same state where
the hole has been filled with value $v$.
%
Establishing $L_l \le L_h$ involves repeatedly applying this relation;
to either $x$ followed by $y$, or $y$ followed by $x$; the
locality of the transformation enables us to reason modularly about both
variables. Naturally, once the reasoning becomes more complex, the ability to
consider a single transformation at a time is crucial.

We remark that these individual transformations are non-directed, and can be
read either left-to-right or right-to-left. In our example, when read
left-to-right, we have an abstraction; when read right-to-left we have a
concretization. This supports our later claim that this methodology works for
both forward and backward simulations.

\subsection{Reasoning over a Superset Language}
\slabel{unionlang}

By defining the state relation $\le$ as the reflexive, transitive closure of
local relations,
we can now attempt to prove forward simulation by a standard induction on the evaluation in $L_h$,
followed by an induction on the $\le$ relation.
However, one problem arises:
intermediate states do not belong to either language, and hence do not have
semantics.
To see why, consider in our proof the induction step corresponding to the transitivity
of the relation, where we have an intermediate state $\Omega_m$ such that
$\Omega_l \le \Omega_m \le \Omega_h$,
and we want to use an induction hypothesis on $\Omega_m \le \Omega_h$. We do so in
order to establish that if $\Omega_h \vdashLtwo e \arrowThree_h (e', \Omega_h')$ (for some
$e'$, $\Omega_h'$) and $\Omega_m \le \Omega_h$, then
there exists $\Omega_m'$ such that $\Omega_m \vdash e \arrowThree (e', \Omega_m')$
 and $\Omega_m' \le \Omega_h'$ -- this is the induction on the size of $\le$
for the
purposes of establishing the forward simulation. This is fine, except reduction
is not defined for hybrid states like $\Omega_m$ which contain a mixture of
high-level and low-level.
%
To address this issue, we instead consider a superset language $L^+$ that contains
semantics from both $L_l$ and $L_h$.

Coming back to our previous example, let us assume that we want to update variable $x$.
If $x$ has already been rewritten, we use the
corresponding semantic rule in $L_h$,
otherwise we execute the program according to the semantics in $L_l$.
For our simplistic example, no further rules are needed, and the strict union
suffices, i.e. $L^+ = L_l \cup L_h$. We will see shortly that this strict union
is not always adequate in the general case.

With this approach, we establish the following theorem for two concrete languages
$L_l$ and $L_h$ (note that in the following $\le$ will always refer to the reflexive,
transitive closure relation):
%

\begin{theorem}[Forward Simulation on Superset Language]
For all $L^+$ states $\Omega_1$, $\Omega_2$, $\Omega_2'$, expressions $e$, $e'$, we have:
  \begin{align*}
  \iflong
  &\Omega_1 \le \Omega_2 \Rightarrow\\
  &
    \Omega_2 \vdashLplus\, e \Downarrow_+ (e', \Omega_2') \Rightarrow\\
  &\exists\; \Omega'_1,\,
    \Omega_1 \vdashLplus\, e \Downarrow_+ (e', \Omega_1') \wedge \Omega'_1 \le \Omega'_2
  \fi
  \end{align*}
\thlabel{fwdsimunify}
\thmvspace
\end{theorem}

Unfortunately, instantiating this theorem with states $\Omega_1 = \Omega_l \in
L_l$ and $\Omega_2 = \Omega_h \in L_h$
does not allow us to derive Theorem~\thref{fwdsim}, which was our initial goal.
Indeed, $\Omega_1$ initially belongs to $L_l$ but reduces with $\Downarrow_+$,
i.e., with the union of the semantics, and so far nothing allows us to conclude
that the resulting $\Omega_1'$ is still in $L_l$.
%
To ensure that this execution is valid with respect to $L_l$, we need to restrict $L^+$ by
excluding a set of rules $R$ from $L_h$ so that $L_l$ and $L^+$ satisfy the
following property.

\begin{definition}[Stability]
\deflabel{stable}
Given two languages $L_l, L^+$ such that $L^+$ is a superset of $L_l$, we say
that $L_l$ is a stable
subset of $L^+$ if, for all $e, e', \Omega \in L_l, \Omega^+ \in L^+$, if $\Omega
\vdashLplus\, e \Downarrow_+ (e', \Omega^+)$,
then $\Omega^+ \in L_l$ and $\Omega \vdashLone\, e \Downarrow_l (e', \Omega^+)$
  \deflabel{stability}
\end{definition}

Combined with stability, Theorem~\thref{fwdsimunify} allows us to directly
derive that $L_l$ and $L^+$ satisfy a forward simulation relation. To conclude,
the last remaining step is to establish the same property between $L^+$ and
$L_h$, which only requires reasoning about the excluded rule set $R$.

The earlier simplistic example of concrete/abstract integers requires no
particular care when defining $L_+$, and the union of the rules suffices (i.e.,
$R = \varnothing$). But
for our real use-case, one can see from
\Figure{architecture} that the semantics of \hlplp exclude some of the rules from
LLBC. We also point out that for our real use-cases, the superset $L_+$ requires
additional administrative rules, to make sure high and low compose (as we will
see shortly).

\myparagraph{Forward \emph{vs.} backward}
While the presentation focused on forward simulations, the approach can be
easily adapted to backward simulations. Leveraging the duality between both
relations, it is sufficient to exclude a set of rules $R$ from $L_l$
instead of $L_h$ so that $L_h$ becomes a stable subset of $L^+$, and to
similarly conclude by reasoning over the rules in $R$.

\section{A Heap-and-Addresses Interpretation of Valued Borrows}
\slabel{refine}

Equipped with our generic proof methodology, we now turn to the Low-Level Borrow Calculus (LLBC), first introduced by~\citet{aeneas},
which aims to provide an operational semantics for a core subset of the Rust language.
To remain conceptually close to Rust, LLBC operates on states containing loans and borrows.
While this helps understanding and explaining the language from the programmer's perspective,
this model departs from standard operational semantics for heap-manipulating programs,
which commonly rely on a low-level model based
on memory addresses and offsets~\cite{leroy2009cacm,leroy2012compcertmodel,jung2017rustbelt}.
In this section, we thus bridge the abstraction gap,
by relating LLBC to a standard, low-level operational semantics, therefore establishing
LLBC as a sound semantic foundation for Rust programs. 
To do so, we introduce the Pointer Language (PL),
a small language explicitly modeling the heap using a model inspired by CompCert's C memory model,
and establish a relation between LLBC and PL that demonstrates that LLBC's borrow-centric view
of memory is compatible with classic pointers.

We start this section by presenting selected rules from LLBC exhibiting the core concepts, before applying
the methodology from \Section{proofmeth} to prove a simulation relation between LLBC and PL.
Due to space constraints, we only present representative rules of the different languages that exhibit
the salient parts of the proofs; the complete presentation is available in the appendix.

\subsection{Background: Low-Level Borrow Calculus (LLBC)}
\slabel{background-llbc}

\begin{figure}[!ht]
  \LlbcSyntax
  \caption{The Low-Level Borrow Calculus (LLBC): Syntax (Selected Constructs).} %
  \flabel{llbc-syntax} %
  \figvspace
\end{figure} %


LLBC aims to model a core subset of the Rust language after desugaring
(\Figure{llbc-syntax}),
where for instance move and copy operations are explicit; conceptually, it is close
to Rust's Mid-level IR (MIR). To provide a quick overview of LLBC, we will rely
on the following running example, inherited from~\citet{aeneas},
which exhibits
several salient Rust features, namely, mutable borrows and reborrows.
At each program point, we annotate this example with the corresponding LLBC state (we use the terms environments and states
interchangeably),
which we present below.

\begin{minted}[xleftmargin=1.5em,mathescape=true,escapeinside=||]{rust}
x  = 0;          // $x \mapsto 0$ $\lnlabel{x-zero}$
px = &mut x;     // $x \mapstoemloan{0},\quad px \mapstoemborrow{0}{0}$ $\lnlabel{borrow}$
px = &mut (*px); // $x \mapstoemloan{0},\quad \_ \mapstoemborrow{0}{(\emloan{\ell_1})},\quad px \mapstoemborrow{1}{0}$$\lnlabel{reborrow}$
assert!(x = 0);  // $x \mapsto 0,\quad \_ \mapsto \bot,\quad px \mapsto \bot$$\lnlabel{assert}$
\end{minted}

\myparagraph{Environments (or States)}
LLBC relies on a borrow-centric view of values, and operates on environments that map variables
to their corresponding values. We present below an excerpt of LLBC's grammar of values; the
full version also includes support for sums, pairs, recursive types, and reserved borrows
which are used to model two-phase borrows~\cite{two-phase-borrows}.
%
$$
v := n~|~\bot~|~\emloan{\ell}~|~\emborrow{\ell}{v}~|~\esloan{\ell}{v}~|~\esborrow{\ell}
$$
The most interesting values consist of borrows and loans, which can either be shared
(annotated with the exponent $s$) or mutable (annotated with the exponent $m$). Both
borrows and loans are associated to a loan identifier $\ell$, which can be seen as
an abstract notion of location. A loan identifier uniquely identifies a \emph{loan} value,
while different borrows may refer to the same identifier in the case of shared borrows.
Additionally, mutable borrows carry the borrowed value $v$.
Other values include constants $n$ (e.g., integers), as well as the $\bot$ value which
represents both uninitialized and invalidated values, and is needed to model terminating borrows
as well as Rust's move semantics.

\begin{figure}
  \centering
  \smaller
  \begin{mathpar}
    \LlbcRvaluesEvalRules
  
    \LlbcCopyRules
  
    \LlbcStatementsEvalRules
  
    \LlbcReorgRules
    
    \LlbcReadRules
    
    \LlbcWriteRules
  \end{mathpar}
  \caption{Operational Semantics for LLBC (Selected Rules)}
  \flabel{llbc-selected}
  \figvspace
\end{figure}

\myparagraph{Semantics}
%
LLBC relies on the operational semantics presented
in \Figure{llbc-selected}. The judgment $\Omega \vdash s \stmtres{r}{\Omega'}$
models a big-step semantics where executing statement $s$ in the initial environment
$\Omega$ yields a new environment $\Omega'$, and a control-flow outcomes
$r \in \{(), \ereturn,\epanic,\econtinue i,\ebreak i\}$.
The judgment $\Omega \vdash rv \exprres{v}{\Omega'}$ is similar,
but operates on assignable rvalues (that we also refer to as \emph{expressions})
and returns values.

Initially, the variable $x$ is declared to be 0, and the corresponding mapping is added to
the environment (line~\lnref{x-zero}). To execute the mutable borrow at line~\lnref{borrow}, we turn to
the rule \Rule{E-MutBorrow}.

The first premise of the rule ($\vdash \Omega(p) \eqmut v$)
retrieves the value $v$ associated to a path $p$ in state $\Omega$.
Accesses (\Rule{Read}) and updates (\Rule{Write}) to environments are annotated
with one of $\kw{mut}, \kw{imm}$ or $\kw{mov}$. These capabilities refine the
behavior of $\Rightarrow$: for instance, the creation of shared borrows
(\Rule{E-SharedBorrow}) allows following existing shared borrow indirections
along path $P$ and uses the $\kw{imm}$ (``immutable'') capability (\Rule{R-Deref-SharedBorrow});
but moving (\Rule{E-Move}) won't follow dereferences and uses the $\kw{mov}$ (``move'') capability:
one can't move a value under a borrow.
%
Once the value is retrieved, LLBC creates a fresh loan identifier $\ell$, and
sets the value associated to $p$ to $\emloan{\ell}$, modeling that $p$ has been
borrowed. The mutable borrow is finally evaluated to $\emborrow{\ell}{v}$

The pattern at line~\lnref{reborrow} is known as a reborrow, and is frequently introduced by the Rust
compiler when desugaring the user-written code.
After inserting a fresh loan identifier (\Rule{E-MutBorrow}), the right-hand side
of the assignment reduces to $\emborrow{\ell_1}{0}$, and $px$
\emph{temporarily} maps to
$\emborrow{\ell_0}{(\emloan{\ell_1})}$ (\Rule{E-Assign}, $\Omega'$).
We now assign $\emborrow{\ell_1}{0}$ to $px$ (\Rule{E-Assign},
$\Omega''$), and doing so, we do not directly override the value at $px$ but save it in a ghost,
anonymous value to remember the reborrow information that we will need to use
later when ending, e.g., $\ell_0$ (\Rule{E-Assign}, $\Omega'''$).

We now evaluate the \li+assert+, which is syntactic sugar for \li+if not ... then panic+.
We need to read $x$, but its value has been mutably
borrowed ($\emloan{\ell_0}$) and is thus not accessible (there is no
corresponding $\Rightarrow$ rule to read from a loan): we need to end the borrow
$\ell_0$. We do this during a reorganization phase, by which we can arbitrarily end
borrows before evaluating any statement (\Rule{E-Reorg}).
One has to note that it is always possible to end a borrow early, but this may lead
to an unnecessarily stuck execution, if, e.g., we later need to dereference this borrow.
We attempt to use \Rule{Reorg-End-MutBorrow} by which we can reinsert a mutably borrowed
value back into its loan. However, this rule requires that we have full access
to the borrowed value (premise $\kw{loan}^{s,m} \not\in v$, which states that $v$
doesn't contain any loans, shared or mutable) which is not
the case because the value was reborrowed through $\ell_1$. We thus apply
\Rule{Reorg-End-MutBorrow} on $\ell_1$ first, which yields the environment
$x \mapstoemloan{0},\; \_ \mapstoemborrow{0}{0},\; p \mapsto\bot$.
We finally apply \Rule{Reorg-End-MutBorrow} again on $\ell_0$, which yields the final environment.

\myparagraph{Shared Borrows}
We briefly describe shared borrows, which are simpler than
mutable borrows and do not pose a challenge in our proof of soundness.
The value of a shared borrow remains attached to the \emph{loan}, as no
modifications are permitted anyhow through the shared borrows.
\Rule{E-SharedBorrow} creates a new fresh loan if
there isn't one at the path $p$ already, and \Rule{R-Deref-SharedBorrow} merely
looks up the value at path $p$ in order to access the corresponding value $v'$.
As mentioned earlier, reading \emph{through} shared loans is permitted since
we don't modify the value, meaning we don't have to end the corresponding borrows to
evaluate the \li+assert+.
\iflong
We updated the example from above to use shared borrows, and annotated each
program point with the resulting LLBC state.
\begin{minted}[xleftmargin=1.5em,mathescape=true,escapeinside=||]{rust}
x  = 0;         // $x \mapsto 0$
px = &x;        // $x \mapsto \esloan{\ell}{0},\quad px \mapsto \esborrow{\ell}$$\lnlabel{shared-borrow}$
px = &(*px);    // $x \mapsto \esloan{\ell}{0},\quad \_ \mapsto \esborrow{\ell},\quad px \mapsto \esborrow{\ell}$$\lnlabel{shared-reborrow}$
assert!(x = 0); // $x \mapsto \esloan{\ell}{0},\quad \_ \mapsto \esborrow{\ell},\quad px \mapsto \esborrow{\ell}$
\end{minted}
\fi

\myparagraph{Earlier Formalism}
A reader familiar with the original presentation of LLBC might have noticed
several minor differences compared to the version presented in this paper.
As part of our work on relating LLBC to a lower-level language, we identified several
improvements to simplify both the formalism and the proofs. Of particular note, instead
of a set of loan identifiers, shared loans now carry a single
identifier.
We found through our formal analysis that the former was not needed, and that the latter
supports a more local style of reasoning that avoids non-local lookup-then-update
operations on the originating loan.
Another improvement concerns the capabilities for the read and write judgments;
not only does this lead to better reuse of rules (no need for separate
judgments),
it also makes some rules more explicit.
%
We also clarified notations to distinguish between reductions operating on expressions
(i.e., rvalues), statements, and environments. Finally, we also extended the grammar
of LLBC and the control-flow outcomes to include support for loops.

\subsection{Simulation Proof}

Our main objective in this section is to prove that LLBC accurately models the Rust semantics, that is,
that every execution at the LLBC level admits a corresponding execution in a more standard, heap-manipulating
language that we dub the Pointer Language (PL).

\myparagraph{Forward \emph{vs} Backward}
\slabel{fw-vs-bw}
In our proof, the source and target language are the same -- we are not doing a
compiler correctness proof. Rather, we reconcile two models of execution over
the same syntax of programs, i.e., given a PL state that
\emph{concretizes} the LLBC state, the program computes in PL the same result as
in LLBC. This is similar to the original circa-2008 style of CompCert proofs,
and qualifies as a forward simulation~\cite{leroy2009cacm}.

Indeed, at this stage, it is not true that every PL execution can be simulated backwards
by an LLBC execution; simply said, a PL program could be safe for reasons that
cannot be explained by LLBC's borrow semantics. We will see in the next section
(\Section{backllbcpl}) how we can obtain the backwards direction, provided the
program is borrow-checked -- a result that is akin to a typing result. For now,
we simply seek to establish that the execution model of LLBC is a correct restriction
of a traditional heap-and-addresses model. To that end, we use the
methodology from \Section{proofmeth}, and for now only use the forward direction
it gives us.

\myparagraph{Difficulties and methodology}
Instead of environments containing borrows and loans, PL operates on an explicit heap
adapted from the CompCert memory model~\cite{leroy2012compcertmodel}. In particular, loan identifiers are replaced by memory addresses,
consisting of a block identifier and an offset, and the memory layout is made semi-explicit,
by including a notion of \emph{size} for each type.

When attempting to relate LLBC to a lower-level language like PL, two main difficulties arise.
First, manipulating an explicit heap requires operating on sequences of words, which need to
be reconciled with more abstract values. Second, one needs to relate borrows and loans
to low-level pointers. The core of the difficulty lies in the fact that in LLBC,
mutable borrows ``carry'' the value they borrow, making it hard to reason about
the provenance of a value in the presence of reborrows.
%
%
%
This is where our methodology comes in (\Section{proofmeth}).
To make the proof simpler and more modular, we proceed in two steps, via the
addition of an intermediary language dubbed HLPL.


\myparagraph{An intermediary language: HLPL (High-Level Pointer Language)}
As before, the program syntax remains the same; what differs now is that HLPL
states are half-way between PL states and LLBC states. HLPL states no longer
feature borrows and loans; but instead of manipulating a heap they retain an abstract notion of \emph{pointers} and
\emph{locations}, denoted respectively $\eptr{\ell}$ and $\eloc{\ell}{v}$ (see below).

The simulation from HLPL to PL is standard and only consists in materializing
locations as a global heap, and mapping pointers to corresponding addresses. We
instead focus on where the challenge lies, namely, going from HLPL (value is
with the location) to LLBC (value is with the borrow). Furthermore, we focus
specifically on the case of mutable borrows: shared borrows have the value
attached to the loan, meaning that the crucial discrepancy only appears in the
case of mutable borrows.

\NewDocumentCommand{\HlplRules}{s}{
  \inferrule[E-Pointer]{
    \begin{array}{l}
    \vdash \Omega(p) \eqimmutmut v \cr
    \vdash \Omega(p) \updtimmutmut v' \eqimmutmut \Omega'
    \end{array}\\
    v' = \begin{cases}
      \eloc {\ell} v'' & \text { if } v = \eloc{\ell} v'' \cr
      \eloc {\ell} v & \ell \text { fresh otherwise}
    \end{cases}
  }{
    \Omega \vdash
      \{\ebrw p,\, \embrw p \IfBooleanT{#1}{,\, \erbrw{p}} \}
    \exprres{\eptr{\ell}}{\Omega'}
  }

  \inferrule[Reorg-End-Pointer]{
  }{
    \Omega[\eptr{\ell}] \hookrightarrow \Omega[\bot]
  }

  \inferrule[Reorg-End-Loc]{
    \eptr\ell \not\in \Omega[\eloc{\ell}{v}]
  }{
    \Omega[\eloc{\ell}{v}] \hookrightarrow \Omega[v]
  }
\IfBooleanT{#1}{

  \inferrule[HLPL-E-Move]{
    \vdash \Omega(p) \eqmove v \\\\
    \bot, \kw{loc}\not\in v \\\\
    \updtstateplace{\Omega}{p}{\bot} \ueqmove \Omega'
  }{
    \Omega \vdash \emove p \exprres{v}{\Omega'}
  }

  \inferrule[HLPL-E-Box-New]{
    \Omega \vdash op \exprres{v}{\Omega'}\\
    l^b \text{ fresh}
  }{
    \Omega \vdash \enew op \exprres{\eptr{\ell}}{(\Omega',\, l^b \rightarrow v)}
  }

  \inferrule[HLPL-E-Box-Free]{
    \vdash (\Omega,\, \ell^b \rightarrow v)(p) \eqmove \eptr{\ell} \\\\
    \text{no locations in } v \\\\
    \eptr{\ell^b}\notin \Omega\\\\
    \vdash \Omega(p) \updtmove \bot \eqmove \Omega'
  }{
    \Omega,\, l \rightarrow v \vdash \efree p \exprres{()}{\Omega',\, \_ \rightarrow v}
  }

  \inferrule[HLPL-E-Assign \hlplonly]{
    \hypertarget{HLPL-E-Assign} 
    \Omega \vdash rv \exprres{v}{\Omega'} \\
    \vdash \Omega'(p) \eqmut v_p : \tau \\
    v_p \text{ has no }\kw{loc} \\\\
    \vdash \Omega'(p) \updtmut v \ueqmut \Omega'' \\
    \Omega''' = \Omega'',\, \_ \rightarrow v_p
  }{
    \Omega \vdash \eassign p {rv} \exprres{()}{\Omega'''}
  }
}

  \inferrule[R-Loc]{
    P \neq [.]\\
    \Omega \vdash P(v) \eqimmut v'
  }{
    \Omega(p) \vdash P(\eloc{\ell}{v}) \eqimmut v'
  }

  \inferrule[R-Deref-Ptr-Loc]{
    \eloc{\ell}{v} \in \Omega\\\\
    \Omega \vdash P(v) \eqimmutmut v'
  }{
    \Omega \vdash P(*(\eptr{\ell})) \eqimmutmut v'
  }
\IfBooleanT{#1}{

  \inferrule[R-Deref-Box-Id]{
    \Omega(\ell^b) = v\\\\
    \Omega \vdash P(v) \eqk v'
  }{
    \Omega \vdash P(*(\eptr{\ell^b})) \eqk v'
  }

  \inferrule[W-Loc]{
    P \neq [.]\\
    \Omega \vdash P[v] \updtimmut w \wimmut{v'}{\Omega'}
  }{
    \Omega(p) \vdash P(\eloc{\ell}{v}) \leftarrow w
      \wimmut{\eloc{\ell}{v'}}{\Omega'}
  }

  \inferrule[W-Deref-Ptr-Loc]{
    \eloc{\ell}{v} \in \Omega\\\\
    \Omega \vdash P(\eloc{\ell}{v}) \updtimmutmut w
      \wimmutmut{v'}{\Omega'[\eloc{\ell}{v''}]}\\\\
    \Omega'' = \Omega'[\eloc{\ell}{v'}]
  }{
    \Omega \vdash P(*(\eptr{\ell})) \updtimmutmut w \ueqimmutmut
      \eptr{\ell} \dashv \Omega''
  }

  \inferrule[W-Deref-Box-Id]{
    \Omega(\ell^b) = v\\\\
    \Omega \vdash P(v) \updtk w \wk{v'}{\Omega'}\\\\
    \Omega'' = (\Omega'(l^b) := v')
  }{
    \Omega \vdash P(*(\eptr{\ell^b})) \updtk w
      \wk{\eptr{\ell^b}}{\Omega''}
  }

  \inferrule[Copy-Loc]{
    \vdash \ecopy v = v'
  }{
    \vdash \ecopy (\eloc{\ell}{v}) = v'
  }

  \inferrule[Copy-Ptr]{
  }{
    \vdash \ecopy (\eptr{\ell}) = \eptr{\ell}
  }
  }
}

\begin{figure}
  \centering
  \smaller
  \begin{mathpar}
    \HlplRules
  \end{mathpar}
  \caption{Selected Rules for HLPL}
  \flabel{hlpl}
  \figvspace
\end{figure}

We define the operational semantics of HLPL in \Figure{hlpl}.
Pointers (resp., locations) behave basically like shared borrows
(resp., loans), in that the value lives with the location. Unlike shared
borrows, we permit modifications through the pointer.
In particular, in HLPL we evaluate borrowing expressions like \li+&mut p+ and \li+&p+ with
\Rule{E-Pointer}, which is very similar to \Rule{E-SharedBorrow}. We remark that
HLPL retains some structure still: one can only update a value
$x \to \eloc\ell0$ via a corresponding pointer $p \to \eptr\ell$, e.g., to obtain
$x \to \eloc\ell1$ (\Rule{W-Deref-Ptr-Loc}, in Appendix). Should one want to update
$x$ \emph{itself}, there must be no outstanding pointers to it
(\Rule{Reorg-End-Pointer}), and the location itself must have been forgotten
(\Rule{Reorg-End-Loc}). Just like in LLBC, these reorganizations may happen at any
time, and as in LLBC, a poor choice of reorganizations may lead to a
stuck execution.

We annotated the reborrowing example with mutable borrows from the previous section to show the
HLPL environments at each program point.
\begin{minted}[xleftmargin=1.5em,mathescape=true,escapeinside=||]{rust}
x  = 0;          // $x \mapsto 0$ $\lnlabel{hlpl-x-zero}$
px = &mut x;     // $x \mapsto \eloc{\ell}{0},\quad px \mapsto \eptr{\ell}$ $\lnlabel{hlpl-borrow}$
px = &mut (*px); // $x \mapsto \eloc{\ell}{0},\quad \_ \mapsto \eptr{\ell},\quad px \mapsto\eptr{\ell}$$\lnlabel{hlpl-reborrow}$
assert!(x == 0); // $x \mapsto \eloc{\ell}{0},\quad \_ \mapsto \eptr{\ell},\quad px \mapsto\eptr{\ell}$$\lnlabel{hlpl-assert}$
\end{minted}

At line~\lnref{hlpl-borrow}, we use \Rule{E-Pointer} to introduce a fresh location
in $x$ and evaluate \li+&mut x+ to a fresh pointer. Contrary to what happens in LLBC,
in HLPL we leave pointed values in place.
At line~\lnref{hlpl-reborrow}, we use \Rule{E-Pointer} again, but this time dereference
$px$ (which maps to $\eptr{\ell}$), which yields $\eloc{\ell}{0}$, and create another
pointer for this location. Because the value we create a pointer to is already a
location value, we do not introduce a fresh location and simply evaluate \li+&mut (*px)+ to
$\eptr{\ell}$.
We now need to evaluate the assignment to $px$ (left hand side of \li+px = &mut (*px)+):
we move the current value of $px$ ($\eptr{\ell}$) to
a fresh anonymous value, then override the value of $px$ with the result of
evaluating the right-hand side ($\eptr{\ell}$). \iflong\\\fi
Finally, at line~\lnref{hlpl-assert},
we need to read $x$. As we can read \emph{through} locations
we do not \emph{need} to end any pointer at this point and leave the environment unchanged;
note that we \emph{could} preemptively end location $\ell$ to get the same
environment as with the LLBC semantics.
\iflong
The same happens with shared borrow in LLBC: we can read \emph{through} shared values.
However, if we were to directly update $x$,
for instance by evaluating \li+x = 1+, we would need to end the location
with \Rule{Reorg-End-Loc}, which would require ending the pointers first by using
\Rule{Reorg-End-Pointer}, yielding the environment:
$x \mapsto 0,\quad \_ \mapsto \bot,\quad px \mapsto\bot$.

It is to be noted that \Rule{E-Pointer} applies both to mutable borrows
and shared borrows: if we update the example above to replace the occurrences
of \li+&mut+ with \li+&+, we get exactly the same environments at each program point.
\fi

We now use the proof methodology of \Section{proofmeth} to show the forward
simulation from LLBC to HLPL.

\newcommand\hlpl{\ensuremath{\text{hlpl}}}
\newcommand\llbc{\ensuremath{\text{llbc}}}

\NewDocumentCommand{\HlplLeRules}{s}{
\IfBooleanT{#1}{
  \inferrule[Le-SharedReserved-To-Ptr]{
  }{
    \Omega[\eptr{\ell}] \le \Omega[\esrborrow{\ell}]
  }

}
\IfBooleanF{#1}{
  \inferrule[Le-Shared-To-Ptr]{
  }{
    \Omega[\eptr{\ell}] \le \Omega[\esborrow{\ell}]
  }

}
  \inferrule[Le-MutBorrow-To-Ptr]{
    \ell \notin \Omega[.,\, .]\\
    \ell \notin v
  }{
    \Omega[\eloc{\ell}{v},\, \eptr{\ell}] \le
    \Omega[\emloan{\ell},\, \emborrow{\ell}{v}]
  }

\IfBooleanT{#1}{
  \inferrule[Le-RemoveAnon]{
    \text{no borrow, loan} \in v
  }{
    \Omega \le
    \Omega, \, \_ \rightarrow v
  }

}
  \inferrule[Le-Merge-Locs]{
    \forall\; v', \eloc{\ell'}{v'} \notin \Omega[\eloc{\ell}{v}]\\
    \Omega' = \left[\ell \Big/ \ell' \right]\left(\Omega[\eloc{\ell}{v}]\right)
  }{
    \Omega' \le
    \Omega[\eloc{\ell}{(\eloc{\ell'}{v}})]
  }

  \inferrule[Le-SharedLoan-To-Loc]{
    \IfBooleanT{#1}{\esrborrow{\ell} \notin \Omega[\eloc{\ell}{v}]}
    \IfBooleanF{#1}{\esborrow{\ell} \notin \Omega[\eloc{\ell}{v}]}
  }{
    \Omega[\eloc{\ell}{v}] \le
    \Omega[\esloan{\ell}{v}]
  }
\IfBooleanT{#1}{

  \inferrule[Le-Box-To-Loc]{
    \ell^b \text{ fresh}
  }{
    \Omega[\eptr{\ell^b}],\, l^b \rightarrow v \le
    \Omega[\ebox{v}]
  }

  \inferrule[Le-Subst]{
    \ell' \not\in \Omega
  }{
    \esubst{\ell}{\ell'}{\Omega} \le \Omega
  }
}
}

\begin{figure}
  \centering
  \smaller
  \begin{mathpar}
    \HlplLeRules
  \end{mathpar}
  \caption{The $\le$ Relation on \hlplp states (Selected Rules)}
  \flabel{hlpl-rel}
  \figvspace
\end{figure}

\myparagraph{The $\le$ relation between HLPL and LLBC states}
Following \Section{proofmeth}, the first step consists in introducing a series of
local rewriting rules (\Figure{hlpl-rel}),
whose transitive closure, written $\le$, relates an HLPL state $\Omega_\hlpl$
to an LLBC state $\Omega_\llbc$. The relation $\Omega_\hlpl \le \Omega_\llbc$
can be read in both directions; but since we are concerned here with a forward
simulation, we read these rules right to left, that is, we
\emph{gradually} transform borrows and loans (from LLBC) into
pointers and locations (from HLPL).

In effect, this amounts to losing information about the nature of
the borrows, in order to only retain an aliasing graph.
Continuing with the right to left intuition,
\Rule{Le-MutBorrow-To-Ptr} states that we can collapse a pair of a mutable loan and its
corresponding borrow to a location and a pointer. Notice how the value
$v$ moves from being attached to the borrow to being attached to the location;
this is the crucial rule that moves from a borrow-centric view to a
location-centric view.

\iflong
Going back to the reborrowing example with mutable borrows from above, we have at line~\lnref{hlpl-borrow}
that the LLBC state ($x \mapstoemloan{0},\; px \mapstoemborrow{0}{0}$)
is related to the HLPL state
($x \mapsto \eloc{\ell}{0},\; px \mapsto \eptr{\ell}$) by \Rule{Le-MutBorrow-To-Ptr}.
The reborrow at line~\lnref{hlpl-reborrow} is more interesting; starting
from the LLBC state we have the following relations (using the notation
$\Omega \ge \Omega' := \Omega' \le \Omega$):
\setcounter{equation}{0}
\begin{align}
\smaller
&x \mapstoemloan{0},\  \_ \mapstoemborrow{0}{(\emloan{\ell_1})},\  px
\mapstoemborrow{1}{0} &\nonumber\\
&\ge x \mapstoemloan{0},\  \_ \mapstoemborrow{0}{(\eloc{\ell_1}{0})},\  px
\mapsto\eptr{\ell_1} & \text{(by \Rule{Le-MutBorrow-ToPtr})}\\
&\ge x \mapsto \eloc{\ell_0}{(\eloc{\ell_1}{0})},\  \_ \mapsto \eptr{\ell_0},\  px
\mapsto\eptr{\ell_1} & \text{(by \Rule{Le-MutBorrow-ToPtr})}\\
&\ge x \mapsto \eloc{\ell_0}{0},\  \_ \mapsto \eptr{\ell_0},\  px
\mapsto\eptr{\ell_0} & \text{(by \Rule{Le-Merge-Locs})}
\end{align}
We convert the pair borrow/loan for $\ell_1$ to a pair pointer/location at step (1) by
\Rule{Le-MutBorrow-ToPtr}. We use \Rule{Le-MutBorrow-ToPtr} again at step (2), this time
for $\ell_0$. In the resulting environment, $x$ contains two stacked locations that we finally ``merge''
together by using \Rule{Le-Merge-Locs} at step (3). This yields the same HLPL
environment as at line~\lnref{hlpl-reborrow}.

We note that the states after the \li+assert+ at line~\lnref{hlpl-assert} are not
related: we can not transform the LLBC state
($x \mapsto 0,\; \_ \mapsto \bot,\; px \mapsto \bot$)
to the HLPL state ($x \mapsto \eloc{\ell}{0},\; \_ \mapsto \eptr{\ell},\; px \mapsto\eptr{\ell}$)
by using the rules for $\le$. The HLPL semantics indeed allows strictly \emph{more} behaviors
than LLBC, as we \emph{don't have} to end any pointer to read $x$.
However, when doing the execution according to the HLPL semantics we \emph{could} have
ended them to yield the HLPL state
$x \mapsto 0,\; \_ \mapsto \bot,\; px \mapsto \bot$, which is related to
the LLBC state at the same program point (as they are actually the same).

The version of the reborrowing example which uses shared borrows is simpler.
This time we convert each shared borrow to a pointer with \Rule{Le-Shared-To-Ptr},
then convert the shared loan to a location with \Rule{Le-SharedLoan-To-Loc};
we have to convert the borrows first because of the premise
$\esborrow{\ell} \notin \Omega$ in \Rule{Le-SharedLoan-To-Loc}.
For instance, after the reborrow at line~\lnref{shared-reborrow}, we have:
\setcounter{equation}{0}
\begin{align}
\smaller
&x \mapsto \esloan{\ell}{0},\quad \_ \mapsto \esborrow{\ell},\quad px \mapsto
\esborrow{\ell} & \nonumber\\
&\ge x \mapsto \esloan{\ell}{0},\quad \_ \mapsto \esborrow{\ell},\quad px \mapsto
\eptr{\ell} & \text{(by \Rule{Le-Shared-To-Ptr})} \\
&\ge x \mapsto \esloan{\ell}{0},\quad \_ \mapsto \eptr{\ell},\quad px \mapsto
\eptr{\ell} & \text{(by \Rule{Le-Shared-To-Ptr})} \\
&\ge x \mapsto \eloc{\ell}{0},\quad \_ \mapsto \eptr{\ell},\quad px \mapsto
\eptr{\ell} & \text{(by \Rule{Le-SharedLoan-To-Loc})}
\end{align}

\fi

\myparagraph{Working in \hlplp}
Naturally, reasoning about $\le$ (in order to establish the forward simulation,
as we do in the next paragraph) is conducted via reasoning by induction.
Specifically, we do the proof by induction on the evaluation derivation, then
in each sub-case do an induction on $\le$.
This is the essence of our proof technique, which
emphasizes local and pointwise reasoning rather than global invariants.
In particular, we took great care to define the rules of $\le$
either as \emph{local} transformations (by defining them with states with holes),
or as \emph{pointwise} transformations (\Rule{Le-Merge-Locs}).

As we explained earlier (\Section{proofmeth}), this leads us to reason about hybrid
states that contain both loans and borrows (like LLBC
states) as well as locations and pointers (like HLPL states).
We call such states \hlplp states, and per \Section{proofmeth} set out to give an
operational semantics to HLPL+.
%
Since \hlplp shares the same syntax as HLPL and LLBC, it suffices to take the union of
the rules from HLPL and LLBC, \emph{adding} \hlplp-specific rules (\Figure{hlplp}), and \emph{excluding} the rules marked \textbf{(LLBC only)}.
The LLBC-only rules introduce new loans and borrows; by excluding those, we
get that HLPL is a stable subset of \hlplp (Definition \defref{stable}).

\NewDocumentCommand{\HlplpRules}{s}{

  \inferrule[HLPL+-E-Pointer]{
    \vdash \Omega(p) \eqimmutmut v \\
    \vdash \Omega(p) \updtimmutmut v' \eqimmutmut \Omega' \\
    v' = \begin{cases}
      \eloc {\ell} v'' & \text { if } v = \eloc{\ell} v'' \cr
      \eloc {\ell} v'' & \text { if } v = \esloan{\ell} v'' \cr
      \eloc {\ell} v & \ell \text { fresh otherwise}
    \end{cases}
  }{
    \Omega \vdash
      \{\ebrw p,\, \embrw p \IfBooleanT{#1}{,\, \erbrw{p}} \}
    \exprres{\eptr{\ell}}{\Omega'}
  }

  \inferrule[HLPL+-E-Assign]{
    \Omega \vdash rv \exprres{v}{\Omega'} \\
    \vdash \Omega'(p) \eqmut v_p \\\\
    v_p \text{ has no outer } \kw{loan}^{s,m}, \text{ no }\kw{loc} \\\\
    \vdash \Omega'(p) \updtmut v \ueqmut \Omega'' \\
    \Omega''' = \Omega'',\; \_ \rightarrow v_p
  }{
    \Omega \vdash \eassign p {rv} \exprres{()}{\Omega'''}
  }

  \inferrule[R-Deref-Ptr-SharedLoan]{
    \esloan{\ell}{v} \in \Omega\\
    \Omega \vdash P(\esloan{\ell}{v}) \eqimmut v'
  }{
    \Omega \vdash P(*(\eptr{\ell})) \eqimmut v'
  }
\IfBooleanT{#1}{
  \inferrule[HLPL+-E-Move]{
    \vdash \Omega(p) \eqmove v \\\\
    \bot, \kw{loan}^{s,m}, \kw{borrow}^r, \kw{loc}\not\in v \\\\
    \updtstateplace{\Omega}{p}{\bot} \ueqmove \Omega'
  }{
    \Omega \vdash \emove p \exprres{v}{\Omega'}
  }

  \inferrule[W-Deref-Ptr-SharedLoan]{
    \esloan{\ell}{v} \in \Omega\\
    \Omega \vdash P(\esloan{\ell}{v}) \updtimmut w
      \wimmut{v'}{\Omega'[\esloan{\ell}{v''}]}\\
    \Omega'' = \Omega'[\esloan{\ell}{v'}]
  }{
    \Omega \vdash P(*(\eptr{\ell})) \updtimmut w
      \wimmut{\eptr{\ell}}{\Omega''}
  }

  \inferrule[HLPL+-Reorg-End-SharedLoan]{
    \esrborrow{\ell},\, \eptr{\ell} \not\in \Omega[\esloan{\ell}{v}]
  }{
    \Omega[\esloan{\ell}{v}] \hookrightarrow \Omega[v]
  }
  }
}

\begin{figure}
  \centering
  \smaller
  \begin{mathpar}
    \HlplpRules
  \end{mathpar}
  \caption{Selected Additional Rules for \hlplp}
  \flabel{hlplp}
  \figvspace
\end{figure}

The \hlplp rules are in \Figure{hlplp}. Their purpose is to replace rules that
would produce new borrows (\Rule{E-MutBorrow}, \Rule{E-SharedBorrow})
with \Rule{HLPL+-E-Pointer}, which directly reduces to a pointer.
In a similar vein, \Rule{R-Deref-Ptr-SharedLoan} deals with a hybrid state where
the state still contains a loan, but the value being reduced is an HLPL pointer,
not a borrow. Finally, \Rule{HLPL+-E-Assign} shows how to add extra
preconditions to extend the ``no outer loan''
condition\footnote{An outer loan is a loan which is not itself inside a borrow;
$\emloan{\ell_0}$ contains one, while $\emborrow{\ell_1}{(\emloan{\ell_0}})$ doesn't.}
(required in
LLBC for soundness) to a hybrid world in which there might be locations too.

Equipped with our individual rewriting rules (which form $\le$) and a semantics
in which those rules operate (\hlplp, the union of HLPL and LLBC, crafted to make HLPL a
stable subset), we now prove that reduction preserves $\le$.

\begin{theorem}[Eval-Preserves-HLPL+-Rel]
  For all $\Omega_l$, $\Omega_r$ \hlplp states, we have:
  \begin{align*}
  \iflong
  &\forall\; s\, r\, \Omega_r',\,\\
  &\Omega_l \le \Omega_r \Rightarrow\\
  &\Omega_r \vdashHlplp\, s \stmtres{r}{\Omega_r'} \Rightarrow
  \exists\; \Omega'_l,\,
    \Omega_l \vdashHlplp\, s \stmtres{r}{\Omega_l'} \wedge \Omega_l' \le \Omega_r'
  \fi
  \end{align*}
\thmvspace
\end{theorem}

The proof is in \iflong\Appendix{hlpl+}\fi;
it consists in a nested case analysis; first, for
each reduction step; then, for each $\le$ step.
\iflong In particular, we\fi use the fact
that if two states are related by one of the $\le$ rules, then the structure
enforced by this rule is generally preserved after the evaluation.

\myparagraph{From \hlplp to HLPL}
\hlplp is merely a proof device; our ultimate goal is to relate LLBC to HLPL, not
\hlplp. Because we excluded (above) from \hlplp those rules that might create new
loans or borrows, we trivially have the fact that the semantics of HLPL and \hlplp
coincide on HLPL states (i.e., \hlplp states that don't contain loans or borrows,
and thus belong to the HLPL subset of \hlplp); that is, that HLPL is a stable
subset of \hlplp in the sense of Definition \defref{stable}.
%
%
%
From this we deduce that there is a forward relation from \hlplp to HLPL:

\begin{theorem}[Forward Relation for \hlplp and HLPL]
  For $\Omega_l$ HLPL state, $\Omega_r$ \hlplp state:
  \begin{align*}
  &\forall\; s\, r\, \Omega_r',\,
   \Omega_l \le \Omega_r \Rightarrow
   \Omega_r \vdashHlplp\, s \stmtres{r}{\Omega_r'} \Rightarrow
  \exists\; \Omega'_l,\,
    \Omega_l \vdashHlpl\, s \stmtres{r}{\Omega_l'} \wedge \Omega'_l
    \le \Omega_r'
  \end{align*}
\thmvspace
\end{theorem}

\myparagraph{From LLBC to \hlplp}
We have just shown that the lower bound, $\Omega_l$, remains in HLPL (by virtue
of stability, Definition~\defref{stability}), and therefore $\le$ can preserve
the relation between HLPL and \hlplp. It now remains to show the link between
\hlplp and LLBC, i.e., the upper bound, $\Omega_h$, reduces in LLBC in a way that
preserves $\le$.

Because \hlplp excludes several rules from LLBC (remember that some rules are
marked \llbconly), LLBC is not a subset of \hlplp, which prevents us from
deriving that result instantly. Instead, we remark that we can still reconstruct the
missing LLBC semantics using well-chosen combinations of evaluation rules from \hlplp and
refinement rules from $\leq$.
For instance, \Rule{E-MutBorrow} states that evaluating $\embrw{p}$ in
LLBC leads to a state with a mutable loan and a mutable borrow. We can build the
same state by using the \hlplp rule \Rule{E-Pointer} to introduce a pointer and a
location, then by using \Rule{Le-MutBorrow-To-Ptr} to transform this state into
a \emph{related} state (in the sense of $\le$), by converting this pointer and
this location to a borrow and a loan.
This means that a reduction in LLBC can always be completed to correspond to a
reduction in \hlplp, which 
allows us to prove
the following theorem stating that, given an LLBC
state $\Omega$, evaluating a statement following the semantics of LLBC leads
to a state \emph{in relation} with the state resulting from the \hlplp semantics.

\begin{theorem}[Eval-LLBC-Preserve-Rel]
  For all $\Omega$ LLBC state we have:
  \begin{align*}
  \forall\; s\, r\, \Omega_r,\,
    \Omega \vdashLlbc\, s \stmtres{r}{\Omega_r} \Rightarrow
  \exists\; \Omega_l,\,
    \Omega \vdashHlplp\, s \stmtres{r}{\Omega_l} \wedge \Omega_l \le \Omega_r
  \end{align*}
\thmvspace
\end{theorem}

Putting \Thm{Eval-Preserves-HLPL-HLPL+-Rel} and \Thm{Eval-LLbc-Preserves-Rel} together
and using the transitivity of $\le$ we finally get the preservation theorem we were
aiming at: LLBC is in forward simulation with HLPL (\Thm{Eval-Preserves-HLPL-LLBC-Rel}).

\myparagraph{Form of our theorems}
%
We take great care to start from \emph{any} initial states
$\Omega_l$ and $\Omega_r$ rather than requiring a program execution with a closed term
and with an empty state (i.e., a ``main'' function). The reason is, as
mentioned before, we see the \llbcs execution as borrow-checking, which we aim
to perform modularly at the function-level granularity. This design for our
theorems will later on allow us, as we connect \llbcs all the way down to PL, to
reason about the execution of a PL function that starts in an initial state
\emph{compatible with the lifetime signature} of the function in LLBC (\Section{backllbcpl}).
%
Should we wish to do so, and using the fact that empty states are in relation with each
other, we can specialize our main theorem to closed programs executing in the
empty state.

\subsection{The Pointer Language (PL)}
\slabel{pl-ref}

So far, we have only proven the central arrow of \Figure{architecture}. We now
describe the simulation between HLPL and PL; we do so briefly, since the
techniques are standard, and do not leverage our new proof methodology. The full
proof is in \iflong\Appendix{pl}\fi.
PL uses an explicit heap adapted from the CompCert memory model;
this is similar to RustBelt, except we have no extra state to account for
concurrency. As before, we retain the same syntax of programs; this is why we
can, ultimately, conclude that LLBC is a sound restriction over the
execution of PL programs.

For the proof of simulation between PL and HLPL, we have no choice but to
introduce a global map from location identifiers to concrete addresses, along
with an explicit notion of heap. In short, we adopt the traditional
global-invariant style of proof; this is the only arrow from \Figure{architecture}
where we cannot apply our methodology.
However, as we took care to design HLPL so that its pointers leave
values in place, the structure of HLPL states and PL states is very close;
as a consequence the proof is technical, but quite straightforward.
It crucially leverages the fact that the rules of HLPL were carefully
crafted so that it is not possible to move a location (see the premises
of \Rule{HLPL-E-Move} or \Rule{HLPL-E-Assign} for instance), as it would break
the relation between the HLPL locations and the PL addresses.
Moving up the hierarchy of languages, this is the reason why we forbid moving
\emph{outer} loans in the LLBC rules, that is loans which are not inside
a borrow (see \Rule{E-Assign} in particular).
We elide the exact statements of the theorems, which can be found in
\iflong\Appendix{pl}\fi.

\subsection{Divergence and Step-Indexed Semantics}

The forward simulation theorem between PL and LLBC relates terminating executions.
Anticipating on the next section where we will consider \llbcs executions
as borrow-checking certificates, this gives us in particular that the
existence of safe executions for LLBC implies that the PL executions are safe.
Unfortunately, as LLBC is defined
in a big-step fashion, one problem arises. Big-step semantics do not allow to reason about
programs that safely diverge, that is, programs that never get stuck or crash, but do not terminate:
these programs cannot be distingushed from stuck programs; in both cases, no evaluation exists.
This is a known problem when studying type systems; previously proposed workarounds include defining
semantics coinductively to model divergence~\cite{leroy2006coinductive,nakata2009trace}.

\begin{figure}
  \centering
  \smaller
  \begin{mathpar}
    \inferrule[E-Step-Zero]{
    }{
      \Omega \vdash s \steparrow{0} \infty
    }

    \inferrule[E-Step-Return]{
    }{
      \Omega \vdash \ereturn \steparrow{n+1} (\ereturn, \Omega)
    }

    \inferrule[E-Step-Seq-Unit]{
      \Omega_0 \vdash s_0 \steparrow{n+1} ((), \Omega_1)\\
      \Omega_1 \vdash s_1 \steparrow{n+1} res
    }{
      \Omega_0 \vdash s_0; s_1 \steparrow{n+1} res
    }

  \inferrule[E-Step-Call]{
    \feqsig{\vec{\_}}\\
    \forall\; j, \Omega^{(j)} \vdash op_j \rightarrow (v_j,\; \Omega^{(j+1)})\\
    \pushstack
      \left([x_{\text{ret}} \rightarrow \bot] \econcat
      [\overrightarrow{x_j \rightarrow v_j}] \econcat
      [\overrightarrow{y_k \rightarrow \bot}]\right) \Omega^{(m)} = \Omega_{0}\\
    \Omega_{0} \vdash body \steparrow{n} res\\
    res' = \begin{cases}
      (\epanic,\, \Omega_{1}) & \text { if } res = (\epanic,\, \Omega_{1}) \cr
      ((),\, \Omega_{3}) &
        \text { if } res = (\ereturn,\, \Omega_{1}) \,\wedge
        \popstack \Omega_{1}\,=\, (v,\, \Omega_{2}) \eand
        \Omega_{2} \vdash \eassign p v \stmtres{()}{\Omega_{3}}
        \cr
      \infty & \text { if } res = \infty \cr
    \end{cases}
  }{
    \Omega^{(0)} \vdash \eassign{p}{f(\overrightarrow{op_j})} \steparrow{n+1} res'
  }
  \end{mathpar}
  \caption{Semantics of LLBC with Step-Indexing (Selected Rules)}
  \flabel{llbc-index}
  \figvspace
\end{figure}

\NewDocumentCommand{\LeToSymbolicRule}{s}{
    \inferrule[Le-ToSymbolic]{
    \IfBooleanF{#1}{\kw{borrow}^{s,m}, \kw{loan}^{s,m}, \bot \not\in v}
    \IfBooleanT{#1}{\kw{borrow}^{s,m,r}, \kw{loan}^{s,m}, \bot \not\in v}\\\\
    \sigma \text{ fresh}
    }{
    \Omega[v] \le \Omega[\sigma]
    }
}
\newcommand\LeToAbsRule{
    \inferrule[Le-ToAbs]{
    \vdash v \toabs \overrightarrow{A}
    }{
    \Omega,\; \_ \rightarrow v \le \Omega,\; \overrightarrow{A}
    }
}
\newcommand\LeMoveValueRule{
    \inferrule[Le-MoveValue]{
    \text{no outer loans in } v\\\\
    \text{hole of $\Omega[.]$ not inside a shared loan or a region abstraction}
    }{
    \Omega[v] \le \Omega[\bot],\; \_ \mapsto v
    }
}
\newcommand\LeMergeAbsRule{
    \inferrule[Le-MergeAbs]{
      \vdash A_0 \MergeAbs A_1 \merge A
    }{
      \Omega,\; A_0,\; A_1 \le \Omega,\; A
    }
}
\newcommand\LeReborrowMutBorrowAbsRule{
    \inferrule[Le-Reborrow-MutBorrow-Abs]{
      \ell_1,\; A \text{ fresh}
    }{
      \Omega[\emborrow{\ell_0} v] \le
      \Omega[\emborrow{\ell_1} v],\; \Abs{A}{\emborrow{\ell_0}\_,\; \emloan\ell_1}
    }
}
\NewDocumentCommand{\LlbcLeRules}{s}{
    \IfBooleanF{#1}{\LeToSymbolicRule}\IfBooleanF{#1}{\LeToSymbolicRule*}
    
    \LeToAbsRule

    \LeMoveValueRule

\IfBooleanT{#1}{
    \inferrule[Le-ClearAbs]{
    }{
       \Omega,\; A\; \{\} \le
       \Omega
    }

}
    \LeMergeAbsRule
\IfBooleanT{#1}{

    \inferrule[Le-Fresh-MutLoan]{
    \ell\text{ fresh}
    }{
    \Omega[v] \le
    \Omega[\emloan{\ell}],\; \_ \rightarrow \emborrow{\ell}{v}
    }

    \inferrule[Le-Fresh-SharedLoan]{
      \ell\text{ fresh}
    }{
      \Omega[v] \le \Omega[\esloan{\ell}{v}]
    }

    \inferrule[Le-Reborrow-MutBorrow]{
      \ell_1\text{ fresh}
    }{
      \Omega[\emborrow{\ell_0}{v}] \le
      \Omega[\emborrow{\ell_1}{v}],\; \_ \rightarrow \emborrow{\ell_0}{(\emloan{\ell_1})}
    }

    \inferrule[Le-Abs-ClearValue]{
      \text{no borrows, loans } \in v
    }{
       \Omega,\; A \cup \{\; v \;\} \le
       \Omega,\; A
    }

    \inferrule[Le-Fresh-SharedBorrow]{
    }{
      \Omega[\esloan{\ell}{v}] \le
      \Omega[\esloan{\ell}{v}],\; \_ \rightarrow \esborrow{\ell}
    }

    \inferrule[Le-Reborrow-SharedLoan]{
      A,\; \ell_1,\; \sigma \text{ fresh}\\
      \bot, \emloanv, \eborrowv \notin v
    }{
       \Omega[\esloan{\ell_0}{v},\, \overrightarrow{\esborrow{\ell_0}}] \le
       \Omega[\esloan{\ell_1}{\sigma},\, \overrightarrow{\esborrow{\ell_1}}],\;
       A\; \{\; \esborrow{\ell_1},\; \esloan{\ell_0}{v} \;\}
    }

    \inferrule[Le-Abs-End-SharedLoan]{
      \Omega = \Omega',\; A[\esloan{\ell}{v}] \\
      \text{no } \esrborrow{\ell} \in \Omega
    }{
      \Omega \le \Omega',\; A[v]
    }

    \inferrule[Le-Abs-End-DupSharedBorrow]{
    }{
      \Omega,\; A \cup \{\; \esborrow{\ell},\; \esborrow{\ell} \;\} \le
      \Omega,\; A \cup \{\; \esborrow{\ell} \;\}
    }

    \inferrule[Le-Reborrow-SharedBorrow]{
      \bot, \emloanv, \eborrowv \notin v\\
      \esloan{\ell_0}{v} \in \Omega[\overrightarrow{\esborrow{\ell_0}}]\\
      \ell_1,\; \sigma,\; A \text{ fresh}
    }{
      \Omega[\overrightarrow{\esborrow{\ell_0}}] \le
      \Omega[\overrightarrow{\esborrow{\ell_1}}],\,
      A\;\{ \esborrow{\ell_0},\; \esloan{\ell_1}{\sigma} \;\}
    }

    \inferrule[Le-Abs-DeconstructPair]{
    }{
       \Omega,\; A \cup \{\; (v_0,\; v_1) \;\} \le
       \Omega,\; A \cup \{\; v_0,\; v_1 \;\}
    }

    \inferrule[Le-Abs-DeconstructSum]{
      C \in \{\kw{Left},\; \kw{Right}\}
    }{
       \Omega,\; A \cup \{\; C\; v \;\} \le
       \Omega,\; A \cup \{\; v \;\}
    }

    \inferrule[Le-AnonValue]{
      \text{ no symbolic values, borrows, loans} \in v
    }{
       \Omega \le \Omega,\, \_ \rightarrow \bot
    }

  }
}

\NewDocumentCommand{\LlbcLeDerivedRules}{s}{
    \LeReborrowMutBorrowAbsRule
\IfBooleanT{#1}{

    \inferrule[Le-Reborrow-MutLoan-Abs]{
      \ell_1,\; A \text{ fresh}
    }{
      \Omega[\emloan{\ell_0}] \le
      \Omega[\emloan{\ell_1}],\; \Abs{A}{ \emborrow{\ell_1}{\_},\; \emloan{\ell_0} }
    }

    \inferrule[Le-Fresh-MutLoan-Abs]{
      \ell,\, A \text{ fresh}
    }{
      \Omega[v] \le
      \Omega[\emloan{\ell}],\; \Abs{A}{\emborrow{\ell}{v}}
    }




    \inferrule[Le-DeconstructSharedLoans]{
      \sigma \text{ fresh symbolic value}
    }{
      \Omega, A \cup \{\; V_0[ \esloan{\ell_0}{(V_1[\esloan{\ell_1}{v}])} ] \;\}
      \le
      \Omega, A \cup \{\; V_0[ \esloan{\ell_0}{(V_1[\sigma])} ],\;
        \esloan{\ell_1}{v} \;\}
    }
  }
}

\newcommand\ToAbsMutBorrowRule{
    \inferrule[ToAbs-MutBorrow]{
    \text{ no borrows, $\bot$} \in v\\
    v \toabs \overrightarrow A
    }{
    \vdash \emborrow\ell v \toabs
    (\cup \overrightarrow A) \cup \{ \emborrow\ell\_ \}
    }
}
\newcommand\ToAbsMutLoanRule{
    \inferrule[ToAbs-MutLoan]{
    A \text{ fresh}
    }{
    \vdash \emloan\ell \toabs A \{ \emloan\ell \}
    }
}
\newcommand\ToAbsPairRule{
    \inferrule[ToAbs-Pair]{
    \vdash v_l \toabs \overrightarrow A_l\\
    \vdash v_r \toabs \overrightarrow A_r
    }{
    \vdash (v_l, v_r) \toabs \overrightarrow A_l,\; \overrightarrow A_r
    }
}
\NewDocumentCommand{\ToAbsRules}{s}{
\IfBooleanT{#1}{
    \inferrule[ToAbs-Empty]{
    \text{no borrows, loans} \in v
    }{
    \vdash v \toabs \emptyset
    }

    \inferrule[ToAbs-Sum]{
    v = \kw{Left}\;v' \vee v = \kw{Right}\;v'\\
    \vdash v' \toabs \overrightarrow A
    }{
    \vdash v \toabs \overrightarrow A
    }

    \inferrule[ToAbs-Box]{
    \vdash v \toabs \overrightarrow A
    }{
    \vdash \ebox{v} \toabs \overrightarrow A
    }

    \inferrule[ToAbs-SharedBorrow]{
    A \text{ fresh}
    }{
    \vdash \esborrow\ell \toabs A \{ \esborrow\ell \}
    }

    \inferrule[ToAbs-SharedLoan]{
    A \text{ fresh}
    }{
    \vdash \esloan{\overrightarrow\ell} v \toabs
      A \{ \esloan{\overrightarrow\ell} v \}
    }

}
    \ToAbsMutBorrowRule

    \ToAbsMutLoanRule

    \ToAbsPairRule
}

\newcommand\MergeAbsUnionRule{
    \inferrule[MergeAbs-Union]{
    }{
      \vdash A_0 \MergeAbs A_1 \merge A_0 \cup A_1
    }
}
\newcommand\MergeAbsMutRule{
    \inferrule[MergeAbs-Mut]{
      \vdash A_0 \MergeAbs A_1 \merge A \\
    }{
      \vdash (A_0 \cup \{ \emloan\ell \}) \MergeAbs (A_1 \cup \{ \emborrow\ell\_ \})
        \merge A
    }
}
\NewDocumentCommand{\MergeAbsRules}{s}{

    \MergeAbsUnionRule

    \MergeAbsMutRule
\IfBooleanT{#1}{

    \inferrule[MergeAbs-Shared]{
      \vdash (A_0 \cup \{ \esloan{\ell} v \}) \MergeAbs A_1 \merge A \\
    }{
      \vdash (A_0 \cup \{ \esloan{\ell} v \}) \MergeAbs
        (A_1 \cup \{ \esborrow{\ell} \})
        \merge A
    }
  }
}

To avoid tricky coinductive reasoning, we instead rely on step-indexed semantics.
We present in \Figure{llbc-index} several of the updated rules for the step-indexed
LLBC. As for related work, we add
a step index to the judgment which evaluates statements (e.g., \Rule{E-Step-Seq}).
This step index can be seen as a standard notion of fuel~\cite{owens2016functional}; when the index is equal to
zero, i.e., the execution is out of fuel, we stop the evaluation and return the $\infty$
value (\Rule{E-Step-Zero}). Otherwise, when evaluating possibly non-terminating statements
(e.g., function calls as described by \Rule{E-Step-Call}), we decrement the step index
in the recursive evaluation (e.g., the evaluation of the function body).

Building on this semantics, we can now define the following evaluation judgment that
hides the step indices, and returns either $\infty$ for diverging computations, or
the previously seen pair of a control-flow outcome and an environment:
\begin{align*}
&\Omega \vdash s \rightsquigarrow \infty :=
  \forall\; n, \Omega \vdash s \steparrow{n} \infty\\
&\Omega \vdash s \rightsquigarrow res :=
  \exists\; n, \Omega \vdash s \steparrow{n} res \eand res \neq \infty
\end{align*}

We follow a similar approach to extend the PL and HLPL languages to model divergence.
Adapting the proofs and theorems from \Section{refine} to the step-indexed semantics
and proving Theorem~\thref{plllbcinf} is straightforward, and we
omit the complete presentation for brevity.
The core idea is that evaluations for the same program have identical step indices in PL,
HLPL, and LLBC: the step index only decrements when entering a function call (or a loop),
which is shared across the three languages.

\begin{theorem}
  For all $\Omega_l$ PL state and $\Omega_h$ LLBC state, we have:
  \begin{align*}
  &\Omega_l \le \Omega_h \Rightarrow
  \forall\; s, \,
    \Omega_h \vdashLlbc\, s \rightsquigarrow \infty \Rightarrow
  \Omega_l \vdashPl s \rightsquigarrow \infty
  \end{align*}
  \thlabel{plllbcinf}
\thmvspace
\end{theorem}

\section{\llbcs is a sound approximation, a.k.a., borrow-checker for LLBC}
\slabel{llbc-sharp}

Building on top of LLBC, \citet{aeneas} also proposed a region-centric shape analysis, which they formalized
as a symbolic (or abstract) interpreter for the LLBC semantics. In their approach, this interpreter
was the backbone of a translation of Rust programs to functional models in different proof assistants, enabling
the verification of safe Rust programs. However, as observed by the authors, this symbolic semantics also
acted as a \emph{borrow checker} for LLBC programs.

In this section, we aim to study the soundness of this borrow checker with respect to the LLBC semantics.
Borrow checking is, conceptually, similar to type checking. A sound borrow checker for LLBC therefore
ensures that, for a given program, if the borrow checker succeeds, then the program
executes safely. In our setting, the definition of the borrow checking rules is however non-standard:
instead of a set of inference rules, the borrow checker is formalized using a symbolic semantics, which
we dub \llbcs. Importantly, this semantics is not deterministic; its implementation relies on several
heuristics to choose the right rules to apply. This is to be contrasted with the
current implementation of borrow-checking in the Rust compiler: while
deterministic, it relies on a mostly lexical lifetime mechanism; we claim that our
approach emphasizes \emph{semantic} rather than \emph{syntactic}
borrow-checking.

We therefore aim to establish a forward simulation from
\llbcs to LLBC, that is, that
for all successful \llbcs evaluations for a given program, there exists a related, valid execution in LLBC.
By composing this property with the forward simulation proven in the previous section, we therefore obtain
that if the borrow-checker (\llbcs) succeeds for a given program, then this program safely executes at the PL level.
The forward simulation from \llbcs to LLBC also allows us to strengthen the relation between PL and LLBC:
by definition of the simulation, successfully borrow checking a program implies the existence of a safe LLBC
evaluation, which then allows us to conclude that, because the semantics of PL are
deterministic, we actually have a bisimulation between PL and LLBC for programs that pass
borrow checking.


In the rest of this section, we focus on applying the proof methodology from \Section{proofmeth}
to prove that LLBC and \llbcs admit a forward simulation. We then show how to
combine this result with the determinism of PL to obtain a bisimulation for PL
and LLBC, under successful borrow-checking.

\subsection{Background: \llbcs}
\slabel{llbcsbackground}

\NewDocumentCommand{\LlbcsRules}{s}{
\IfBooleanT{#1}{
  \inferrule[\llbcs-Reorg-End-MutBorrow]{
    \text{hole of } \Omega[\emloan{\ell},\, .] \text{ not inside a borrowed value or an abs.}\\\\
    \kw{loan}^{s,m}, $\erborrowv$ \not\in v
  }{
    \Omega[\emloan{\ell},\, \emborrow{\ell}{v}] \hookrightarrow \Omega[v,\bot]
  }

  \inferrule[\llbcs-Reorg-End-SharedReservedBorrow]{
    \text{hole of } \Omega[.] \text{ not inside a borrowed value or an abs.}
  }{
    \Omega[\esrborrow{\ell}] \hookrightarrow \Omega[\bot]
  }



  \inferrule[Reorg-End-Abstraction]{
    \text{no borrows, loans} \in \overrightarrow{v},\, \overrightarrow{v'}\\
    \overrightarrow{\sigma} \text{ fresh}
  }{
     \Omega,\, \Abs{A}{
       \overrightarrow{v},\,
       \overrightarrow{\esborrow{\ell}},\,
       \overrightarrow{\emborrow{\ell'}{(v' : \tau)}}
       }
     \hookrightarrow
     \Omega,\,
       \overrightarrow{\_ \rightarrow \esborrow{\ell}},\,
       \overrightarrow{\_ \rightarrow \emborrow{\ell'}{(\sigma : \tau)}}
  }

  \inferrule[E-IfThenElse-Symbolic]{
   \Omega \vdash \kop \exprres{v}{\Omega'}\\
   v = (\sigma : \bool) \vee v = \esloan{\ell}{(\sigma : \bool)}\\
   \Omega_0 = \Omega'[\ktrue \Big/ \sigma]\\
   \Omega_1 = \Omega'[\kfalse \Big/ \sigma]\\
   \Omega_0 \vdash s_0 \steparrow{n+1} S_0^\#\\
   \Omega_1 \vdash s_1 \steparrow{n+1} S_1^\#
  }{
    \Omega \vdash \eite{\kop}{s_0}{s_1} \steparrow{n+1} S_0^\# \cup S_1^\#
  }

  \inferrule[E-Match-Symbolic]{
   \vdash \Omega(p) \eqimmut v\\
   v = (\sigma : \tau_0 + \tau_1) \vee v = \esloan{\ell}{(\sigma : \tau_0 + \tau_1)}\\
   \sigma_0,\, \sigma_1 \text{ fresh}\\
   \Omega_0 = \Omega'[\eleft{(\sigma_0 : \tau_0)} \Big/ \sigma]\\
   \Omega_1 = \Omega'[\eright{(\sigma_0 : \tau_0)} \Big/ \sigma]\\
   \Omega_0 \vdash s_0 \rightsquigarrow S_0^\#\\
   \Omega_1 \vdash s_1 \rightsquigarrow S_1^\#
  }{
    \Omega \vdash (\ematchsum{p}{s_0}{s_1}) \rightsquigarrow S_0^\# \cup S_1^\#
  }
}

  \inferrule[E-Call-Symbolic \llbcsonly]{
    \hypertarget{E-Call-Symbolic}\\ 
    \feqsignoindices{\vec{\rho}}\\
        \Omega_j \vdash op_j \exprres{v_j}{\Omega_{j+1}}\\
    \overrightarrow{\rho}\text{ fresh}\\
    \overrightarrow{\abssigrho}, \; \vout = \kw{inst\_sig}(\Omega_n, \vec\rho, \vec v, \tau_\text{ret}) \\
    \Omega_n,\; \overrightarrow{\abssigrho} \vdash \eassign p \vout \stmtres{()}{\Omega'}\\
  }{
    \Omega_0 \vdash \eassign{p}{\fcallnoindices} \stmtres{()}{\Omega'}
  }
\IfBooleanT{#1}{

  \inferrule[Copy-Symbolic]{
    \sigma' \text{ fresh}
  }{
    \vdash \ecopy \sigma = \sigma'
  }

  \inferrule[Reorg-SymbolicBox]{
    \sigma' \text{ fresh}
  }{
    \Omega[\sigma : \ebox{\tau}] \hookrightarrow \Omega[\ebox \sigma']
  }

  \inferrule[Reorg-SymbolicPair]{
    \sigma_0,\, \sigma_1 \text{ fresh}
  }{
    \Omega[\sigma : (\tau_0,\, \tau_1)] \hookrightarrow \Omega[(\sigma_0,\, \sigma_1)]
  }
}
}

\newcommand{\mmath}[1]{$#1$} 
We start by providing some background
on \llbcs before delving into the proof.
As before, (\Section{background-llbc}), we pick an example from
\citet{aeneas}, this time to illustrate \llbcs.
The difficulty of borrow-checking
this example lies in the fact that we don't know which borrow is returned by \li+choose+.
One has to note that the assert may or may not succeed (one could show that it does with Aeneas);
either way, the program remains in safe Rust.
%
\iflong
\begin{minted}[xleftmargin=1.5em,mathescape,escapeinside=||]{rust}
fn choose<'a, T>(b : bool, x : &'a mut T, y : &'a mut T) -> &'a mut T { ... }

let mut x = 0; let mut y = 0; let px = &mut x; let py = &mut y;
// $ x \mapsto \emloan\ell_x,\quad  y \mapsto \emloan \ell_y,\quad  px \mapsto \emborrow{\ell_x}0,\quad  py \mapsto \emborrow{\ell_y}0$$\lnlabel{choose:env0}$
let pz = choose(true, move px, move py); |\mmath{\lnlabel{choose-call}}|
// $x \mapsto \emloan\ell_x,\quad y \mapsto \emloan\ell_y,\quad px \mapsto \bot,\quad py \mapsto \bot,\quad pz \mapsto \emborrow{\ell_r}{\sigma},$ $\lnlabel{call:0}$
// $A(\rho) \ \{\quad \emborrow{\ell_x}\_,\quad \emborrow{\ell_y}\_,\quad \emloan\ell_r \quad \}$ $\lnlabel{call:1}$
*pz = 1; |\lnlabel{choose:assign}|
// $x \mapsto \emloan\ell_x,\quad y \mapsto \emloan\ell_y,\quad px \mapsto \bot,\quad py \mapsto \bot,\quad pz \mapsto \emborrow{\ell_r}{\sigma'},\;\qquad\quad\quad\;$         $\hspace{.9ex}$step 0$\lnlabel{step0:0}$
// $A(\rho) \ \{\quad \emborrow{\ell_x}\_,\quad \emborrow{\ell_y}\_,\quad \emloan\ell_r \quad \}$ $\lnlabel{step0:1}$
//
// $x \mapsto \emloan\ell_x,\quad y \mapsto \emloan\ell_y,\quad px \mapsto \bot,\quad py \mapsto \bot,\quad pz \mapsto \bot,$                             $\,$step 1$\lnlabel{step1:0}$
// $A(\rho) \ \{\quad \emborrow{\ell_x}\_,\quad \emborrow{\ell_y}\_,\quad \sigma' \quad \}$ $\lnlabel{step1:1}$
//
// $x \mapsto \emloan\ell_x,\quad y \mapsto \emloan\ell_y,\quad px \mapsto \bot,\quad py \mapsto \bot,\quad pz \mapsto \bot,$                             $\,$step 2$\lnlabel{step2:0}$
// $\_ \mapsto \emborrow{\ell_x}{\sigma_x},\quad \_ \mapsto \emborrow{\ell_y}{\sigma_y}$ $\lnlabel{step2:1}$
//
// $x \mapsto \sigma_x,\quad y \mapsto \emloan\ell_y,\quad px \mapsto \bot,\quad py \mapsto \bot,\quad pz \mapsto \bot,\quad \_ \mapsto \bot,\quad \_ \mapsto \emborrow{\ell_y}{\sigma_y}$       step 3$\lnlabel{step3}$
assert!(x >= 0);
\end{minted}
\fi
Up to line~\lnref{choose:env0}, \llbcs coincides with LLBC; we have local knowledge regarding
variables in scope, and symbolic execution coincides with concrete execution.
Line~\lnref{choose-call} is where the action happens: the caller invokes \li+choose+ without
any knowledge of its definition; \llbcs is a modular analysis. The only
information is from the type of \li+choose+,
which states that upon calling
\li+choose+, the caller relinquishes ownership of the arguments (here,
$px$ and $py$); in exchange, the caller obtains a new borrow
(here, $pz$).
The region (also called lifetime) annotations, which state that the input
and outputs borrow have the same lifetime \li+'a+, have to be understood as follows:
for as long as the output borrow, i.e., $pz$, is alive, we have to consider that the input
borrows are alive, i.e., that $x$ and $y$ are borrowed and
thus not accessible.

To account for this call, \llbcs uses several mechanisms. First,
$pz$ points to a symbolic value, $\sigma$. Second,
$px$ and $py$ now point to $\bot$, to account for the fact that
they have been moved out (in practice, the Rust compiler inserts reborrows that
we elide here for readability).
Third, for each lifetime, a fresh region
$\rho$ is introduced together with $A(\rho)$ (``the region abstraction of $\rho$''),
a device which encodes the
relationship between the input and output values. Namely, $A(\rho)$
encodes both that the callee owns the arguments provided by the caller
($A(\rho)$ now borrows $x$ and $y$ through $\ell_x$ and $\ell_y$; the borrowed values
are replaced with $\_$, because remembering them is not useful), and that the caller
owns a borrow that belongs to the region of the callee ($A(\rho)$ loans out
the value $\sigma$ to the caller, through $\ell_r$). The content of the fresh region abstraction ($A(\rho)$)
and the value assigned to $pz$ are derived from the signature
of \li+choose+ (\Rule{E-Call-Symbolic}). We leave $\kw{inst\_sig}$ to the
\iflong
Appendix;
\fi
suffices to say that, in order to instantiate a signature, one has to
project the borrows inside each argument (here, $\ell_x$ and $\ell_y$) onto the region
abstraction they belong to; dually, borrows within the output (here, $\ell_r$)
need to be associated to loans in their corresponding region abstractions.

Symbolic values and region abstractions are the two key ingredients that turn
LLBC into \llbcs, its symbolic counterpart. Symbolic values behave as expected:
as we increment $pz$ (line~\lnref{choose:assign}), we simply learn that $pz$ points to
a fresh symbolic value $\sigma'$ (step 0).
Region abstractions can be transformed according
to specific rules; we show a few intuitive ones, namely: 
a loan inside a region abstraction can be ended like any regular loan
(step 1, where we invalidate $pz$);
a borrow inside a region abstraction cannot be directly terminated,
however region abstraction themselves can be
terminated provided they don't contain loans anymore (\Rule{Reorg-End-Abs}),
handing all the borrows being held back
to the caller, inside fresh (ghost) anonymous variables (step 2, where we get access back
to the borrows of $x$ and $y$);
as the values of $x$ and $y$ might have been modified by \li+choose+ or through the
borrow $pz$, we introduce fresh symbolic values ($\sigma_x$ and $\sigma_y$) when
putting those borrows back in the environment.
Previous LLBC rules also apply over symbolic values, meaning a borrow
can be terminated and the loan originator regains the value, albeit a symbolic
one (step 3, where we regain access to $x$ by ending $\ell_x$), thus allowing the
assert to borrow-check.
We could also regain access to $y$ by ending $\ell_y$; we do not do so as it is not necessary.

The rules applying to region abstractions encode the contracts enforced by function
signatures; in the present case with \li+choose+ we get that:
for as long as the borrow $pz$ is alive, $x$ and $y$ are borrowed and thus
inaccessible; if we want to retrieve access to $x$ or $y$, we have to end $pz$
first (step 1), which gives us access back to \emph{both} the borrows of $x$ and $y$ at the same
time (step 2), in return allowing us to get access back to $x$ and $y$ (step 3),
albeit with (potentially) updated values.



\subsection{Simulation Relation}
\slabel{llbcssim}

Considering \llbcs as a borrow checker, we now aim to establish a property akin to type safety.
As \llbcs is defined as a semantics instead of a set of inference rules, this corresponds to
a forward simulation. 
We assume that programs are executing in an environment $\mathcal{P}$,
which consists of a set of function definitions alongside their signature.
Formally, we aim to prove the following property: 
\begin{theorem}
  For all states $\Omega$ and $\Omegas$, statement $s$, and $\Ss$ set of
  states with outcomes (i.e., pairs of a control-flow outcome and a state), we have:
  \begin{align*}
  &(\forall\; f \in \mathcal{P},\; \borrowchecks{f}) \Rightarrow
  \Omega \le \Omegas \Rightarrow
  \Omegas \vdashLlbcs\, s \rightsquigarrow \Ss \Rightarrow\\
  &(\Omega \vdashLlbc s \rightsquigarrow \infty)
   \eor (\exists\; \Omega_1,\, \Omega \vdashLlbc s \stmtres{\epanic}{\Omega_1}) \eor\\
  &(\exists\, \Omega_1\, \Omegas_1\,
    r \in \{(),\,\ereturn,\,\ebreak i,\, \econtinue i \},\,
    \Omega \vdashLlbc\, s \stmtres{r}{\Omega_1} \eand \Omega_1 \le \Omegas_1
    \eand (r,\; \Omegas_1) \in \Ss)
  \end{align*}
  \thlabel{borrow-checks}
\thmvspace
\end{theorem}
This property should be understood as follows. We assume that all functions appearing
in the environment have been borrow-checked to match their signature, as represented
by the predicate $\borrowchecks{f}$; in practice, this can be done independently for
each function, relying on the modularity of borrow-checking (see \Section{borrow-checks} below).
Then for all states
$\Omega$ in LLBC, $\Omegas$ in \llbcs initially in relation, for all evaluations of
a program $s$ starting from $\Omegas$ and returning a set of abstract states $\Ss$,
there exists a related execution of $s$ in LLBC that either diverges, panics (panicking is
safe), or yields
a result related to one of the states in $\Ss$.
We note that, as in our case empty environments are trivially in relation, we get the
standard typing result, that is: given some entry point to the program, say a \li+fn main()+
function, an evaluation of the program starting in the empty state is safe.

\myparagraph{Local Transformations}
Similarly to the previous section, this proof will rely on the methodology outlined
in \Section{proofmeth}. We describe below its different components.
The first step in our methodology is to define a set of local transformations, whose
reflexive transitive closure will allow turning a concrete LLBC state into its abstract
\llbcs counterpart; we present selected rules in \Figure{llbcp-rel-selected}. Similarly
to how transformations between HLPL and LLBC led to the \hlplp language, transformations
between LLBC and \llbcs commonly span hybrid states that do not belong to either language.
We dub this hybrid, union language \llbcp, and define transformations as operating on \llbcp
states. The semantics of \llbcp is almost exactly the union of LLBC and \llbcs. To ensure
that LLBC remains a stable subset of \llbcp, a key ingredient of our approach, we exclude
\Rule{E-Call-Symbolic}, the only rule introducing symbolic values and region abstractions;
similarly to our handling of loans and borrows in \hlplp, they will instead be
introduced through the state transformations. We now detail several of the rules
that induce the $\le$ relation on \llbcp states.
Contrary to \hlplp where we explained the transformations from right to left, in
the case of \llbcp it is more natural to go from left to right, from concrete
to abstract; we adopt this convention in the rest of the section.

\begin{figure}
  \centering
  \smaller
  \begin{mathpar}
    \LeToSymbolicRule
    
    \LeToAbsRule

    \LeMergeAbsRule

    \MergeAbsUnionRule
    
    \LeMoveValueRule

    \ToAbsPairRule
    
    \LeReborrowMutBorrowAbsRule

    \ToAbsMutLoanRule
    
    
    \MergeAbsMutRule

    \ToAbsMutBorrowRule
  \end{mathpar}
  \caption{The Relation $\le$ about \llbcp States (Selected Rules)}
  \flabel{llbcp-rel-selected}
  \figvspace
\end{figure}

We designed the rules of $\le$ so that they allow us to lose information about
the state.
\Rule{Le-ToSymbolic} is one of the simplest transformation rules. If we have a plain,
concrete value (i.e., that does not contain any loans, borrows, or $\bot$), then we can
forget its precise value by transforming it into a fresh symbolic value.
\Rule{Le-MoveValue} implements a move: so long as no one else relies on $v$ (as
captured by the premises), $v$ can be moved out into an anonymous variable; in effect,
this allows us to lose the information that the owner of $v$ has access to it.
\Rule{Le-ToAbs} captures the core rewriting to go from LLBC to \llbcs:
it allows abstracting away parts of the borrow graph by moving borrows and loans associated
to anonymous variables into fresh region abstractions, thus forgetting their precise relationships.
To do so, it relies on $\MergeAbs$, which we explain below,
and on the $\toabs$ judgement,
which transforms a (possibly complex) value $v$ into a set of fresh region
abstractions $\overrightarrow{A}$.

The $\toabs$ judgment relies on $\cup$, which is 
plain set union.
\Rule{ToAbs-MutLoan} simply transfers the loan to a fresh region abstraction.
Some values, such as pairs, may contain several loan identifiers, and as such,
give rise to several, independent region abstractions (\Rule{ToAbs-Pair}).
\Rule{ToAbs-MutBorrow} converts the inner borrowed value into a set of
region abstractions, computes their union, then adds the outer mutable borrow to
the result. It is particularly useful to abstract away reborrow patterns: going back to the
example of \Section{background-llbc}, after the reborrow (line~\lnref{reborrow})
we have the state:
$x \mapstoemloan{0},\; \_ \mapstoemborrow{0}{(\emloan{\ell_1})},\; px \mapstoemborrow{1}{0}$.
We can abstract away the link between $px$ and $x$ by using \Rule{Le-ToAbs},
resulting in the state:
$x \mapstoemloan{0},\; \Abs{A}{\emborrow{\ell_0}{\_},\, \emloan{\ell_1}},\; px \mapstoemborrow{1}{0}$.
The presence of $A$ enforces the constraint that: if we want to end $\ell_0$ to
retrieve access to $x$, then we first need to invalidate $px$ by ending $\ell_1$
(by \Rule{Reorg-End-Abs} we can end a region abstraction only if it doesn't
contain any loans); this constraint is very similar to what we had in the original state,
albeit the fact that we don't know anymore that $\ell_1$ is \emph{exactly} a reborrow
of $\ell_0$.

The non-deterministic $\MergeAbs$ operator \emph{merges} two different region
abstractions using semantic criteria.
When interpreting Rust programs,
a region abstraction can be understood as a set of values associated to a given lifetime.
Merging two region abstractions therefore corresponds to a notion of lifetime weakening:
if we have two distinct lifetimes, Rust allows adding lifetime constraints to guarantee
that borrows associated to both lifetimes will be ended at the same time. This pattern
frequently occurs when typechecking a function body against a more restrictive lifetime
signature.

A naive, but sound interpretation of merging regions $A_0$ and $A_1$ consists of
taking the union of all values in both regions using rule \Rule{MergeAbs-Union}. 
Consider for instance the state:
\begin{math}
  x \mapstoemloan{x},\,
  y \mapstoemloan{y},\,
  \AbsI{0}{\emborrow{\ell_x}{\_},\,\emloan{\ell_p}},\,
  \AbsI{1}{\emborrow{\ell_y}{\_}},\,
  p\mapstoemborrow{p}{\sigma}
\end{math}.
In this state, we have to invalidate $p$ to get access back to $x$, but can
independently get access back to $y$ by ending $A_1$ then $\ell_y$.
If we merge $A_0$ and $A_1$ we produce
$\AbsI{2}{\emborrow{\ell_x}{\_},\, \emborrow{\ell_y}{\_},\, \emloan{\ell_p}}$;
in the new state we can no longer retrieve access to $y$ without invalidating $p$.

However, we can also perform more precise transformations, for instance by removing
a mutable loan and its associated borrow (\Rule{MergeAbs-Mut}).
Intuitively, this rule allows hiding them in the internals
of the region abstraction: ending the merged abstraction amounts, in the original
state, to ending the first abstraction, all the borrows and loans that were hidden
by means of \Rule{MergeAbs-Mut}, then the second abstraction; in the state
resulting from the merge, we simply abstract all those steps away.
For instance, we can merge two region abstractions
$\AbsI{0}{\emborrow{\ell_0}{\_},\; \emloan{\ell_1}}$ and
$\AbsI{1}{\emborrow{\ell_1}{\_},\; \emloan{\ell_2}}$ into
$\AbsI{2}{\emborrow{\ell_0}{\_},\; \emloan{\ell_2}}$; this makes $\ell_1$ disappear.

In practice, we devise an algorithm to judiciously apply the $\MergeAbs$ rules. Indeed,
greedily applying \Rule{MergeAbs-Union}
instead of leveraging \Rule{MergeAbs-Mut} creates an abstraction which contains
both a loan and its associated borrow; we can never end such abstractions because of a
cyclic dependency between ending the abstraction and ending the borrow, eventually
leading the symbolic evaluation to get stuck.
We emphasize that $\MergeAbs$ is not symmetric: \Rule{MergeAbs-Mut} is directed, and
there is no version of it with a borrow on the left, and a loan on the right.
This is necessary for soundness.
Consider environments
$A_0 = \{\emborrow{\ell_2}{v},\; \emloan{\ell_0},\; \emborrow{\ell_1}{v_1}\}$ and
$A_1 = \{\emloan{\ell_1},\; \emborrow{\ell_0}{v_0}\}$. This symbolic state
features a cyclic dependency (perhaps, because of poorly chosen uses
of \Rule{Le-MergeAbs}) -- it is thus crucial, for our proof of forward
simulation, to make sure that such a symbolic state remains stuck. The directionality
of \Rule{MergeAbs-Mut} guarantees just that: if $A_0$ is on the left of the $\MergeAbs$
we can eliminate $\ell_0$ but not $\ell_1$; if $A_0$ is on the right
we can eliminate $\ell_1$ but not $\ell_0$;
we thus rightfully prevent borrow-checking from succeeding.


Building on the transformations previously presented, we now focus on the simulation
proof itself. As outlined in \Section{proofmeth}, the proof can be broken down in three
steps: we need to establish a forward simulation on \llbcp, prove that LLBC is a stable
subset of \llbcp, and conclude by reasoning on the remaining rules in \llbcs that
were excluded from \llbcp.

The second point can be easily obtained by induction on an \llbcp evaluation. The first
point is tedious, but straightforward and similar to the proof between HLPL and LLBC.
The core idea is that the local transformations turn a state into a more
abstract version, thus allowing fewer execution steps.
Compared to \hlplp, the novel part of the proof is the third point, which requires
relating an abstract, modular execution of a function call
(\Rule{E-Call-Symbolic}, \Figure{llbcs-selected}) to its
concrete counterpart which enters the function body (\Rule{E-Step-Call},
\Figure{llbc-index}).
Before doing so, we focus briefly on how exactly symbolic execution works, and
show how to modularly borrow-check a function in \llbcs.

\subsection{Borrow-Checking a Program}
\slabel{borrow-checks}

\begin{figure}
  \centering
  \smaller
  \begin{mathpar}
    \LlbcsRules
  \end{mathpar}
  \caption{Operational Semantics for \llbcs (Selected Rules)}
  \flabel{llbcs-selected}
  \figvspace
\end{figure}

\begin{figure}
\centering
\smaller
\[
  \begin{array}{l}
    \kw{borrow\_checks}\;(\fsignoindices{\overrightarrow{\rho}})\;:= \\
    \qquad \kw{let}\ \vec v,\;\vecabsinputrho = \kw{init}(\vec\rho,\vec\tau) \\
    \qquad \kw{let}\ \Omegas_0 = \vecabsinputrho,\; \overrightarrow{x \rightarrow v},\;
      \overrightarrow{y\rightarrow\bot},\;\xret\rightarrow\bot \\
    \qquad \kw{let}\ \vout,\;\vecabssigrho = \kw{final}(\vec\rho,\vec\tau, \tau_\kw{ret}) \\
    \qquad \kw{let}\ \Omegas_1 = \overrightarrow{\abssigrho},\; \overrightarrow{x\rightarrow\bot},\;
      \overrightarrow{y\rightarrow\bot},\;\xret\rightarrow\vout \\
    \qquad \exists\; \Ss,\;
      \Omegas_0\vdashLlbcs s \stmtarrow \Ss \eand\\
    \qquad \quad\;\forall\; res \in \Ss,\;\exists\; \Omegas,\;
    res = (\epanic,\; \Omegas) \eor (res = (\ereturn,\; \Omegas) \eand \Omegas \le \Omegas_1)
\end{array}
\]
\caption{The Borrow Checking Predicate For Functions}
\flabel{borrow-checks}
\figvspace
\end{figure}

The actual setup of the symbolic execution is performed by
\kw{borrow\_checks} (\Figure{borrow-checks}). The $\kw{init}$ function
\iflong
(in Appendix),
\fi
given the types of the arguments $\vec\tau$, initializes a set of input values
$\vec v$ by allocating fresh symbolic values and borrows, along with an initial set of region abstractions
$\vecabsinputrho$ which contain their associated loans, and materialize the signature and its lifetimes;
this then serves to create the initial symbolic environment
$\Omegas_0$, where the input parameters are initialized with $\vec v$, and
where the remaining local variables and the special return variable are uninitialized.
Symmetrically, $\Omegas_1$ captures the expected region abstractions upon
exiting the function, and all local variables must be uninitialized except the return
variable which must contain some value $\vout$; note that this value might contain borrows which refer
to loans appearing in the region abstractions $\overrightarrow{\abssigrho}$, as is
actually the case for \li+choose+ below.
The function then borrow-checks if
executing the body $s$ in $\Omegas_0$ is safe, and leads to states
which either panic or are in relation with $\Omegas_1$.

We illustrate borrow-checking on the \li+choose+ function, below.
The $\kw{init}$ function computes an initial state following
the function signature, lines 2-3. The function has no local variables, so
the $\vec y$ from $\kw{borrow\_checks}$ are absent. The return value is
uninitialized (line 3).
In this state, the variable $b$ contains a symbolic value $\sigma_b$,
while $x$ and $y$ borrow some other symbolic values, their associated loans
being placed in a region abstraction $A_{\text{in}}$ which holds all the loans
of lifetime \li+'a+; as \li+choose+ has only one lifetime, the initial state holds
a unique region abstraction.
Importantly, and this is a minor difference with the original formalism,
$A_{\text{in}}$ also contains some borrows $\ell^{(0)}_x$ and $\ell^{(0)}_y$,
whose corresponding loans are in an \emph{unspecified} place (i.e., not in the current
state).
As such, this initial state is a partial state that can be composed with other
partial states to form a complete state, where in particular all borrows have an
associated loan; we will use this in the proof of the forward simulation, by
applying framing lemmas in the same spirit as the frame rule in separation logic~\cite{ohearn2001frame}.
In this context, $A_{\text{in}}$ really acts as an abstraction barrier between the local state
(the callee) and some external state (the caller); intuitively, $\ell^{(0)}_x$ should
be exactly $\ell_x$ while $\ell^{(0)}_y$ should be exactly $\ell_y$, but we abstracted
this information away.

\begin{minted}[xleftmargin=1.5em,mathescape=true,escapeinside=||]{rust}
fn choose<'a, T>(b : bool, x : &'a mut T, y : &'a mut T) -> &'a mut T {
  // $\AbsI{\text{in}}{\emborrow{\ell^{(0)}_x}{\_},\,\emborrow{\ell^{(0)}_y}{\_},\,\emloan{\ell_x},\,\emloan{\ell_y}},$
  // $b\mapsto\sigma_b,\quad x\mapsto\emborrow{\ell_x}{\sigma_x},\quad y\mapsto\emborrow{\ell_y}{\sigma_y},\quad \xret\mapsto\bot$
  if b { ret = move x;
    // $\AbsI{\text{in}}{\emborrow{\ell^{(0)}_x}{\_},\,\emborrow{\ell^{(0)}_y}{\_}\emloan{\ell_x},\,\emloan{\ell_y}},$
    // $b\mapsto\kw{true},\quad x\mapsto\bot,\quad y\mapsto\emborrow{\ell_y}{\sigma_y},\quad \xret\mapsto\emborrow{\ell_x}{\sigma_x}$
    return; // $\le \AbsI{\text{out}}{\emborrow{\ell^{(0)}_x}{\_},\,\emborrow{\ell^{(0)}_y}{\_},\,\emloan{\ell}},\, b\mapsto\bot,\, x\mapsto\bot,\, y\mapsto\bot,\, \xret\mapsto\emborrow{\ell}{\sigma}$
  } else { ret = move y; return; } }
\end{minted}

Upon reaching the \li+return+, symbolic execution yields the state at lines 5-6, where
the special return value variable $ret$ now contains the
borrow coming from $x$.
We show at line 7 the target output state, computed via
$\kw{final}$: the goal is now to establish that the final
environment (as computed by the symbolic execution) refines the output
environment (as determined by the function signature).

We first reorganize the context by ending
all the outer loans (loans which are themselves not inside borrows)
which we see in local variables.
We then repeatedly compare the two states while applying
$\le$ rules until they match: we move the values contained by the local variables (except
$ret$) into
fresh region abstractions (\Rule{Le-MoveValue}, \Rule{Le-ToAbs}), transform the values contained by the
return variable $ret$ into symbolic values (\Rule{Le-ToSymbolic}), and merge region
abstractions together (\Rule{Le-MergeAbs}). We end up in the target state of line 7;
we then do the same for the \li+else+ branch (omitted), which allows us to conclude that
\li+choose+ satisfies its signature. In other words, \li+choose+ borrow-checks.

\myparagraph{Forward Simulation Between \llbcp and \llbcs}
We now resume the presentation of the proof of
the simulation between LLBC and \llbcs: there remains to show
that we can replace \Rule{E-Call} with
\Rule{E-Call-Symbolic}.
There are several crucial points.
First, evaluation rules are local, which means the
evaluation relation is also defined for partial states in
which borrows don't necessarily have an associated loan; equipped
with partial states, we define and prove a framing lemma in the spirit of separation
logic, which states that if state $\Omegas_0$ evaluates to
$\Omegas_1$, we can compose $\Omegas_0$ with a disjoint frame $\Omegas_f$ which doesn't
get modified during the evaluation.
As the $\le$ rules are also local, we define a similar framing
property for the $\le$ relation.
Finally, we carefully designed \Rule{E-Call}, \Rule{E-Call-Symbolic} and the
$\kw{borrow\_checks}$ predicate so that: 1. we can always turn a part of
the concrete input state into a more abstract partial state that is in relation
with $\Omegas_0$; 2. applying the frame rules to the state appearing in the
conclusion of $\kw{borrow\_checks}$ produces the state resulting from
\Rule{E-Call-Symbolic}.
There are some other technicalities; we refer the interested reader to the
\iflong Appendix\fi.

\subsection{Backward Simulation From LLBC to PL}
\slabel{backllbcpl}

The forward simulation from \llbcs to LLBC allows us to conclude about the soundness of \llbcs
as a borrow-checker. However, it also guarantees the existence of a backward simulation between
LLBC and PL for programs that successfully borrow-check; we formalize this in Theorem~\thref{llbcplbackward}.
This theorem states that, assuming that all functions in program $P$ borrow-check, and that the state
$\Omegallbc$ is in relation with the initial borrow-checking state $\Omegas_0$ for function $g$,
then any PL execution of $g$ starting from a related state $\Omegapl$ has a related LLBC execution.
Combined with the forward simulation from LLBC to PL proven in \Section{refine}, this provides
the main result of our paper: we have a bisimulation relation between LLBC and PL for programs
that borrow-check.

\begin{theorem}[Backward Simulation Between PL and LLBC]
For all $\Omegapl$ PL state and $\Omegallbc$ LLBC state, for all function
$g\langle\vec{\rho},\,\vec{\tau}\rangle$, we have:
\begin{align*}
  \smaller
& \kw{let}\ \vec v,\;\vecabsinputrho = \kw{init}(\vec\rho,\vec\tau); \; \;
  \kw{let}\ \Omegas_0 = \vecabsinputrho,\; \overrightarrow{x \rightarrow v},\;
      \overrightarrow{y\rightarrow\bot},\;\xret\rightarrow\bot; \\
&(\forall\; f \in \mathcal{P},\; \borrowchecks{f}) \Rightarrow
\Omegapl\le\Omegallbc\Rightarrow \Omegallbc \le \Omegas_0 \Rightarrow
\forall\; res,\;
\Omegapl \vdash g.\kw{body} \stmtarrow res \Rightarrow\\
&\exists\; res',\;
\Omegallbc \vdash g.\kw{body} \stmtarrow res' \eand\\
&\quad(res = res' = \infty) \eor (\exists\; r\; \Omegapl_1\; \Omegallbc_1,\;
res = (r,\, \Omegapl_1) \eand res' = (r,\, \Omegallbc_1) \eand
\Omegapl_1 \le \Omegallbc_1)
\end{align*}
\thlabel{llbcplbackward}
\thmvspace
\end{theorem}

Building on the previous sections, the proof is straightforward:
if all functions in program $\mathcal{P}$ borrow-check, so does
$g$, which by definition means that there exists an \llbcs evaluation of $g.\kw{body}$ starting from $\Omegas_0$.
By forward simulation, this implies the existence of an LLBC evaluation starting from $\Omegallbc$
and returning a result $res_{llbc}$, which
in turn implies the existence of a related PL evaluation starting from $\Omegapl$ and returning $res_{pl}$.
As PL is deterministic, we derive $res = res_{pl}$ for any PL execution starting from $\Omegapl$, and
conclude by exhibiting the witness $res' = res_{llbc}$.

\section{Merging Abstract Environments}

As we saw in the previous section, \llbcs offers a sound, modular borrow checker for the
LLBC semantics. However, its approach based on a symbolic collecting semantics
struggles with scalability when considering disjunctive control flow. Consider the rule
\Rule{E-IfThenElse-Symbolic}, which defines the \llbcs semantics for evaluating
common if-then-else constructs. This rule evaluates both branches independently,
yielding sets of states with outcomes $\Ss_0$ and $\Ss_1$, and finally returns their union;
each state in the resulting set is then considered as a starting point to evaluate
the rest of the program. When a program contains many branching constructs, this
leads to a combinatorial explosion known to symbolic execution practitioners as
the path explosion problem. This issue becomes even worse when considering loops,
as the number of paths to analyze can be infinite.

To circumvent this issue, one possible solution is to merge control flow paths at different
program points, typically after the end of a branching statement. Doing so requires soundly
merging environments or, borrowing terminology from abstract interpretation, ``computing
a join''~\cite{cousot77}.
In this section, we show how to define such an operator for
abstract, borrow-centric environments. Leveraging the forward simulation from \llbcs to LLBC,
we then prove its soundness with respect to the LLBC semantics.

\subsection{Joining Environments}
\slabel{joins}

We present in \Figure{join-collapse} representative rules for our join operator.
To make the presentation easier to follow, we will rely on the following small example,
which is a standard Rust pattern after desugaring. We annotate this program with the \llbcs
environments at relevant program points.
Our goal is to compute the join of the environments after evaluating both branches;
we show the resulting environment below,
and explain in the rest of this section how it is computed.

\iflong
\begin{minted}[linenos=false,mathescape=true,escapeinside=||]{rust}
// $b \mapsto \sigma_0$
x = 0; y = 1; px = &mut x; py = &mut y;
// $b \mapsto \sigma_0,\; x \mapstoemloan{0},\; y \mapstoemloan{1},\; px \mapstoemborrow{0}{0},\; py \mapstoemborrow{1}{1}$
if b {
  p = move px; // $b \mapsto \kw{true},\, x \mapstoemloan{0},\, y \mapstoemloan{1},\, px\mapsto\bot,\, py \mapstoemborrow{1}{1},\,p\mapstoemborrow{0}{0}$
}
else {
  p = move py; // $b \mapsto \kw{false},\, x \mapstoemloan{0},\, y \mapstoemloan{1},\, px \mapstoemborrow{0}{0},\, py \mapsto\bot,\,p\mapstoemborrow{1}{1}$
}
// Result of the join:
// $b\mapsto\sigma_1,\;x \mapstoemloan{0},\; y \mapstoemloan{1},\; px \mapsto \bot,\; py\mapsto\bot,$
// $p \mapstoemborrow{2}{\sigma},\;\AbsI{}{ \emborrow{\ell_0}{\_},\, \emborrow{\ell_1}{\_},\, \emloan{\ell_2} }\lnlabel{choose-join}$
\end{minted}
\fi

We start with some high-level explanations before giving a formal description of the
join operation.
Intuitively, our goal is to compute an environment $\Omega_2$ which is
more general than the environments $\Omega_0$ and $\Omega_1$ resulting from
evaluating the \li+then+ and \li+else+ branches, respectively.
Our target property is that the resulting environment is in relation with the environments
we join, that is: $\Omega_{0, 1} \le \Omega_2$; we can then resume evaluation
with $\Omega_2$ instead of $\Omega_0$ and $\Omega_1$, and still get a forward simulation
between LLBC and \llbcs.

A join is naturally computed as a pointwise operation over the variables in both
environments: for each variable, we compute the join of its associated values.
If we have the same value on both sides we keep it unchanged (e.g., $x$ and $y$ above).
If the values differ but don't contain loans, borrows, or $\bot$ (e.g., $b$ which
maps to either \kw{true} or \kw{false}), we simply introduce a fresh symbolic value to
account for the fact that both environments have access to a valid value, but we don't
have more information about it.
An interesting case happens when we have $\bot$ on one side but not on the other (e.g.,
$px$ and $py$): if one environment doesn't have ownership of the value at a given
place, we have to consider that the joined environment doesn't have either, and thus
contains $\bot$. Generally speaking, when the permissions to access a place differ in both
environments, we have to be conservative by taking their greatest lower bound; for instance, if a value
is loaned in one environment, we have to consider it as loaned after the join.
The last interesting case happens when borrows are involved, for instance with $p$
which borrows either $x$ (through $\ell_0$) or $y$ (through $\ell_1$).
The device which allows encoding a loss of information in the borrow graph
is the region abstraction, which was initially introduced
to handle function calls (\Section{llbcsbackground}), but works for joins as well.
In the present case, we can encode the constraint
that $p$ borrows either $x$ or $y$ by introducing a fresh borrow
$\ell_2$ linked to $\ell_0$ and $\ell_1$ through the fresh region abstraction $A$;
in the resulting environment, we get that $p$ is a valid borrow, $x$ and $y$ are loaned,
and recovering access to $x$ \emph{or} $y$ requires invalidating $p$.

In practice, we perform the join in two phases. We first compute a pointwise join of the
variables in both environments. As this pointwise join may produce an
ill-formed environment where some borrows and loans are duplicated (see below), we
need to chain it with a \emph{collapse} phase, which removes those duplications
to produce a well-formed environment. The join then collapse operation satisfies the
target property; that is, if successful, it produces an environment in relation with
the environments before the join (\Theorem{join-collapse-le}).\\

Formally, we write \joinenv{\Omega}{\Omega'}{\Omega_0}{\Omega_1}{\Omega_2} to denote
that the join of environments $\Omega_0$ and $\Omega_1$ yields a new environment $\Omega_2$.
We define the join operator inductively on the environments. The states $\Omega$ and $\Omega'$
on the left of the turnstile correspond to the top-level environments being joined, which might
differ from $\Omega_0$ and $\Omega_1$ inside recursive derivations. A few specific rules
in our complete presentation require access to them, but they can be safely ignored
in this section.

\myparagraph{Joining Values}
Joins are computed pointwise: if a variable $x$ is present in both environments,
we perform a join on its associated value (\Rule{Join-Var}). Value joins are formally
defined using the judgement \joinval{\Omega_0}{\Omega_1}{v_0}{v_1}{v}{\overrightarrow{A}}.
This judgment is similar to the environment join above: it states that merging values $v_0$
and $v_1$ yields a new value $v$, and creates a set of region abstractions $\overrightarrow{A}$
to be added to the current environment.
When values are the same, for instance $x$ and $y$ in our example above, the join is the identity (\Rule{Join-Same}). When values differ but do not contain borrows, loans, or $\bot$, e.g.,
variable $b$, a fresh symbolic value is returned (\Rule{Join-Symbolic}).

The more interesting cases occur when borrows or loans are involved. In the example above,
let us look at the variable $p$, which corresponds to a mutable borrow in both branches,
although associated to loan identifiers $\ell_0, \ell_1$ and values $0, 1$ respectively.
A naive way to join these borrows would be to create a new borrow associated to a fresh
loan identifier $\ell_2$ and a fresh symbolic value $\sigma$, and to constrain $\ell_2$ to end whenever trying to end either
$\ell_0$ or $\ell_1$; this can be done by creating a fresh region abstraction containing
$\emborrow{\ell_0}{0},\,\emborrow{\ell_1}{1}$, and $\emloan{\ell_2}$. This attempt
unfortunately does not work in practice, as it might lead to invalid states containing
duplicated mutable borrows: coming back to our example, the left environment still contains
$\emborrow{\ell_1}{1}$ associated to $y$. To fix this issue, we additionally keep track of
the origin of values (\Rule{Join-MutBorrows}).
For presentation purposes, we will denote a value $v$ coming from the
left (resp. right) environment as $\lbox{v}$ (resp. \rbox{v}). 
We follow a similar approach to join a value with $\bot$, e.g., for variables $px$ and $py$
(\Rule{Join-Bottom-Other}, \Rule{Join-Other-Bottom}), ultimately
leading to the joined environment below.
%
%
{\smaller
\begin{align*}
&b\mapsto\sigma_1,\qquad x \mapstoemloan{0},\qquad y \mapstoemloan{1},\\
&px \mapsto \bot,\qquad \AbsI{0}{ \rbox{\emborrow{\ell_0}{\_}} },\qquad
py \mapsto \bot,\qquad \AbsI{1}{ \lbox{\emborrow{\ell_1}{\_}} },\\
&p \mapstoemborrow{2}{\sigma},\qquad
\AbsI{2}{ \lbox{\emborrow{\ell_0}{\_}},\, \rbox{\emborrow{\ell_1}{\_}},\, \emloan{\ell_2} }
\end{align*}}

\myparagraph{Collapsing Environments}
While keeping track of the origin of values avoids inconsistencies due to duplicated
borrows, it gives rise to new environments that do not belong to \llbcs. Instead
of extending the semantics, we propose a set of local transformations that gradually turns
an environment with marked values (e.g., $\lbox{v}$) back into an \llbcs state. We dub
their transitive closure the \emph{collapse} operator, which we denote $\collapse$,
and show several rules in \Figure{join-collapse}.

Coming back to our running example, we aim to collapse the joined environment to remove
all markers. To do so, we first apply \Rule{Collapse-Merge-Abs} twice to merge abstractions
$A_0$, $A_1$, and $A_2$, using the merge rules seen in \Section{llbcssim}.
We finally use \Rule{Collapse-Dup-MutBorrow} twice to simplify
$\rbox{\emborrow{\ell_0}{\_}}$ and $\lbox{\emborrow{\ell_0}{\_}}$
into $\emborrow{\ell_0}{\_}$, then $\lbox{\emborrow{\ell_1}{\_}}$ and
$\rbox{\emborrow{\ell_1}{\_}}$ into $\emborrow{\ell_1}{\_}$ to obtain the following
\llbcs environment.
{\smaller
\begin{align*}
&b\mapsto\sigma_1,\qquad x \mapstoemloan{0},\quad y \mapstoemloan{1},\quad
px \mapsto \bot,\quad
py \mapsto \bot,\quad\\
&p \mapstoemborrow{2}{\sigma},\quad
\AbsI{3}{ \emborrow{\ell_0}{\_},\, \emborrow{\ell_1}{\_},\, \emloan{\ell_2} }
\end{align*}}

\myparagraph{Soundness}
We now set our sights on proving the soundness of the join and collapse operators.
Formally, we aim to prove the following theorem, which states that for any \llbcs
states $\Omega_l$, $\Omega_r$, if the composition of join and collapse yields an
\llbcs state $\Omega_c$, that is, a state with no marked values, then this state
is related to both $\Omega_l$ and $\Omega_r$.
We can then resume the evaluation with the joined state, instead of the set of
states resulting from the two branches.

\begin{theorem}[Join-Collapse-Le]
  \thlabel{join-collapse-le}
  For all $\Omega_l$, $\Omega_r$, $\Omega_j$, $\Omega_c$ we have:
  \begin{align*}
  &\joinenv{\Omega_l}{\Omega_r}{\Omega_l}{\Omega_r}{\Omega_j} \Rightarrow
  \;\vdash \Omega_j \collapse \Omega_c \Rightarrow
  \text{no marked value in } \Omega_c \Rightarrow
  \forall\; m \in \{l,\,r\},\, \Omega_m \le \Omega_c
  \end{align*}
\thmvspace
\end{theorem}

The proof relies on an induction on the reductions for join and collapse.
To explain the intuition behind the proof, we will consider state projections keeping
marked values from only one side. For presentation purposes, we will focus on the environment
on the left. The state projection is then defined as discarding all values from the right side
(e.g., $\rbox{v}$) and removing the left markers (e.g., replacing $\lbox{v}$ by $v$).
Then, the intuition is that the left environment will always be in
relation with the left projection of the join.
Using this notion of projection, there is almost a one-to-one mapping between the rules
defining join and collapse on one side, and the rules defining the $\le$ relation on the
other.
For instance,
applying the left projection to the conclusion of \Rule{Join-MutBorrows} yields exactly
\Rule{Le-Reborrow-MutBorrow-Abs}.

One important point of this soundness theorem is that it requires that the result of
collapse does not contain any marked value; in our implementation of these rules,
we rely on several heuristics to find a derivation satisfying this condition, and
raise an error when unsuccessful. As we will see in \Section{evaluation},
while incomplete, these heuristics are sufficient to cover a large subset of Rust.

%

\subsection{Extending Support to Loops}
\slabel{loops}

The join and collapse operator we presented allows us to handle disjunctive control flow without demultiplying
the number of states to consider. While the presentation focused on simple branching, i.e., merging two environments
after an if-then-else statement, this approach also applies to more complex constructs, such as loops. 
As an example, consider the toy program below which iteratively increments variable $x$
through its mutable borrow $p$. While this program is purposedly simple,
its reborrow inside a loop is a characteristic pattern when iterating
over recursive data structures in Rust; we provide a more realistic example
in \iflong\Appendix{loop-list}\fi.

\begin{minted}[linenos=false,mathescape=true,escapeinside=||]{rust}
x  = 0; p = &mut x; // $x \mapstoemloan{0},\; p \mapstoemborrow{0}{0}$
loop {
  p = &mut (*p); // $x \mapstoemloan{0},\; \_ \mapstoemborrow{0}{(\emloan{\ell_1})},\; p \mapstoemborrow{1}{0}$
  *p += 1;       // $x \mapstoemloan{0},\; \_ \mapstoemborrow{0}{(\emloan{\ell_1})},\; p\mapstoemborrow{1}{1}$
  continue; }
\end{minted}

To borrow-check this loop, our goal is to derive a state general
enough to encompass all possible states upon entering the loop;
this is known as computing a fixpoint in program analysis.
Unfortunately, our example shows that using state inclusion to determine
if a given state is a fixpoint is not sufficient. Starting from a hypothetical
fixpoint, executing the loop body will create a new loan identifier and
an anonymous mapping during the reborrow, which after joining with the initial environment
will yield new region abstractions. Our main observation is that it is actually sufficient
to consider a fixpoint up to loan and region identifier substitution. The intuition is that,
seeing \llbcs states as shape graphs~\cite{chang2008relational,laviron2010shape},
the substitutions correspond to graph isomorphisms, which preserve the semantics of a program.

Additionally, compared to arbitrary joins, loops follow a generic pattern. The loop body
possibly introduces fresh borrows and anonymous mappings, before being merged with
an earlier snapshot. To compute a fixpoint according to the \llbcs semantics, we therefore
rely on several heuristics. First, we convert all fresh anonymous mappings
to region abstractions using \Rule{Le-ToAbs}. Second, we flatten the environment
by merging freshly introduced region abstractions that contain related borrows and loans,
that is, values associated to the same loan identifier. We finally apply the join operator
presented in the previous section to the resulting state and the initial environment, and
repeat this approach until we find a fixpoint. In our example, we get the environment
below, which remains the same after executing the loop body, up to substitution of $\ell_2$ and
$A_2$ by fresh identifiers.
\[
\smaller
x \mapstoemloan{0},\; \AbsI{2}{\emborrow{\ell_0}{\_},\, \emloan{\ell_2}},\; p \mapstoemborrow{2}{\sigma}
\]

Note that, while conceptually similar to widening operators in abstract interpretation,
we do not claim that our approach terminates. Our implementation of \llbcs fails if
the fixpoint computation does not converge after a fixed number of steps.
In practice, we however observe that these heuristics are sufficient to handle a wide range
of examples, as we demonstrate in the next section; in those examples the
computation actually converges in one step.

\NewDocumentCommand{\JoinValueRules}{s}{
  \inferrule[Join-Same]{
  }{
    \joinval{\Omega_0}{\Omega_1}{v}{v}{v}{\emptyset}
  }

  \inferrule[Join-Symbolic]{
    \text{no borrows, loans, $\bot$} \in v_0,\, v_1\\\\
    \sigma \text{ fresh}
  }{
    \joinval{\Omega_0}{\Omega_1}{v_0}{v_1}{\sigma}{\emptyset}
  }

  \inferrule[Join-Bottom-Other]{
    \text{no outer loan} \in v\\\\
    \vdash v \toabs \overrightarrow{A}
  }{
    \joinval{\Omega_0}{\Omega_1}{\bot}{v}{\bot}{\overrightarrow{\rbox{A}}}
  }
\IfBooleanT{#1}{

  \inferrule[Join-Other-Bottom]{
    \text{no outer loan} \in v\\\\
    \vdash v \toabs \overrightarrow{A}
  }{
    \joinval{\Omega_0}{\Omega_1}{v}{\bot}{\bot}{\overrightarrow{\lbox{A}}}
  }
}

  \inferrule[Join-MutBorrows]{
    \ell_2,\, A' \text{ fresh}\\
    \joinval{\Omega_0}{\Omega_1}{v_0}{v_1}{v_2}{\overrightarrow{A}}
  }{
    \Omega_0,\, \Omega_1 \vdash \joinvals{(\emborrow{\ell_0}{v_0})}{(\emborrow{\ell_1}{v_1})}
    \Downarrow
      \emborrow{\ell_2}{v_2} \joinres\\\\
      \Abs{A'}{
        \lbox{\emborrow{\ell_0}\_},\,
        \rbox{\emborrow{\ell_1}{\_}},\,
        \emloan{\ell_2}
      },\,
      \overrightarrow{A}
  }
\IfBooleanT{#1}{

  \inferrule[Join-SharedBorrows]{
  \ell_2,\, \sigma,\, A \text{ fresh}\\
  \text{no mut loan, borrow, $\bot$} \in v_0,\, v_1\\\\
  \esloan{\ell_0}{v_0} \in \Omega_0\\
  \esloan{\ell_1}{v_1} \in \Omega_1
  }{
  \Omega_0,\, \Omega_1 \vdash \joinvals{(\esborrow{\ell_0})}{(\esborrow{\ell_1})}
  \Downarrow\\\\ \esborrow{\ell_2} \joinres\\\\
    \Abs{A}{
      \lbox{\esborrow{\ell_0}},\,
      \rbox{\esborrow{\ell_1}},\,
      \esloan{\ell_2}\sigma }
  }

  \inferrule[Join-MutLoans]{
  \ell_2 \text{ fresh}
  }{
  \Omega_0,\, \Omega_1 \vdash \joinvals{(\emloan{\ell_0})}{(\emloan{\ell_1})}
    \Downarrow
    \emloan{\ell_2}\\\\ \joinres
    \Abs{A}{
      \emborrow{\ell_2}\_,\,
      \lbox{\emloan{\ell_0}},\,
      \rbox{\emloan{\ell_1}}
    }
  }

  \inferrule[Join-SharedLoans]{
    \text{no mut loan, borrow, $\bot$} \in v_0,\, v_1\\
    \ell_2,\, \sigma,\, A \text{ fresh}
  }{
  \Omega_0,\, \Omega_1 \vdash \joinvals{(\esloan{\ell_0}{v_0})}{(\esloan{\ell_1}{v_1})}
    \Downarrow\\\\
    \esloan{\ell_2}{\sigma} \joinres
    \Abs{A}{
      \esborrow{\ell_2},\,
      \lbox{\esloan{\ell_0}{v_0}},\,
      \rbox{\esloan{\ell_1}{v_1}}
    }
  }

  \inferrule[Join-MutLoan-Other]{
    \ell' \text{ fresh}\\
    \vdash \emborrow{\ell'}{v} \toabs \overrightarrow{A}\\\\
    \joinval{\Omega_0}{\Omega_1}{(\emloan{\ell})}{(\emloan{\ell'})}{v'}{\overrightarrow{A'}}
  }{
    \Omega_0,\, \Omega_1 \vdash \joinvals{(\emloan{\ell})}{v} \Downarrow
      v' \joinres
        \rbox{\overrightarrow{A}},\,
        \overrightarrow{A'}
  }

  \inferrule[Join-Other-MutLoan]{
    \ell' \text{ fresh}\\
    \vdash \emborrow{\ell'}{v} \toabs \overrightarrow{A}\\\\
    \joinval{\Omega_0}{\Omega_1}{(\emloan{\ell'})}{(\emloan{\ell})}{v'}{\overrightarrow{A'}}
  }{
    \Omega_0,\, \Omega_1 \vdash \joinvals{v}{(\emloan{\ell})} \Downarrow
      v' \joinres
        \lbox{\overrightarrow{A}},\,
        \overrightarrow{A'}
  }

  \inferrule[Join-SharedLoan-Other]{
    \ell' \text{ fresh}\\\\
    \joinval{\Omega_0}{\Omega_1}{(\esloan{\ell}{v_0})}{(\esloan{\ell'}{v_1})}{v_2}{\overrightarrow{A}}
  }{
    \joinval{\Omega_0}{\Omega_1}{(\esloan{\ell}{v_0})}{v_1}{v_2}{\overrightarrow{A}}
  }

  \inferrule[Join-Other-SharedLoan]{
    \ell' \text{ fresh}\\\\
    \joinval{\Omega_0}{\Omega_1}{(\esloan{\ell'}{v_0})}{(\esloan{\ell}{v_1})}{v_2}{\overrightarrow{A}}
  }{
    \joinval{\Omega_0}{\Omega_1}{v_0}{(\esloan{\ell}{v_1})}{v_2}{\overrightarrow{A}}
  }

  \inferrule[Join-Tuple]{
    \joinval{\Omega_0}{\Omega_1}{v_0}{v_1}{v}{\overrightarrow{A_0}}\\\\
    \joinval{\Omega_0}{\Omega_1}{w_0}{w_1}{w}{\overrightarrow{A_1}}
  }{
    \Omega_0,\, \Omega_1 \vdash \joinvals{(v_0, w_0)}{(v_1, w_1)} \Downarrow\\\\
    (v, w) \joinres
    \overrightarrow A_0, \overrightarrow A_1
  }

  \inferrule[Join-Sum]{
    \Omega_0,\, \Omega_1 \vdash \joinvals{v_0}{v_1} \Downarrow v \joinres
      \overrightarrow{A}\\\\
    C = \kw{Left} \vee C = \kw{Right}
  }{
    \Omega_0,\, \Omega_1 \vdash \joinvals{(C\,v_0)}{(C\,v_1)} \Downarrow C\,v \joinres \overrightarrow{A}\\\\
  }

  \inferrule[Join-Box]{
    \Omega_0,\, \Omega_1 \vdash \joinvals{v_0}{v_1} \Downarrow v \joinres
      \overrightarrow{A}
  }{
    \Omega_0,\, \Omega_1 \vdash \joinvals{(\ebox{v_0})}{(\ebox{v_1})} \Downarrow \ebox{v}
      \joinres \overrightarrow{A}
  }


  \inferrule[Join-Same-MutBorrow]{
    \joinval{\Omega_0}{\Omega_1}{v_0}{v_1}{v_2}{\overrightarrow{A}}
  }{
    \Omega_0,\, \Omega_1 \vdash \joinvals{(\emborrow{\ell}{v_0})}{(\emborrow{\ell}{v_1})}\\\\
      \Downarrow
      \emborrow{\ell}{v_2} \joinres \overrightarrow{A}
  }

  \inferrule[Join-Same-SharedLoan]{
    \text{no $\bot$} \in v_0,\, v_1\\\\
    \joinval{\Omega_0}{\Omega_1}{v_0}{v_1}{v_2}{\overrightarrow{A}}
  }{
    \Omega_0,\, \Omega_1 \vdash \joinvals{(\esloan{\ell}{v_0})}{(\esloan{\ell}{v_1})}\\\\
      \Downarrow
      \esloan{\ell}{v_2} \joinres \overrightarrow{A}
  }
}
} 

\newcommand\JoinValueRulesPrecise{
    \inferrule[Join-MutBorrows-Precise]{
      \joinval{\Omega_0}{\Omega_1}{v_0}{v_1}{v_2}{\overrightarrow A} \\
      l_0',\, l_1',\, l_2,\, A_0,\, A_1,\, A_2 \text{ fresh}
    }{
      \Omega_0,\, \Omega_1 \vdash
        \joinvals{(\emborrow{\ell_0}{v_0})}{(\emborrow{\ell_1}{v_1})}
        \Downarrow
        \emborrow{\ell_2}{v_2} \joinres\\\\
        \AbsI{0}{ \lbox{\emborrow{\ell_0}{\_}},\, \emloan{\ell_0'} },\;
        \AbsI{1}{ \rbox{\emborrow{\ell_1}{\_}},\, \emloan{\ell_1'} },\;
        \AbsI{2}{ \emborrow{\ell_0'}{\_},\,
        \emborrow{\ell_1'}\_,\,
        \emloan{\ell_2} },\,
        \overrightarrow A
    }

    \inferrule[Join-SharedBorrows-Precise]{
      \esloan{\ell_0}{v_0} \in \Omega_0\\
      \esloan{\ell_1}{v_1} \in \Omega_1\\\\
      \text{no borrows, loans, $\bot$} \in v_0,\, v_1\\
      \ell_0',\, \ell_1',\, \ell_2,\, s_0,\, s_1,\, s_2,\, A_0,\, A_1,\, A_2 \text{ fresh}
    }{
      \Omega_0,\, \Omega_1 \vdash \joinvals{(\esborrow{\ell_0})}{(\esborrow{\ell_1})}
      \Downarrow \esborrow{\ell_2} \joinres\\\\
        \AbsI{0}{ \lbox{\esborrow{\ell_0}},\, \esloan{\ell_0'}{s_0} },\,
        \AbsI{1}{ \rbox{\esborrow{\ell_1}},\, \esloan{\ell_1'}{s_1} },\,
        \AbsI{2}{ \esborrow{\ell_0'},\, \esborrow{\ell_1'},\, \esloan{\ell_2}{s_2} }
    }
}

\NewDocumentCommand\JoinStateRules{s}{
\IfBooleanT{#1}{
  \inferrule[Join-Same-Abs]{
    \joinenv{\Omega_0}{\Omega_1}{\Omega'_0}{\Omega'_1}{\Omega_2}
  }{
    \joinenv{\Omega_0}{\Omega_1}{(A,\, \Omega'_0)}{(A,\, \Omega'_1)}{ A,\, \Omega_2}
  }

  \inferrule[Join-AbsLeft]{
    A \notin \Omega_1\\
    \joinenv{\Omega_0}{\Omega_1}{\Omega'_0}{\Omega'_1}{\Omega_2}
  }{
    \joinenv{\Omega_0}{\Omega_1}{(A,\, \Omega'_0)}{\Omega'_1}{\lbox{A},\, \Omega_2}
  }

  \inferrule[Join-Same-Anon]{
    \joinenv{\Omega_0}{\Omega_1}{\Omega'_0}{\Omega'_1}{\Omega_2}
  }{
    \joinenv{\Omega_0}{\Omega_1}{(\_ \rightarrow v,\, \Omega'_0)}{
      (\_ \rightarrow v,\, \Omega'_1)}{\_ \rightarrow v,\, \Omega_2}
  }

  \inferrule[Join-AbsRight]{
    A \notin \Omega_0\\
    \joinenv{\Omega_0}{\Omega_1}{\Omega'_0}{\Omega'_1}{\Omega_2}
  }{
    \joinenv{\Omega_0}{\Omega_1}{\Omega'_0}{(A,\, \Omega'_1)}{\rbox{A},\, \Omega_2}
  }
}

  \inferrule[Join-Var]{
    \joinval{\Omega_0}{\Omega_1}{v_0}{v_1}{v_2}{\overrightarrow{A}}\\
    \joinenv{\Omega_0}{\Omega_1}{\Omega'_0}{\Omega'_1}{\Omega_2} 
  }{
    \joinenv{\Omega_0}{\Omega_1}{(x\rightarrow v_0,\, \Omega'_0)}{(x \rightarrow v_1,\, \Omega'_1)}
      {x \rightarrow v_2,\, \overrightarrow A,\, \Omega_2}
  }

\IfBooleanT{#1}{
  \inferrule[Join-Empty]{
  }{
    \joinenv{\Omega_0}{\Omega_1}{\emptyset}{\emptyset}{\emptyset}
  }
}
}

\NewDocumentCommand{\CollapseRules}{s}{
    \inferrule[Collapse-Merge-Abs]{
      \vdash A_0 \MergeAbs A_1 \merge A
    }{
      \Omega,\, A_0,\, A_1 \collapse \Omega,\, A
    }

    \inferrule[Collapse-Dup-MutBorrow]{
    }{
      \Omega,\, A \cup \{\,
        \lbox{\emborrow{\ell}{\_}},\,
        \rbox{\emborrow{\ell}{\_}}
      \,\}\\\\
      \collapse
      \Omega,\, A \cup \{\, \emborrow{\ell}{\_} \,\}
    }

\IfBooleanT{#1}{

    \inferrule[Collapse-Dup-MutLoan]{
    }{
      \Omega,\, A \cup \{\,
        \lbox{\emloan{\ell}},\,
        \rbox{\emloan{\ell}}
      \,\}\\\\
      \collapse
      \Omega,\, A \cup \{\, \emloan{\ell} \,\}
    }

    \inferrule[Collapse-Dup-SharedBorrow]{
    }{
      \Omega,\, A \cup \{\, \lbox{\emborrow{\ell}},\,\rbox{\emborrow{\ell}} \,\}\\\\
      \collapse
      \Omega,\, A \cup \{\, \emborrow{\ell} \,\}
    }

    \inferrule[Collapse-Dup-SharedLoan]{
      \text{no borrows, loans, $\bot$ } \in v_0,\, v_1\\\\
      v_2 = \begin{cases}
        v_0 & \text { if } v_1 = v_0 \cr
        \sigma & \text{ where $\sigma$ fresh otherwise}
      \end{cases}
    }{
      \Omega,\, A \cup \{\, \lbox{\esloan{\ell}{v_0}},\,\rbox{\esloan{\ell}{v_1}} \,\}\\\\
      \collapse
      \Omega,\, A \cup \{\, \esloan{\ell}{v_2} \,\}
    }

}
    \inferrule[MergeAbs-Mut-MarkedLeft]{
      \vdash A_0 \MergeAbs A_1 \merge A \\
    }{
      \vdash (A_0 \cup \{ \lbox{\emloan\ell} \}) \MergeAbs
             (A_1 \cup \{ \lbox{\emborrow\ell\_} \})
        \merge A
    }
\IfBooleanT{#1}{

    \inferrule[MergeAbs-Shared-MarkedLeft]{
      \vdash (A_0 \cup \{ \esloan{\ell} v \}) \MergeAbs A_1 \merge A \\
    }{
      \vdash (A_0 \cup \{ \lbox{\esloan{\ell}} v \}) \MergeAbs
        (A_1 \cup \{ \lbox{\esborrow{\ell}} \})
        \merge A
    }

    \inferrule[MergeAbs-Mut-MarkedRight]{
      \vdash A_0 \MergeAbs A_1 \merge A \\
    }{
      \vdash (A_0 \cup \{ \rbox{\emloan\ell} \}) \MergeAbs
             (A_1 \cup \{ \rbox{\emborrow\ell\_} \})
        \merge A
    }

    \inferrule[MergeAbs-Shared-MarkedRight]{
      \vdash (A_0 \cup \{ \esloan{\ell} v \}) \MergeAbs A_1 \merge A \\
    }{
      \vdash (A_0 \cup \{ \rbox{\esloan{\ell}} v \}) \MergeAbs
        (A_1 \cup \{ \rbox{\esborrow{\ell}} \})
        \merge A
    }
  }
}

\begin{figure}
  \centering
  \smaller
  \begin{mathpar}
  \JoinValueRules
  
  \JoinStateRules
  
  \CollapseRules
  
  \end{mathpar}
  \caption{Join and Collapse (Selected Rules)}
  \flabel{join-collapse}
  \figvspace
\end{figure}

\section{Implementation and Evaluation}
\slabel{evaluation}

In the previous sections, we focused on establishing simulations to demonstrate
the soundness of \llbcs with respect to PL, a low-level, heap-manipulating semantics.
Our theorems establish that, for any execution in \llbcs, there exists a related execution
in LLBC, and hence in PL by composing simulations. As PL is deterministic, it gives
us that all PL states in relation with the initial \llbcs state safely execute.
One key question remains however:
seeing \llbcs as a borrow-checker for LLBC, are we able to construct \llbcs derivations
in order to apply our theoretical results?

To this end, we implement an \llbcs interpreter, and evalute it on a set of Rust programs.
We start from Aeneas' symbolic execution~\cite{aeneas}, which implements
the \llbcs semantics presented in \Section{llbcsbackground}.
We extend this interpreter with our novel support for merging environments and handling
loops, which represents about 3000 lines of OCaml code.

Our test suite consists of two categories of programs. First, we implement a collection of
22 micro-tests (totalling 300 LoCs, without blanks and comments) spanning various Rust patterns based on loops, such as
incrementing counters, updating values in a vector (for instance,
to reinitialize it), summing the elements in an array or a slice, reversing a list
in place, or retrieving a shared or mutable borrow to the n-th element of a list.
Second, we evaluate our approach on the hashmap and the b-$\epsilon$ tree implemented and presented in the
original Aeneas paper, consisting of 867 LoCs in total.
Due to Aeneas' previous limitations, those implementations
were based on recursive functions to, e.g., insert an element or resize the hashmap.
Here, we update them to use more idiomatic loops for recursive data structure
traversals.
Leveraging our extension for joins and loops, the \llbcs interpreter successfully
borrow-checks all the examples, requiring less than 1s for the whole test suite.
Interestingly, the precedent paper noted that some functions of the b-$\epsilon$ tree are
not accepted by Rust's current borrow checker. Similarly, we observe that only Polonius
and Aeneas are able to borrow check the updated version; we show one such (minimized) function
below, with the errors given by rustc.
We provide the code and the tests in the supplementary material~\cite{artifact}.

\begin{minted}[linenos=false,mathescape=true,escapeinside=||]{rust}
enum List<T> { Cons(T, Box<List<T>>), Nil }
// error[E0499]: cannot borrow `*ls` as mutable more than once at a time
fn get_suffix_at_x<'a>(mut ls: &'a mut List<u32>, x: u32) -> &'a mut List<u32> {
  while  let List::Cons(hd, tl) = ls {
   if *hd == x { break } // ^^ first mutable borrow occurs here
   else { ls = tl } }
  ls } // < second mutable borrow occurs here
\end{minted}

%
%

\section{Related Work}
\slabel{relatedwork}

\son{TODO: quote \cite{pearce-lightweight} and Oxide}

\myparagraph{RustHorn} RustHorn~\cite{matsushita2020rusthorn} operates on
the Calculus of Ownership and Reference (COR), a Rust-like core
calculus inspired by $\lambdar$. Given a COR program, RustHorn computes a
first-order logical encoding that can then serve as a basis for
reasoning upon the COR program. The RustHorn paper
provides a proof of soundness and completeness of the encoding.
Specifically, the authors establish a proof of bisimulation between COR, and the
execution of a custom resolution procedure (dubbed SDLC) that
mimics program execution when executed over the logical encoding.

The main difference with our work is that COR already takes for granted the
ownership discipline of Rust, and materializes lifetimes within its program
syntax: COR features instructions for creating or ending a lifetime, and
asserting that a lifetime outlives another. One consequence is that COR cannot
exhibit more behaviors than SDLC, hence why the bisimulation can be established.

In contrast, our low-level language, PL, does not
feature lifetimes, and as such exhibits more behaviors than LLBC. This means we
can target the underlying execution model that Rust programs run on, rather than
taking lifetimes as an immutable, granted analysis that we have to trust. A
drawback is that we can only establish a forward simulation in the general case.
Second, we do not commit to lifetimes, nor to any other particular
implementation strategy of borrow-checking (e.g., Polonius~\cite{polonius}). Instead, we
declaratively state what ownership-related operations may be performed in LLBC.
It is then up to a particular borrow-checker implementation (e.g., Aeneas)
to be
proven sound with regards to this semantics. Notably, LLBC is not
deterministic, and several executions may be valid simultaneously, e.g., by
terminating borrows at different points (eagerly or lazily).

To summarize, RustHorn focuses on establishing the soundness of a logical
encoding with regards to a model of the Rust semantics that assumes
borrow-checking has been performed and can be trusted; here, we establish that
LLBC is a correct model of execution for Rust programs, that \llbcs is a
valid borrow-checker for the LLBC semantics, and that \llbcs's borrow-checking
does indeed guarantee soundness of execution for LLBC programs.

\myparagraph{RustBelt} In a similar fashion, RustBelt~\cite{jung2017rustbelt}
is built atop $\lambdar$, a core calculus that is already annotated with
operations to create and end lifetimes. The operational semantics of $\lambdar$
itself is given by translation; the lifetime operations are assumed to
be given by the Rust compiler. RustBelt focuses on a proof of semantic typing,
with two chief goals: first, prove the lifetime-based type system sound with
regards to the (lifetime-annotated) core language; second, use the semantic
typing relation to establish that pieces of unsafe code do satisfy the type they
export.

Our work differs from RustBelt in several ways. From the technical standpoint,
RustBelt assumes lifetimes are given. Whether the lifetimes annotations are
correct, and whether they lead to a successful execution is irrelevant -- if the
input program features improper lifetime annotations, this is outside of
RustBelt's purview. In contrast, we attempt to determine what needs to be
established from a semantic perspective in order to borrow-check a Rust program,
and prove that successful borrow-checking entails safety of execution. From
a goals standpoint, RustBelt attempts to understand the expected behavior of a
Rust program that features unsafe blocks, using semantic typing. We do not
consider unsafe code, but we intend to tackle this in future
work.

\myparagraph{RustHornBelt} The combination of RustHorn and RustBelt,
RustHornBelt~\cite{RustHornBelt},
aims to establish that the logical encoding of RustHorn is sound with regards to
$\lambdar$. RustHornBelt extends the methodology of RustBelt; at a very
high-level, RustHornBelt proves the encoding of RustHorn, but with $\lambdar$
instead of COR, and with a machine-checked proof instead of pen-and-paper. For
the same reasons as above, we see this endeavor as addressing a different
problem than ours: RustHornBelt is concerned with a logical encoding that
leverages lifetimes as a central piece of information, rather than giving a
functional, semantic account of the borrow-checking and execution of Rust
programs.

\myparagraph{Tree Borrows} Tree Borrows~\cite{treeborrows2023},
the successor of Stacked Borrows~\cite{jung2019stacked},
attempts to provide a semantics for correct borrow handling in the presence of
unsafe code. This allows detecting, at run-time, violations of the contract
(undefined behavior). Tree Borrows, unlike this work, operates at runtime by
tracking permissions at the level of memory cells. We operate statically, and
focus on safe code, proposing a new notion of borrow-checking that we prove to
be semantically sound.

\myparagraph{Shape Analysis} Perhaps more closely connected to this work is the
field of shape
analysis~\cite{calcagno2009compositional,berdine2007shape,distefano2006local,chang2008relational}. Very
active in the 2000s, the goal was to design abstract domains that would be able to 
infer shape predicates for pointer languages. Using familiar notions of
concrete and symbolic executions, the analysis would then be able to identify
bugs in programs via abstract interpretation. This has led to industrial tools
such as Meta (née Facebook)'s Infer~\cite{calcagno2015infer}.

We differ from these works in several ways.
First, we operate in a much more structured language
than, say, C; these works traditionally operate over pointer languages, with
NULL pointers, and untagged unions (anonymous sums). In our setting, we can enforce much
more discipline onto the original language, and benefit from a lot more
structure than languages like C may exhibit. However, this requires reasoning
about and proving the correctness of a non-standard, borrow-centric semantics,
and developing novel borrow-centric shape analyses, i.e., \llbcs.
Second, our analysis does not
exactly fit within the static analysis framework, and is merely inspired by it.
We exhibit similarities in the design of our join operation,
which just like in shape analysis involves reconciling competing shapes, folding
inductive predicates, and abstracting over differing concrete
values~\cite{chang2008relational,illous2017relational}.

\myparagraph{Mezzo} The Mezzo programming language~\cite{pottier2013programming} blends type
system, ownership and shape analysis. Mezzo is equipped with a syntactic proof
of type soundness~\cite{balabonski2016design}, but for an operational semantics
à la ML. The ``merge operation''~\cite{protzenko2014mezzo} is akin to our join
operation, and served to some degree as inspiration. It is, however, much more
involved, supporting singleton types, substructural typing, and
multiple types for a given variable $x$ (top, ``dynamic'', ``singleton $x$'',
and other degrees of folding/unfolding of a substructural type).
Furthermore, the Mezzo merge algorithm is much more sophisticated: it
interleaves backtracking, quantifier instantiation strategies, and folding of
existential predicates. We perform none of these.

\myparagraph{Other Rust verification tools}
Creusot~\cite{creusot} loosely followed the RustHornBelt approach, and as such,
benefits from its formalization and mechanization. As mentioned by
\citet{RustHornBelt}, there remain some discrepancies, namely that RustHornBelt
operates on a core language (instead of surface Rust), and that Creusot does not
use the predicate transformers RustHornBelt relies on.
Verus~\cite{lattuada2023verus} contains a pen-and-paper formalization, not about Rust
itself, lifetimes, or borrow-checking, but rather about the soundness and
termination of their approach to specifications relying on ghost permissions. As
such, the proof for Verus answers a different question, namely, whether their
design on top of existing Rust is sound. The proof remains of limited scope,
only taking into account two possible lifetimes.
Prusti~\cite{prusti} translates Rust programs into Viper's core logic; the
soundness of verifying Rust code then depends on the soundness of Viper, and of
the translation itself. To the best of our knowledge, no formal argument exists
as to the soundness of the translation.

\section*{Acknowledgements}

Sidney Congard contributed to a preliminary exploration of the problem of joining
environments in LLBC in 2022; the join operation introduced in the present work is a
complete rewrite of this first attempt. We thank François Pottier for various
advice and comments about working with simulations and step-indexing.
We thank Raphaël Monat for helping us clarify several key points of abstract
interpretation and shape analysis. Finally, we thank Ralf Jung for many insightful
discussions about the semantics of Rust, and for suggesting to not use sets of
loan identifiers to handle shared loans in the semantics of LLBC.

\bibliographystyle{ACM-Reference-Format}
\bibliography{paper.bib}

\iflong
\appendix

\section{Forward Simulation Between HLPL and LLBC}
\appendixlabel{hlpl+}

\Figure{llbc-syntax-full} describes the syntax for LLBC programs.
Figures \fref{llbc-rvalue}, \fref{llbc-copy}, \fref{llbc-statements}, \fref{llbc-reorg},
\fref{llbc-read} and \fref{llbc-write} give the operational semantics of LLBC.
This semantics is the big-step semantics without step-indexing; we omit the version
with step-indexing, which just consists in adding a step index to all the rules
(the index is the same in the conclusion and in the premises, see \Rule{E-Step-Seq-Unit})
at the exception of the rules to evaluate loops and function calls (see
\Rule{E-Step-Call}), and finally adding the rule \Rule{E-Step-Zero}.
\Figure{llbc-loops} introduces the additional rules to evaluate loop statements.
Figures \fref{hlpl-full} and \fref{hlplp-full} describe the operational semantics
for HLPL and \hlplp. We omit some rules for HLPL because it shares most of its
semantics with LLBC.
The semantics of \hlplp is actually not \emph{exactly} a superset of the semantics of HLPL, because
we need to replace \Rule{HLPL-E-Assign} with \Rule{HLPL+-E-Assign},
\Rule{HLPL-E-Pointer} with \Rule{HLPL+-E-Pointer} and
\Rule{HLPL-E-Move} with \Rule{HLPL+-E-Move}.
For \Rule{HLPL+-E-Assign}, we indeed need to prevent overriding $v_p$ if it contains
locations as well as \emph{outer loans}.
Without this additional restriction, we can not prove the forward simulation for states
related by \Rule{Le-MutBorrow-To-Ptr} or \Rule{Le-SharedLoan-To-Loc}, because they allow
turning outer loans into locations (we might get into a situation where we are allowed to
overwrite a value in the right state but not the left state).
We note that because loan values can only exist in \hlplp states, \Rule{HLPL-E-Assign} and
\Rule{HLPL+-E-Assign} coincide on HLPL states. As a consequence we still get
the crucial property we need for the proof of the forward simulation, that is that HLPL is
a stable subset of \hlplp.
Similar reasonings apply for \Rule{HLPL+-E-Pointer} and \Rule{HLPL+-E-Move}.
We need the additional restrictions for the \Rule{HLPL+-E-Pointer} because otherwise
we may insert locations where there are already shared loans, which causes issues
when combined with \Rule{Le-SharedLoan-To-Loc}.

\begin{figure}
  \LlbcSyntax*
  \caption{The Low-Level Borrow Calculus (LLBC): Syntax} %
  \flabel{llbc-syntax-full} %
\end{figure} %

\begin{figure}
  \smaller
  \arraycolsep=1pt %
  \centering

  \begin{array}[t]{llll}
    v & ::= & & \text{value} \\
    && \ktrue \mid \kfalse \mid n_{\mathsf{i32}} \mid n_{\mathsf{u32}} \mid \ldots &
      \text{literal constants} \\
    && \eleft{v} \mid \eright{v} & \text{sum value} \\
    && (v_0,\, v_1) & \text{pairs}\\
    && \bot & \text{bottom (invalid) value}\\
    && \emloan{\ell} & \text{mutable loan}\\
    && \emborrow{\ell}{v} & \text{mutable borrow}\\
    && \esloan{\ell}{v} & \text{shared loan}\\
    && \esborrow{\ell} & \text{shared borrow}\\
    && \erborrow{\ell} & \text{reserved borrow}
    \\[1ex]

  id & ::= & & \text{environment binding identifier}\\
    && x & \text{variable identifier}\\
    && \_x & \text{anonymous (ghost) variable identifier}
    \\[1ex]

  \Omega^{\text{LLBC}} & ::= & \{\;
    \kw{env} : id \underset {\text{partial}} \longrightarrow v,\;
    \kw{stack} : [[x]]\; \}\; & \text{state}\\[1ex]
  \end{array}
  \flabel{llbc-grammar}
  \caption{Grammar of LLBC States and Values}
\end{figure}

\begin{figure}
  \centering
  \smaller
  \begin{mathpar}
    \LlbcRvaluesEvalRules*
  \end{mathpar}
  \caption{Rules to Evaluate Rvalues (LLBC)}
  \flabel{llbc-rvalue}
\end{figure}

\begin{figure}
  \centering
  \smaller
  \begin{mathpar}
    \LlbcCopyRules*
  \end{mathpar}
  \caption{Rules to Evaluate Copy (LLBC)}
  \flabel{llbc-copy}
\end{figure}

\begin{figure}
  \centering
  \smaller
  \begin{mathpar}
    \LlbcStatementsEvalRules*
  \end{mathpar}
  \caption{Rules to Evaluate Statements (LLBC)}
  \flabel{llbc-statements}
\end{figure}

\begin{figure}
  \centering
  \smaller
  \begin{mathpar}
    \LlbcLoopsEvalRules*
  \end{mathpar}
  \caption{Additional Rules to Evaluate Loops (LLBC)}
  \flabel{llbc-loops}
\end{figure}

\begin{figure}
  \centering
  \smaller
  \begin{mathpar}
    \LlbcReorgRules*
  \end{mathpar}
  \caption{Reorganization Rules (LLBC)}
  \flabel{llbc-reorg}
\end{figure}

\begin{figure}
  \centering
  \smaller
  \begin{mathpar}
    \LlbcReadRules*
  \end{mathpar}
  \caption{Read Rules (LLBC)}
  \flabel{llbc-read}
\end{figure}

\begin{figure}
  \centering
  \smaller
  \begin{mathpar}
    \LlbcWriteRules*
  \end{mathpar}
  \caption{Write Rules (LLBC)}
  \flabel{llbc-write}
\end{figure}

\begin{figure}
  \smaller
  \arraycolsep=1pt %
  \centering

  \begin{array}[t]{llll}
    v & ::= & & \text{value} \\
    && \ktrue \mid \kfalse \mid n_{\mathsf{i32}} \mid n_{\mathsf{u32}} \mid \ldots &
      \text{literal constants} \\
    && \eleft{v} \mid \eright{v} & \text{sum value} \\
    && (v_0,\, v_1) & \text{pairs}\\
    && \bot & \text{bottom (invalid) value}\\
    && \eloc{\ell}{v} & \text{location (value with an address)}\\
    && \eptr{\ell} & \text{pointer}
    \\[1ex]

  \Omega^{\text{HLPL}} & ::= & \{\;
    \kw{env} : x \underset {\text{partial}} \longrightarrow v,\;
    \kw{stack} : [[x]]\; \} & \text{state}\\[1ex]
  \end{array}
  \flabel{hlpl-grammar}
  \caption{Grammar of HLPL States and Values}
\end{figure}

\begin{figure}
  \centering
  \smaller
  \begin{mathpar}
    \HlplRules*
  \end{mathpar}
  \caption{Selected Rules for HLPL}
  \flabel{hlpl-full}
\end{figure}

\begin{figure}
  \centering
  \smaller
  \begin{mathpar}
    \HlplpRules*
  \end{mathpar}
  \caption{Additional Rules for \hlplp}
  \flabel{hlplp-full}
\end{figure}

\begin{figure}
  \centering
  \smaller
  \begin{mathpar}
    \HlplLeRules*
  \end{mathpar}
  \caption{The $\le$ Relation on \hlplp states}
  \flabel{hlpl-rel-long}
\end{figure}

We now turn to the proof of \Thm{Eval-Preserves-HLPL+-Rel}. We need several
auxiliary lemmas, to show that evaluating rvalues (i.e., expressions - Lemma
\thref{hlpl+-rvalue-rel}), evaluating assignments (Lemma \thref{hlpl+-assign-rel}) and
finally reorganizing states (Lemma \thref{hlpl+-reorg-rel}) preserves the relation $\le$.

\begin{lemma}[Rvalue-Preserves-HLPL+-Rel]
\thlabel{hlpl+-rvalue-rel}

For all $\Omega_l$ and $\Omega_r$ \hlplp states and $rv$ right-value we have:
\begin{align*}
&\Omega_l \le \Omega_r \Rightarrow
\forall\; v_r\, \Omega'_r, \ \Omega_r \vdashHlplp rv \exprres{v_r}{\Omega'_r} \Rightarrow\\
&\exists\; v_l\, \Omega'_l,\ \Omega_l \vdashHlplp rv \exprres{v_l}{\Omega'_l} \wedge
(v_l,\, \Omega'_l) \le (v_r,\, \Omega'_r)
\end{align*}

where we define $(v_l,\, \Omega'_l) \le (v_r,\, \Omega'_r)$ as:
\begin{align*}
(v_l,\, \Omega'_l) \le (v_r,\, \Omega'_r) := (\Omega'_l,\, \_ \rightarrow v_l) \le (\Omega'_r,\, \_ \rightarrow v_r)
\end{align*}
\end{lemma}

\noindent\textbf{Proof}\\
We do the proof by induction on $\Omega_r \vdashHlplp rv \exprres{v_r}{\Omega'_r}$.
Then, in (most of) the subcases we do the proof by induction on $\Omega_l \le \Omega_r$.
The high-level idea is to show that, if $\Omega_l$ is related to $\Omega_r$ in some specific
manner (for instance, $\Omega_l$ is $\Omega_r$ where we replaced a shared borrow by
a pointer by using \Rule{Le-SharedReservedBorrow-To-Ptr}) then in most
situations they remain related excatly the same way (that is, they remain related
by \Rule{Le-SharedReservedBorrow-To-Ptr}). For instance, if the left state is
the right state where we replaced a shared borrow by a pointer, then after
a move operation, the left state is still exactly the right state where we replaced
a shared borrow by a pointer; of course, we need to reason about wether the borrow
was moved or not, so that we can instantiate
\Rule{Le-SharedReservedBorrow-To-Ptr} with the proper state with a hole to show
that the resulting left state and right state are related.
\begin{itemize}
\item Case $\ecopy p$.
  We have (premises of \Rule{E-Copy}):
  \begin{align*}
  &\vdash \Omega_r(p) \eqimmut v_r \eand\\
  &\bot, \emloan\relax, \kw{borrow}^{m,r} \not\in v \eand\\
  &\vdash \ecopy v_r = v'_r
  \end{align*}

  By induction on $\Omega_l \le \Omega_r$.
  \begin{itemize}
  \item Reflexive case. Trivial.
  \item Transitive case. Trivial by the induction hypotheses.
  \item Case \Rule{Le-SharedReservedBorrow-To-Ptr}.

    By the premises of \Rule{Le-SharedReservedBorrow-To-Ptr}, there exists $\Omega_1[.]$ such that
    $\Omega_l = \Omega_1[\eptr{\ell}] \le \Omega_1[\esborrow{\ell}] = \Omega_r$
    (doing the $\erborrow{\ell}$ case later).

    We have to reason about whether the hole of $\Omega_1[.]$ is inside the value we read
    at path $p$ or not, that is: either the hole is not at path $p$, in which case
    we read the same value in $\Omega_l$ and $\Omega_r$, or the hole is inside,
    in which case the value we read differs in exactly one place, where we have
    $\esborrow\ell$ for $\Omega_r$ and $\eptr\ell$ for $\Omega_l$.
    We formalize this in the auxiliary lemma \thref{hplp-shared-aux} below.
    \begin{lemma}{Auxiliary Lemma}
    \thlabel{hplp-shared-aux}
    \begin{align*}
    &\forall\; p\, v_r, \,\\
    &\quad(\vdash \Omega_r(p) \eqimmut v_r \Rightarrow)\\
    &\quad\Omega_l = \Omega_1[\eptr{\ell}] \wedge \Omega_1[\esborrow{\ell}] = \Omega_r \Rightarrow\\
    &\quad((\vdash \Omega_l(p) \eqimmut v_r) \eor\\
    &\quad\ \ (\exists\; V[.],\, (\vdash \Omega_l(p) \eqimmut V[\eptr\ell]) \eand
      (\vdash \Omega_r(p) \eqimmut V[\esborrow\ell])))
    \end{align*}
    \end{lemma}
    
    The proof of this lemma is straightforward by induction on $p$: we simply have to show
    that if we can read through one path element on the right (for instance, we can reduce
    a projection) then we can do the same on the left.
    The tricky case happens when dereferencing (i.e., $*$). We have to pay
    attention to two elements.
    \begin{itemize}
    \item We might dereference $\esborrow\ell$ on the right and $\eptr\ell$ on the left. In this
      case we have to use the fact that dereferencing $\eptr\ell$ is the same as dereferencing
      $\esborrow\ell$ (because the read rules are defined so that it is the case; see
      \Rule{R-Deref-SharedBorrow} and \Rule{R-Deref-Ptr-SharedLoan}).
    \item Dereferencing a value allows us to ``jump'' to a value elsewhere in the
      environment (a shared loan, a location, or an HLPL box). We then have to make a case
      disjunction on whether the hole is inside the value we jump to or not. We made
      the theorem statement general enough so that we can handle this case.
    \end{itemize}
    
    Given lemma \thref{hplp-shared-aux}, we instantiate it on $p$
    and $v_r$ then do a case disjunction on its conclusion.
    \begin{itemize}
    \item Case 1 (we read the same value on the left and the right).
      We have:
      \begin{align*}
      &\vdash \Omega_r(p) \eqimmut v_r \eand\\
      &\vdash \Omega_l(p) \eqimmut v_r
      \end{align*}
      Because $(v'_r,\, \Omega_l) \le (v'_r,\, \Omega_r)$ is defined
      as $(\Omega_l,\, \_ \rightarrow v'_r) \le (\Omega_r,\, \_ \rightarrow v'_r)$,
      we can conclude by using \Rule{Le-SharedReservedBorrow-To-Ptr}, which gives us:
      $(v'_r,\, \Omega_l) \le (v'_r,\, \Omega_r)$.
    \item Case 2 (the hole is inside the value we read).
      There exists $V[.]$ such that:
      \begin{align*}
      &\vdash \Omega_l(p) \eqimmut V[\eptr\ell] \eand\\
      &\vdash \Omega_r(p) \eqimmut V[\esborrow\ell]
      \end{align*}
      By induction on $V[.]$ we prove that the copy differs in exactly one place
      as well, that is, there exists $V'[.]$ such that:
      $\vdash \ecopy V[\eptr\ell] = V'[\eptr\ell]$ and
      $\vdash \ecopy V[\esborrow\ell] = V'[\esborrow\ell]$.
      
      This time we have to apply \Rule{Le-SharedReservedBorrow-To-Ptr} twice,
      once for the original borrow, once for the borrow inside the copied value.
      \begin{align*}
      \Omega_l,\, \_ \rightarrow V'[\eptr\ell] &=\
        \Omega_1[\eptr\ell], \_ \rightarrow V'[\eptr\ell]\\
      &\le\ \Omega_1[\esborrow\ell], \_ \rightarrow V'[\eptr\ell]\\
      &=\ \Omega_r, \_ \rightarrow V'[\eptr\ell]\\
      &\le\ \Omega_r, \_ \rightarrow V'[\esborrow\ell]
      \end{align*}
    \end{itemize}

    The case $\erborrow{\ell}$ is similar, and actually even simpler because:
    1. in the proof of \thref{hplp-shared-aux} we can't dereference a reserved
    borrow; 2. for the end of the proof we can't copy a reserved borrow.
  \item Case \Rule{Le-MutBorrow-To-Ptr}. same as case
    \Rule{Le-SharedReservedBorrow-To-Ptr}. We prove an auxiliary lemma
    which is similar to \thref{hplp-shared-aux}, and use in the proof
    the fact that pointers and mutable borrows are dereferenced in ways
    compatible with \Rule{Le-MutBorrow-To-Ptr} (\Rule{R-Deref-MutBorrow},
    \Rule{R-Deref-Ptr-Loc}). For the end of the proof, in the case we don't
    read the same value on the left and on the right, we use the fact
    that we can't copy mutable borrows or mutable loans.
  \item Case \Rule{Le-RemoveAnon}. Similar to \Rule{Le-SharedReservedBorrow-To-Ptr}
  but the auxiliary lemma is even simpler because we read the same value on the
  left and on the right (the presence or absence of an anonymous value \emph{without}
  borrows or loans doesn't have any impact on evaluation).
  \item Case \Rule{Le-Merge-Locs}.
  By the premises of the rule, there exists $\Omega_1[.]$ $\ell_0$, $\ell_1$ and $v$ such
  that:
  $\Omega_r = \Omega_1[\eloc{\ell_0}{(\eloc{\ell_1}{v})}]$,
  $\forall\; v', \eloc{\ell_1}{v'} \notin \Omega_1[\ell_0]$,
  and
  $\Omega_l = \left[\ell_0 \Big/ \ell_1 \right]\left(\Omega_1[\eloc{\ell_0}{v}]\right)$.
  
  We prove that, depending on whether the hole of $\Omega_1[.]$ is inside the value we read
  at $p$ or not, then reading in $\Omega_l$ along $p$ is well defined, and the read value is
  the same as in $\Omega_r$ modulo two things: 1. we substitute $\ell_0$ for $\ell_1$;
  2. we may have collapsed the collapsed the locations for $\ell_0$ and $\ell_1$.
  Formally, we prove the auxiliary lemma below.
  
  \begin{lemma}{Auxiliary Lemma}
    \thlabel{hplp-mergeloc-aux}
    \begin{align*}
    &\Omega_r = \Omega_1[\eloc{\ell_0}{(\eloc{\ell_1}{v})}] \Rightarrow
    (\forall\; v', \eloc{\ell_1}{v'} \notin \Omega_1[\ell_0]) \Rightarrow\\
    &\Omega_l = \left[\ell_0 \Big/ \ell_1 \right]\left(\Omega_1[\eloc{\ell_0}{v}]\right) \Rightarrow
    \forall\; p\, v_r, \, \vdash \Omega_r(p) \eqimmut v_r \Rightarrow\\
    &\quad(\vdash \Omega_l(p) \eqimmut \left[\ell_0 \Big/ \ell_1 \right](v_r) \;\vee\; \\
     &\quad\ (\exists\; V[.],\ \vdash \Omega_l(p) \eqimmut \left[\ell_0 \Big/ \ell_1
      \right](V[\eloc{\ell_0}{v}]) \wedge v_r = V[\eloc{\ell_0}{(\eloc{\ell_1}{v})}]))
    \end{align*}
  \end{lemma}
  We do the proof of \thref{hplp-mergeloc-aux} by induction on
  $\vdash \Omega_r(p) \eqimmut v_r$. A crucial point which makes the proof
  works is that the substitution applies a \emph{pointwise} transformation.
  The difficult case is the dereference ($*$).
  If we dereference $\ell_0$ to read $\eloc{\ell_1}{v}$ on the right,
  we read $\esubst{\ell_1}{\ell_0}{v}$ on the left. We then have to make a case
  disjunction on whether the path is empty (in which case we stop) or not
  (in which case we dive into the shared loan by \Rule{R-SharedLoan}). Also note that because
  we do not enforce that states are well-formed, there may be
  several $\eloc{\ell_0}{\ldots}$ in $\Omega_r$, meaning we don't have to read
  $\eloc{\ell_1}{v}$
  (but this doesn't have much impact on the proof, because then a similar
  $\eloc{\ell_0}{\ldots}$ will appear on the left, and we have, again, to make a case
  disjunction on whether the hole appears inside the pointed value or not).

  Given \thref{hplp-mergeloc-aux}, we can easily conclude the proof (we relate $\Omega'_l$ to $\Omega'_r$ with
  \Rule{Le-Merge-Locs} again).
  \item Case \Rule{Le-SharedLoan-To-Loc}.
    Same as above. We also use the fact that copying a location (respectively,
    a shared loan) removes the location (respectively, the shared loan)
    wrapper (\Rule{Copy-Loc}, \Rule{Copy-SharedLoan}).

  \item Case \Rule{Le-Box-To-Loc}.
    We note that $\eptr{\ell}$ and $\ebox{v}$ are dereferenced in ways compatible
    with \Rule{Le-Box-To-Loc}.
    
    By using \thref{hplp-shared-aux} as model, we prove the auxiliary
    below, and conclude by using \Rule{Le-Box-To-Loc}.
    
    \begin{lemma}{Auxiliary Lemma}
    \begin{align*}
    &\forall\; p\, v_r, \, \vdash \Omega_r(p) \eqimmut v_r \Rightarrow\\
    &\Omega_l = (\Omega_1[\eptr{\ell}],\, \_ \ell \rightarrow v) \eand
     \Omega_1[\ebox{v}] = \Omega_r \Rightarrow\\
    &((\vdash \Omega_l(p) \eqimmut v_r) \eor\\
    &\quad (\exists\; V[.],\, (\vdash \Omega_l(p) \eqimmut V[\eptr\ell]) \eand
      (\vdash \Omega_r(p) \eqimmut V[\ebox v])))
    \end{align*}
    \end{lemma}

  \item Case \Rule{Le-Subst}. The read judgement is not affected by the substitution,
    and we use the fact that $\vdash \ecopy v$ commutes with identifier substitutions.
  \end{itemize}
\item Case $\emove p$.
 \begin{itemize}
 \item Reflexive case. Trivial.
 \item Transitive case. Trivial by the induction hypotheses.
 \item Case \Rule{Le-SharedReservedBorrow-To-Ptr}.
   There exists $\Omega_1[.]$ such that $\Omega_l = \Omega_1[\eptr\ell]$ and
   $\Omega_1[\esrborrow\ell]$.
   By using the assumption $\vdash \Omega_r(p) \eqmove v_r$ and the fact that
   the move capability doesn't allow to dereference a shared
   borrow (\Rule{R-Deref-SharedBorrow}, \Rule{W-Deref-SharedBorrow}) then when reading or updating
   along path $p$ we don't have to consider the case where we might dereference
   $\esrborrow{\ell}$ in the right environment and $\eptr{\ell}$ in the left
   environment. This allows us to prove by induction on $p$ the following auxiliary lemma
   (either the hole of $\Omega_2$ is independent of the moved value, or it is inside).
   
   \begin{lemma}{Auxiliary Lemma}
    \begin{align*}
    &\forall\; p\, v_r, \, \vdash \Omega_r(p) \eqmove v_r \Rightarrow\\
    &\Omega_l = \Omega_1[\eptr{\ell}] \wedge \Omega_1[\esrborrow{\ell}] = \Omega_r \Rightarrow\\
    &(\exists\; \Omega_2[.,\,.],\
      \Omega_1[.] = \Omega_2[.,\ v_r] \eand
      (\forall\; v,\ \vdash \Omega_2[.,\ v](p) \eqmove v) \eand\\
     &\quad  (\forall\; v,\ \updtstateplace{\Omega_2[.,\, .]}{p}{v} \ueqmove \Omega_2[.,\ v])
      ) \eor\\
    &(\exists\; \Omega_2[.]\ V[.],\,
      \Omega_1[.] = \Omega_2[V[.]] \eand
      (\forall\; v,\ \vdash \Omega_2[v](p) \eqmove v) \eand\\
     &\quad (\forall\; v,\ \updtstateplace{\Omega_2[.]}{p}{v} \ueqmove \Omega_2[v])
      )
    \end{align*}
  \end{lemma}   
   
   Given this lemma, we conclude the proof by using \Rule{Le-SharedReservedBorrow-To-Ptr}.

 \item Case \Rule{Le-MutBorrow-To-Ptr}.
   There exists $\ell$, $v$, $\Omega_1[.,\ .]$ such that
   $\Omega_l = \Omega_1[\eloc{\ell}{v}]$ and $\Omega_r = \Omega_1[\emloan{\ell},\ \emborrow{\ell}{v}]$.
   The proof is similar to the case \Rule{Le-SharedReservedBorrow-To-Ptr} but this time we
   have two holes.
   
   Because it is not possible to dive into mutable borrows (\Rule{R-Deref-MutBorrow},
   \Rule{W-Deref-MutBorrow}) or mutable loans when moving values,
   we prove that when reading or updating along path
   $p$ we can't dereference $\ell$ and can't go through $\emloan{\ell}$.
   Also, following the premise of \Rule{E-Move}, we can't move mutable loans.
   This allows us to prove the following theorem by induction on $p$.
   
    \begin{lemma}{Auxiliary Lemma}
    \begin{align*}
    &\forall\; p\, v_r,\,\\
    &\vdash \Omega_r(p) \eqmove v_r \Rightarrow\\
    &(\exists\; \threeholes{\Omega_2} . . .,\,\\
    &\quad  \twoholes{\Omega_1} . . = \threeholes{\Omega_2} . . {v_r} \eand\\
    &\quad  (\forall\; v, \vdash \threeholes{\Omega_2} . . v (p) \eqmove v) \eand\\
    &\quad  (\forall\; v, \updtstateplace{\threeholes{\Omega_2} . . .} p v
        \ueqmove \twoholes{\Omega_2} . v)) \eor\\
    &(\exists\; \twoholes{\Omega_2} . .\, \onehole{V} .,\,\\
      &\quad \twoholes{\Omega_1} . . = \twoholes{\Omega_2} . {\onehole{V} .} \eand\\
      &\quad (\forall\; v,\, \vdash \twoholes{\Omega_2} . v (p) \eqmove v) \eand\\
      &\quad (\forall\; v,\, \updtstateplace{\twoholes{\Omega_2} . .} p v \ueqmove
          \twoholes{\Omega_2} . v))
    \end{align*}
  \end{lemma}

  We do a case disjunction on the conclusion of the auxiliary lemma.
  \begin{itemize}
  \item Case 1. There exists $\threeholes{\Omega_2} . . .$ such that:
    \begin{align*}
    &\vdash \Omega_r(p) \eqmove v_r\\
    &\updtstateplace{\Omega_r} p \bot \ueqmove
     \threeholes{\Omega_2}{\emloan{\ell}}{\emborrow{\ell}{v}} \bot\\
    &\vdash \Omega_l(p) \eqmove v_r\\
    &\updtstateplace{\Omega_l} p \bot \ueqmove
     \threeholes{\Omega_2}{\eloc{\ell}{v}}{\eptr\ell} \bot
    \end{align*}
    
    Thus:
    \begin{align*}
    &\Omega_r(p) \vdash \emove p \exprres{v_r}{\threeholes{\Omega_2}{\emloan\ell}{\emborrow{\ell}{v}}{\bot}}\\
    &\Omega_l(p) \vdash \emove p \exprres{v_r}{\threeholes{\Omega_2}{\eloc\ell v}{\eptr{\ell}}{\bot}}
    \end{align*}

    Posing:
    \begin{align*}
    &\Omega'_r := \threeholes{\Omega_2}{\emloan\ell}{\emborrow{\ell}{v}}{\bot}\\
    &\Omega'_l := \threeholes{\Omega_2}{\eloc\ell v}{\eptr{\ell}}{\bot}
    \end{align*}

   By instantiating \Rule{Le-MutBorrow-To-Ptr} with the state with holes
   $\threeholes{\Omega_2} . . \bot$) we get
   $(v_r,\, \Omega'_l) \le (v_r,\, \Omega'_r)$.

  \item Case 2. There exists $\twoholes{\Omega_2} . .$, $\onehole V .$ such that:
    \begin{align*}
    &\twoholes{\Omega_1} . . = \omegatwoholes 2 . {\onehole V .}\\
    &\forall\; v,\, \vdash \omegatwoholes 2 . v (p) \eqmove v\\
    &\forall\; v,\, \updtstateplace{\omegatwoholes 2 . .} p v \ueqmove \omegatwoholes 2 . v
    \end{align*}
  
    This gives us:
    \begin{align*}
    &\vdash \Omega_r(p) \eqmove \onehole V {\emborrow\ell v}\\
    &\updtstateplace{\Omega_r} p \bot \ueqmove \omegatwoholes 2 {\emloan\ell}{\bot}\\
    &\vdash \Omega_l(p) \eqmove \onehole V {\eptr\ell}\\
    &\updtstateplace{\Omega_l} p \bot \ueqmove \omegatwoholes 2 {\eloc 2 v} \bot
    \end{align*}
  
  \end{itemize}

  By instantiating \Rule{Le-MutBorrow-To-Ptr} with the state with holes
  $\omegatwoholes 2 . \bot,\, \_ \rightarrow \onehole V .$ we get:
  $(v_r,\, \Omega'_l) \le (v_r,\, \Omega'_r)$.
 \item Case \Rule{Le-RemoveAnon}. Trivial, for the same arguments as in the $\ecopy p$ case.
 \item Case \Rule{Le-Merge-Locs}.
   There exists $\ell_1$, $\ell_2$, $v$, $\Omega_1[.,\ .]$ such that
    $\Omega_l = \esubst{\ell_2}{\ell_1}{(\Omega_1[\eloc{\ell_1}{v}])}$,
    $\Omega_r = \Omega_1[\eloc{\ell_1}{(\eloc{\ell_2}{v})}]$.

  The crucial point of the proof is that the read and write rules are such that it is not
  possible to move the inner location (i.e., $\eloc{\ell_2}{v}$) elsewhere (\Rule{R-Loc}
  requires the $\mathsf{immut}$ capability, while here we use the $\mathsf{mov}$
  capability). If it were the case, we could not apply \Rule{Le-Merge-Locs} to conclude
  the proof because after the move the locations might not be nested anymore (i.e.,
  we might have lost the fact that $\ell_1$ and $\ell_2$ point to the same value).
  This is actually the
  reason why we forbid moving a value through a pointer which points to a location
  (while we allow moving a value through a pointer which points to an HLPL box).


 \item Case \Rule{Le-SharedLoan-To-Loc}. Similar to the case \Rule{Le-Mut-Borrow-To-Ptr}
  but simpler because we only have to consider the loan, which can't be moved.

 \item Case \Rule{Le-Box-To-Loc}. Also similar to the cases above.
 \item Case \Rule{Le-Subst}. We easily get that the read and write judgements
   commute with identifier substitutions.
 \end{itemize}

 \item Case \Rule{HLPL+-E-Pointer} ($\ebrw p$, $\erbrw p$, $\embrw p$).
  \begin{itemize}
  \item Reflexive case. Trivial.
  \item Transitive case. Trivial by the inductive hypotheses.
  \item Case \Rule{Le-SharedReservedBorrow-To-Ptr}.
    By the premises of \Rule{Le-SharedReservedBorrow-To-Ptr},
    there exists $\Omega_1[.]$ such that
    $\Omega_l = \Omega_1[\eptr{\ell}] \le \Omega_1[\esborrow{\ell}] = \Omega_r$
  
    We prove by induction on $p$ the auxiliary lemma.
    \begin{align*}
    &\forall\; p\, v_r,\\
    &\vdash \Omega_r(p) \eqmove v_r \Rightarrow\\
    &(\exists\; \omegatwoholes 2 . . \, v,\,\\
      &\quad  \omegaonehole 1 . = \omegatwoholes 2 . v \eand\\
      &\quad  (\forall\; v,\, \omegatwoholes 2 . v (p) \eqmove v) \eand\\
      &\quad  (\forall\; v,\, \updtstateplace{\omegatwoholes 2 . .} p v \ueqmove v)) \eor\\
    &(\exists\; \omegaonehole 2 . \, \onehole V .,\,\\
      &\quad \omegaonehole 1 . = \omegaonehole 2 . {\onehole V .}\\
      &\quad  (\forall\; v,\, \omegaonehole 2 v (p) \eqmove v) \eand\\
      &\quad  (\forall\; v,\, \updtstateplace{\omegaonehole 2 .} p v \ueqmove v))
    \end{align*}

    We conclude by doing a case disjunction.

  \item Case \Rule{Le-MutBorrow-To-Ptr}. Similar to above.
  \item Case \Rule{Le-RemoveAnon}. Similar to above (adding an anonymous value
    with no borrows or loans doesn't have any effect on the evaluation).
  \item Case \Rule{Le-Merge-Locs}.
    Similar to the move case. We note that we can't introduce a location between
    two shared locations, meaning that we can apply \Rule{Le-Merge-Locs} to conclude.
  \item Case \Rule{Le-SharedLoan-To-Loc}. Similar. The important case is the case
    where we transform the shared loan to which we create a new pointer.
  \item Case \Rule{Le-Box-To-Loc}. Similar to the \Rule{Le-MutBorrow-To-Ptr} case.
  \item Case \Rule{Le-Subst}. We easily prove by induction that the read and write
    judgements commute with substitutions.
     However, we have to consider the cases where the substitution is applied to the
     (potentially fresh) location that the new pointer points to. If the location
     is fresh, we can use the substituted identifier to apply \Rule{HLPL+-E-Pointer}
     on the left, making the two states equal, which allows us to conclude by
     reflexivity of $\le$. If it is not fresh, we conclude by \Rule{Le-Subst}.
  \end{itemize}
\item Case $\enew$. The premises of the rule give us that:
  \begin{align*}
  &\Omega_r \vdash op \exprres{v'_r}{\Omega''_r}\\
  &v_r = \eptr\ell\\
  &\Omega'_r = \Omega''_r,\, \ell \rightarrow v'_r
  \end{align*}

  By the induction hypothesis we get:
  \begin{align*}
  &\Omega_l \vdash op \exprres{v'_l}{\Omega''_l}\\
  &(v'_l,\, \Omega''_l) \le (v'_r,\, \Omega''_r)
  \end{align*}
  
  By \Rule{HLPL-E-Box-New} we get:
  \begin{align*}
  &\Omega_l \vdash \enew op \exprres{\eptr\ell}{\Omega''_l},\, \_ \rightarrow v'_l
  \end{align*}
  
  All we have to show is that (for some $\ell'$ fresh, that we choose so that
  it doesn't appear in the left but also not in the right environment - this
  makes reasoning about \Rule{Le-Subst} easier, and we can always use
  \Rule{Le-Subst} to map $\ell$ to $\ell'$):
  \begin{align*}
  &(\eptr\ell',\, (\Omega''_l,\, \ell' \rightarrow v'_l)) \le
   (\eptr\ell,\, (\Omega''_r,\, \ell \rightarrow v'_r))
  \end{align*}
  
  We prove it by induction on $\le$.
  
\item Case constants. Trivial, because the states are unchanged, and the result
  of evaluating the constants is independent of the states.
\item Case adt constructor. Trivial by the induction hypotheses (we conclude
  in a manner similar to the case $\enew$).
\item Case unary/binary operations ($\neg$, +, -, etc.). Trivial
  by the induction hypotheses.
\end{itemize}

\medskip
We now prove the following lemma about assignments.

\begin{lemma}[Assign-Preserves-HLPL+-Rel]
\thlabel{hlpl+-assign-rel}
For all $\Omega_l$ and $\Omega_r$ \hlplp states, $rv$ right-value and $p$ place we have:
\begin{align*}
&\Omega_l \le \Omega_r \Rightarrow
\forall\, \Omega'_r, \ \Omega_r \vdashHlplp \eassign p rv \stmtres{()}{\Omega'_r} \Rightarrow\\
&\exists\, \Omega'_l,\ \Omega_l \vdashHlplp \eassign p rv \stmtres{()}{\Omega'_l} \wedge
\Omega'_l \le \Omega'_r
\end{align*}
\end{lemma}

\noindent\textbf{Proof}\\
Lemma \thref{hlpl+-rvalue-rel} gives us that there exists $v_r$, $\Omega''_r$,
$v_l$, $\Omega''_l$ such that:
\begin{align*}
&\Omega_r \vdashHlplp rv \exprres{v_r}{\Omega''_r} \eand\\
&\Omega_l \vdashHlplp rv \exprres{v_l}{\Omega''_l} \eand\\
&(v_l,\, \Omega''_l) \le (v_r,\, \Omega''_r)
\end{align*}

We do the proof by induction on $(v_l,\, \Omega''_l) \le (v_r,\, \Omega''_r)$, then
on the path $p$. The reasoning is very similar to what we saw in the proof of lemma
\thref{hlpl+-rvalue-rel}. We focus on the important parts of the proofs.

\begin{itemize}
\item Reflexive case. Trivial.
\item Transitive case. Trivial by the induction hypotheses.
\item Case \Rule{Le-SharedReservedBorrow-To-Ptr}. We cannot update through a
  shared or a reserved borrow. We have to reason about three cases:
  1. the hole is the right-value and moved to the place at $p$; 2. the hole is
  in the overwritten value, in which case we use the fact that it will be moved to an
  anonymous variable; 3. the hole is neither in the right-value or the overwritten value.
  In all cases we conclude by using \Rule{Le-SharedReservedBorrow-To-Ptr}.
\item Case \Rule{Le-MutBorrow-To-Ptr}.
  Same as \Rule{Le-SharedReservedBorrow-To-Ptr} but we have to consider more cases:
  1. the right-value might contain the mutable borrow that gets transformed to
  a pointer; 2. the overwritten value might contain the mutable borrow \emph{and/or}
  the mutable loan, but they will be moved to an anonymous value.
  In all cases we conclude by using \Rule{Le-MutBorrow-To-Ptr}.
\item Case \Rule{Le-RemoveAnon}.
  Similar to the cases in \thref{hlpl+-rvalue-rel}: trivial by using the fact that an
  anonymous value with no loans or borrows doesn't have any influence on the evaluation.
\item Case \Rule{Le-Merge-Locs}.
  We have to consider the fact that the two locations may be in the overwritten value.
  If it is the case, they are both moved to an anonymous variable. In particular,
  they can't get separated because we can't dive into a location by using the
  move capability (\Rule{W-Loc}), meaning we can conclude by applying
  \Rule{Le-Merge-Locs}.
\item Case \Rule{Le-SharedLoan-To-Loc}.
  Similar to above cases. The shared loan might be in the overwritten value,
  in which case it is moved to an anonymous value.
\item Case \Rule{Le-Box-To-Loc}.
  Similar to above cases. We have to consider the following cases:
  1. the box value may be in the right-value; 2. we may move the (inner) boxed
  value but not the outer box; 3. we may overwrite the box value; 4. we may
  overwrite the (inner) boxed value but not the outer box value.
  In all cases, we conclude by using \Rule{Le-Box-To-Loc}.
\item Case \Rule{Le-Subst}. Trivial by using the fact that the read and write
  judgements commute with identifier substitutions.
\end{itemize}

\medskip
We now prove that reorganizations preserve the relation between \hlplp states.
\begin{lemma}[Reorg-Preserves-HLPL+-Rel]
\thlabel{hlpl+-reorg-rel}
For all $\Omega_l$ and $\Omega_r$ \hlplp states we have:
\begin{align*}
&\Omega_l \le \Omega_r \Rightarrow
\forall\; \Omega'_r, \ \Omega_r \hookrightarrow \Omega'_r \Rightarrow\\
&\exists\; \Omega'_l,\ \Omega_l \hookrightarrow \Omega'_l \wedge
\Omega'_l \le \Omega'_r
\end{align*}
\end{lemma}
\noindent\textbf{Proof}\\
By induction on $\Omega_r \hookrightarrow \Omega'_r$.
\begin{itemize}
\item Case \Rule{Reorg-None}. Trivial.

\item Case \Rule{Reorg-Seq}. Trivial by the induction hypotheses.

\item Case \Rule{Reorg-End-MutBorrow}.
  By induction on $\Omega_l \le \Omega'_r$.
  \begin{itemize}
  \item Reflexive case. Trivial.
  \item Transitive case. Trivial by the induction hypotheses.
  \item Case \Rule{Le-SharedReservedBorrow-To-Ptr}.
    We have to make a case disjunction on whether the shared borrow we convert
    to a pointer is inside the mutably borrowed value or not.
    We apply \Rule{Reorg-End-MutBorrow} to the left environment and conclude
    with \Rule{Le-SharedReservedBorrow-To-Ptr}.
  \item Case \Rule{Le-MutBorrow-To-Ptr}.
    By using the premises of \Rule{Le-MutBorrow-To-Ptr} we get that there can't
    be other mutable borrows or loans with the same identifier as the one
    being transformed (note again that we don't have a well-formedness assumption
    about the states).
    We have to consider the following cases: 1. the ended mutable borrow
    may contain the borrow we convert to a pointer; 2. the ended mutable borrow
    and the transformed borrow may be the same.
    In all cases, we end the pointer on the left, then the location.
    If we are in case 2., we conclude by using reflexivity of $\le$ (the states on the
    left and on the right are now the same). Otherwise, we conclude by using
    \Rule{Le-MutBorrow-To-Ptr}.
  \item Case \Rule{Le-RemoveAnon}. Trivial.
  \item Case \Rule{Le-Merge-Locs}.
    Following the premise of \Rule{Le-Merge-Locs}, the borrow we end is necessarily
    independent of the inner location we get rid of. This allows us apply
    \Rule{Reorg-End-MutBorrow} to the left state and conclude
    by using \Rule{Le-Merge-Locs}.
  \item Case \Rule{Le-SharedLoan-To-Loc}.
    The shared loan doesn't have the same loan identifier as the mutable borrow and the
    mutable loan we end. We apply \Rule{Reorg-End-MutBorrow} to the left state
    and conclude by \Rule{Le-SharedLoan-To-Loc}.
  \item Case \Rule{Le-Box-To-Loc}. Straightforward. We have to reason about
    the respective places of the box and the borrow and the loan. We apply
    \Rule{Reorg-End-MutBorrow} to the left state and conclude
    by \Rule{Le-Box-To-Loc}.
  \item Case \Rule{Le-Subst}. We apply \Rule{Reorg-End-MutBorrow} to the left
    state, potentially on the substituted identifier.
  \end{itemize}

\item Case \Rule{Reorg-End-SharedReservedBorrow}.
  By induction on $\Omega_l \le \Omega'_r$.
  \begin{itemize}
  \item Reflexive case. Trivial.
  \item Transitive case. Trivial by the induction hypotheses.
  \item Case \Rule{Le-SharedReservedBorrow-To-Ptr}.
    The borrow we end may be the one we transform to a pointer. If it is the case,
    we end the corresponding pointer (\Rule{Reorg-End-Pointer} and we conclude by using
    reflexivity of $\le$.  Otherwise, we use \Rule{Le-SharedReservedBorrow-To-Ptr}.
  \item Case \Rule{Le-MutBorrow-To-Ptr}.
    By using the premise of \Rule{Le-MutBorrow-To-Ptr} we get that the mutable
    borrow (and its loan) can't have the same loan identifier as the shared borrow
    we convert to a pointer.
    This allows us to conclude by using \Rule{Reorg-End-SharedReservedBorrow} on the left
    environment then \Rule{Le-MutBorrow-To-Ptr}.
  \item Case \Rule{Le-RemoveAnon}. Trivial because the anonymous variables
    can't contain borrows or loans.
  \item Case \Rule{Le-Merge-Locs}.
    The borrow we convert to a pointer may be affected by the substitution.
    In all cases, we conclude with \Rule{Reorg-End-SharedReservedBorrow} then
    \Rule{Le-Merge-Locs}.
  \item Case \Rule{Le-SharedLoan-To-Loc}.
    We note that the shared loan we end and the borrow we convert are necessarily
    independent.
    We conclude by \Rule{Reorg-End-SharedReservedBorrow} applied to the left
    environment then \Rule{Le-SharedLoan-To-Loc}.
  \item Case \Rule{Le-Box-To-Loc}.
    We apply \Rule{Reorg-End-SharedReservedBorrow} to the left environment then
    conclude by \Rule{Le-Box-To-Loc}.
  \item Case \Rule{Le-Subst}. We apply \Rule{Reorg-End-SharedReservedBorrow} to the left
    state, potentially on the substituted identifier.
  \end{itemize}

\item Case \Rule{HLPL+-Reorg-End-SharedLoan}.
  By induction on $\Omega_l \le \Omega'_r$.
  \begin{itemize}
  \item Reflexive case. Trivial.
  \item Transitive case. Trivial by the induction hypotheses.
  \item Case \Rule{Le-SharedReservedBorrow-To-Ptr}.
    By the premise of \Rule{Reorg-End-SharedLoan}, the borrow we convert
    can't have the same identifier as the shared loan we end.
    We apply \Rule{Reorg-End-SharedLoan} to the left state and conclude
    by \Rule{Le-SharedReservedBorrow-To-Ptr}.
  \item Case \Rule{Le-MutBorrow-To-Ptr}.
    By the premise of \Rule{Le-MutBorrow-To-Ptr} the borrow (and the loan) we
    convert can't have the same identifier as the shared loan we end.
    We apply \Rule{Reorg-End-SharedLoan} to the left environment and
    conclude by \Rule{Le-MutBorrow-To-Ptr}.
  \item Case \Rule{Le-RemoveAnon}. Trivial because the anonymous variable
    can't contain loans.
  \item Case \Rule{Le-Merge-Locs}.
    By the premise of \Rule{Le-Merge-Locs}, the shared loan we end on the right
    can't have the same identifier as the inner location that we remove.
    This means that it is not affected by the substitution, and as a result
    the premises of \Rule{Reorg-End-SharedLoan} are also satisfied in the left
    state; we apply this rule and conclude by \Rule{Le-Merge-Locs}.
  \item Case \Rule{Le-SharedLoan-To-Loc}.
    We have to make a case disjunction on whether the shared loan we convert
    is also the shared loan we end. If it is the case, we apply \Rule{Reorg-End-Loc}
    to the left state and conclude by reflexivity of $\le$,
    otherwise we apply \Rule{Reorg-End-Shared-Loan} and conclude by
    \Rule{Le-SharedLoan-To-Loc}.
  \item Case \Rule{Le-Box-To-Loc}. Straightforward. We conclude by \Rule{Le-Box-To-Loc}.
  \item Case \Rule{Le-Subst}. We apply \Rule{Reorg-End-SharedLoan} to the left
    state, potentially on the substituted identifier.
  \end{itemize}

\item Case \Rule{Reorg-End-Pointer}.
  By induction on $\Omega_l \le \Omega'_r$.
  \begin{itemize}
  \item Reflexive case. Trivial.
  \item Transitive case. Trivial by the induction hypotheses.
  \item Case \Rule{Le-SharedReservedBorrow-To-Ptr}.
    We apply \Rule{Reorg-End-Pointer} in the left environment
    (beware that in the right state we have a \emph{borrow}, so there must be
    another pointer, possibly with the same identifier, that we can also find
    in the left state) and conclude by \Rule{Le-SharedReservedBorrow-To-Ptr}.
  \item Case \Rule{Le-MutBorrow-To-Ptr}. Same as the case \Rule{Le-SharedReservedBorrow-To-Ptr}.
  \item Case \Rule{Le-RemoveAnon}. Trivial.
  \item Case \Rule{Le-Merge-Locs}. The pointer may be affected by the substitution,
    but we can also end it in the left state and conclude by \Rule{Le-Merge-Locs}.
  \item Case \Rule{Le-SharedLoan-To-Loc}.
    We apply \Rule{Reorg-End-Pointer} in the left state and conclude by
    \Rule{Le-SharedLoan-To-Loc}.
  \item Case \Rule{Le-Box-To-Loc}.
    We apply \Rule{Reorg-End-Pointer} in the left state and conclude by
    \Rule{Le-Box-To-Loc}.
  \item Case \Rule{Le-Subst}. We apply \Rule{Reorg-End-Pointer} to the left
    state, potentially on the substituted identifier.
  \end{itemize}
  
\item Case \Rule{Reorg-End-Loc}.
  By induction on $\Omega_l \le \Omega'_r$.
  \begin{itemize}
  \item Reflexive case. Trivial.
  \item Transitive case. Trivial by the induction hypotheses.
  \item Case \Rule{Le-SharedReservedBorrow-To-Ptr}.
    By the premises of \Rule{Reorg-End-Loc}, the location we end and the
    borrow we convert don't have the same identifier. We also end the location
    on the left and conclude by \Rule{Le-SharedReservedBorrow-To-Ptr}.
  \item Case \Rule{Le-MutBorrow-To-Ptr}. Same as case \Rule{Le-SharedReservedBorrow-To-Ptr}.
  \item Case \Rule{Le-RemoveAnon}. Trivial.
  \item Case \Rule{Le-Merge-Locs}.
    This one is subtle.
    Let's name $\ell_0$ the location we end, $\ell_1$ the outer location
    and $\ell_2$ the inner location we merge with $\ell_1$.
    We might have that $\ell_2 = \ell_0$. If this happens, it means that
    there are no borrows or loans with identifier $\ell_2$, meaning that
    the substitution applied by \Rule{Le-Merge-Locs} has no effect.
    In particular, if we end $\ell_2$ in the left state by \Rule{Reorg-End}
    then we get the same state as on the right and conclude by reflexivity of
    $\le$.
    We might also have that $\ell_1 = \ell_0$. In this case, we also end
    the outer location $\ell_1$ in the left state then conclude by applying
    \Rule{Le-Subst} to substitute $\ell_2$ (the inner location identifier)
    with $\ell_1$.
    If we are in one of the above cases, we apply \Rule{Reorg-End-Loc}
    and conclude by \Rule{Le-Merge-Locs}.
  \item Case \Rule{Le-SharedLoan-To-Loc}.
    We apply \Rule{Reorg-End-Loc} to the left state (there must be a location
    which corresponds to the location we end on the right - the shared loan
    and this location can not be related) and conclude by
    \Rule{Le-SharedLoan-To-Loc}.
  \item Case \Rule{Le-Box-To-Loc}. Straightforward.
    We apply \Rule{Reorg-End-Loc} to the left state and conclude by \Rule{Le-Box-To-Loc}.
  \item Case \Rule{Le-Subst}. We apply \Rule{Reorg-End-Loc} to the left
    state, potentially on the substituted identifier, and conclude either
    by reflexivity or by \Rule{Le-Subst}.
  \end{itemize}

\item Case \Rule{Reorg-Activate-Reserved}.
  By induction on $\Omega_l \le \Omega'_r$.
  \begin{itemize}
  \item Reflexive case. Trivial.
  \item Transitive case. Trivial by the induction hypotheses.
  \item Case \Rule{Le-SharedReservedBorrow-To-Ptr}.
    If the borrow we activate is not the one we convert to a pointer,
    we apply \Rule{Reorg-Activate-Reserved} on the left state and conclude by
    \Rule{Le-SharedReservedBorrow-To-Ptr}.
    If it is the same, we do nothing on the left state and conclude
    by \Rule{Le-MutBorrow-To-Ptr} (the premises of \Rule{Reorg-Activate-Reserved}
    give us the premises we need for \Rule{Le-MutBorrow-To-Ptr}).
  \item Case \Rule{Le-MutBorrow-To-Ptr}.
    The borrow we activate can not be the same as the borrow we convert to
    a pointer. This means we can apply \Rule{Reorg-Activate-Reserved} to the left
    environment and conclude by \Rule{Le-MutBorrow-To-Ptr}.
  \item Case \Rule{Le-RemoveAnon}. Trivial.
  \item Case \Rule{Le-Merge-Locs}.
    The locations we merge can't have the same identifier as the borrow we activate.
    we can apply \Rule{Reorg-Activate-Reserved} to the left
    environment and conclude by \Rule{Le-Merge-Locs}.
  \item Case \Rule{Le-SharedLoan-To-Loc}.
    Similar to \Rule{Le-SharedReservedBorrow-To-Ptr}.
  \item Case \Rule{Le-Box-To-Loc}. We apply \Rule{Reorg-Activate-Reserved} to the left
    state and conclude by \Rule{Le-Box-To-Loc}.
  \item Case \Rule{Le-Subst}.
    We make a case disjunction on whether the borrow we activate is the one on
    which we apply the substitution to use \Rule{Reorg-Activate-Reserved} on the proper
    borrow in the left state and conclude by \Rule{Le-Subst}.
  \end{itemize}
\end{itemize}

\medskip
We finally turn to the proof of the target theorem, that is \Thm{Eval-Preserves-HLPL+-Rel}.\\

\noindent\textbf{Proof}\\
By induction on $\Omega_r \vdashHlplp\, s \rightsquigarrow \stmtres{r}{\Omega_r'}$.

\begin{itemize}
\item Case empty statement. Trivial.
\item Case \Rule{E-Reorg}. By Lemma \thref{hlpl+-reorg-rel}.
\item Case $s_0;\; s_1$. Trivial by the induction hypotheses.
\item Case $\eassign p :rv$. By Lemma \thref{hlpl+-assign-rel}.
\item Case \li+if then else+. Trivial by the induction hypotheses.
\item Case \li+match+. Trivial by the induction hypotheses.
\item Case $\efree p$. The reasoning is similar to the $\emove p$ and $\eassign p$
  cases in Lemmas \thref{hlpl+-rvalue-rel} \thref{hlpl+-assign-rel}.
  The proof is straightforward by induction on $\Omega_l \le \Omega_r$.
  The interesting case is the combination of \Rule{E-Box-Free} in the right
  state and \Rule{Le-Box-To-Loc} as the relation between the two states.
  By applying \Rule{Reorg-End-Pointer} then \Rule{HLPL-E-Box-Free} to the left
  state we get equal states and
  conclude by reflexivity of $\le$. Importantly, the rule \Rule{E-Box-Free} was
  written so that only the \emph{content} of the box gets moved to an anonymous
  value, so that we can relate \Rule{E-Box-Free} to \Rule{HLPL-E-Box-Free}.
\item Case $\ereturn$. Trivial.
\item Cases $\epanic$, $\ebreak i$, $\econtinue i$. Trivial.
\item Case loop. Trivial by the induction hypotheses.
\item Case function call.
  The rule is heavy but the proof is straightforward.
  We use the Lemma \thref{hlpl+-rvalue-rel} to relate the states resulting
  from evaluating the operands given as inputs to the function (trivial induction
  on the number of inputs).
  We easily get that the states resulting from $\pushstack$ are
  related (again, induction on the number of inputs then on the number of
  local variables which are not used as inputs).
  The induction hypothesis gives us that the states are related after evaluation
  of $body$. If the tag is $\epanic$ we are done.
  If it is $\ereturn$, we prove that the states resulting from $\popstack$
  are related (induction on the number of local variables, and we use
  \thref{hlpl+-assign-rel} for the assignments which ``drop'' the local variables).
  Finally, we do a reasoning similar to the one we did in
  \thref{hlpl+-assign-rel} to show that the states are still related after
  assigning to the destination.
\end{itemize}

\section{Forward Simulation Between PL and HLPL}
\appendixlabel{pl}

The Pointer Language (PL) uses an explicit heap adapted from the CompCert memory model,
where the memory is a map from block id to sequence of memory cells, and each memory cell
contains a word.
We do not fix the architecture (that is, the width of a word) as this is not relevant
to the proof.
An address is a block id and an offset inside this block.
In this model, values are sequences of cells;
for instance, a pair is the concatenation of the cells of its sub-values.
Similarly to HLPL, PL doesn't make any distinctions between shared borrows and mutables borrows,
which are modeled as pointers.
Because PL operates over the same AST as LLBC, the evaluation of places is still
relatively high-level and requires typing information to compute the proper
offsets; for this purpose, memory blocks are typed, and those types are used
for the sole purpose of guiding those projections.
Also note that the simulation proof relies on the fact that the rules for HLPL and LLBC
preserve the types (e.g., see \Rule{Write-Base} in \Figure{llbc-read} - we made the
types implicit in most of the rules, in particular in the body of the paper, for the
purpose of clarity) so it is not possible to update a value with a value of a different
type, in order to enforce the fact that the \emph{size} of the values is preserved.
Similarly to what CompCert does, every variable has its own memory block, every
allocation inserts a fresh memory block in the heap, while deallocation
removes the corresponding block.
We always deallocate a complete block at once (we do not deallocate sub-blocks).
We do not distinguish allocations on the stack and allocations on the heap; however, as we
forbid double frees, deallocating a block allocated on the stack (i.e., a variable) leads
to the program eventually being stuck as we deallocate variables when popping the stack.

We show the grammar of values and states in \Figure{pl-grammar}.
Values are addresses (a block id and an offset), literals (which must fit into a word), or undefined
values. We use the symbol $\eundef$ for undefined values to distinguish
them from $\bot$ values that we use in LLBC.
While we use $\bot$ to keep track of non-initialized values and lost
permissions such as moved values or ended borrows, in PL $\eundef$
is simply a non-initialized, poison value~\cite{llvm-poison}.
Similarly to $\bot$, it causes the program to get stuck if it ever attempts to read them.
But at the difference of LLBC, moving a value (\Rule{PL-E-Move}) leaves it
unchanged by behaving like a copy, while there is no such thing
as ending a borrow or a pointer by replacing it with a $\eundef$.

\begin{figure}
  \smaller
  \arraycolsep=1pt %
  \centering

  \begin{array}[t]{llll}
    v & ::= & & \text{value} \\
    && addr & \text{address} \\
    && \ktrue \mid \kfalse \mid n_{\mathsf{i32}} \mid n_{\mathsf{u32}} \mid \ldots &
      \text{literal constants} \\
    && \eundef & \text{undefined value}
    \\[1ex]

    addr & ::= && \text{address}\\
    && (bi, \, n) & \text{block id and offset}
    \\[1ex]

    \Omega^{\text{LLBC}} & ::= & \{\;
      \kw{env} : x \underset {\text{partial, inj}} \longrightarrow bi,\;
      \kw{heap} : bi \underset {\text{partial}} \longrightarrow [v] : \tau,\;
      \kw{stack} : [[x]]\; \} \; & \text{state}\\[1ex]
  \end{array}
  \caption{Grammar of PL States and Values}
  \flabel{pl-grammar}
\end{figure}

As with our low-level memory model values are modeled as sequences of words, we need to
introduce a notion of type size to properly relate PL values and HLPL values (\Figure{sizeof}).

\begin{figure}
\smaller
\centering
\begin{align*}
&\esizeof{\tau} := 1 \text{ if $\tau$ is a literal type}\\
&\esizeof{(\tau_0, \tau_1)} := \esizeof{\tau_0} + \esizeof{\tau_1} \\
&\esizeof{(\tau_0 + \tau_1)} := 1 + \kw{max}(\esizeof{\tau_0},\; \esizeof{\tau_1}) \quad\text{  (we need an
    integer for the tag)}\\
&\esizeof{(\ebox{\tau})} := 1\\
&\esizeof{(\ebrw{\tau})} := 1\\
&\esizeof{(\embrw{\tau})} := 1\\
&\esizeof{(\mu X.\, \tau)} := \esizeof{(\tau[\mu X.\, \tau \Big/ X])}
\end{align*}
\caption{The \kw{sizeof} Function}
\flabel{sizeof}
\end{figure}

We use the following notations.
If $s$ is a sequence, we use the standard notations about subsequences by using intervals.
For instance, $(\Omega.\emem\; bi)[n;\, m[$ is the sub-sequence of $\Omega.\emem\; bi$ covering
the cells from index $n$ (included) to index $m$ (excluded).
We also use the standard index notation: $(\Omega.\emem\, bi)[i]$ is the cell at
index $i$ in sequence $\Omega.\emem\; bi$.
We define an ``update'' notation for sequences: $(\Omega.\emem\; bi)[n;\, m[ := [\vec{v}]$ is $\Omega$
where the sub-sequence of $\Omega.\emem\; bi$ of indices $[n;\, m[$ has been udpated
with $[\vec{v}]$.
The notation $[\vec{v}] : \tau$ simply means that the sequence $[\vec{v}]$ has length
$\esizeof{\tau}$, and \emph{nothing more}.
In particular, we do not check any well-typedness property of those values.

We introduce read and write judgments, like in HLPL and LLBC.
Those reading and writing judgments are guided by types, which we use to compute
address offsets when reducing projections, and to compute the length of the sequence
of cells we have to read.
We use this typing information only for the projections and nothing
else. In particular, we do not check any well-typedness property of the states.

For reading in the environment, we introduce two judgments (\Figure{pl-read}).
The judgment
$\vdash \Omega(p : \tau) \readarrow [\vec{v}]$
states that reading $\esizeof{\tau}$ cells at place $p$ gives
$[\vec{v}]$. The judgment $\vdash \&\Omega(p : \tau) \readarrow addr$,
states that $addr$ is the address of the (sequence of) value(s) at place $p$. Note
that this judgment does not enforce that there are $\esizeof{\tau}$ cells available at
$addr$.
Also note that for sum values, we reserve the first cell for the tag (see \Rule{PL-ReadAddress-ProjSum}).

For writing in environments we introduce the judgment
$\vdash \Omega(p) \updtoperator [\vec{v}] \updtarrow \Omega'$ (\Figure{pl-write}), which states
that updating the cells at place $p$ in $\Omega$ with the sequence of values
$[\vec{v}]$ yields state $\Omega'$. The full judgment is in \Figure{pl-write}.

\NewDocumentCommand{\PlReadRules}{s}{
  \inferrule[PL-ReadAddr-Var]{
    \Omega.\eenv\, x = bi
  }{
    \vdash \&\Omega(x : \tau) \readarrow (bi,\, 0)
  }

  \inferrule[PL-ReadAddress-Deref]{
    \vdash \&\Omega(p : \{\&,\, \&\mathsf{mut},\, *\}\tau) \readarrow (bi,\, n)\\
    (\Omega.\emem\, bi)[n] = (bi',\, n')
  }{
    \vdash \&\Omega(*p : \tau) \readarrow (bi',\, n')
  }

  \inferrule[PL-ReadAddress-ProjPairLeft]{
    \vdash \&\Omega(p : (\tau_0,\, \tau_1)) \readarrow (bi,\, n)\\
  }{
    \vdash \&\Omega(p.0 : \tau_0) \readarrow (bi,\, n)
  }

  \inferrule[PL-ReadAddress-ProjPairRight]{
    \vdash \&\Omega(p : (\tau_0,\, \tau_1)) \readarrow (bi,\, n)\\
    n + \esizeof{\tau_0} < \elength{\Omega.\emem\, bi}
  }{
    \vdash \&\Omega(p.1 : \tau_1) \readarrow (bi,\, n + \esizeof{\tau_0})
  }

  \inferrule[PL-ReadAddress-ProjSum]{
    \vdash \&\Omega(p : \tau_0 + \tau_1) \readarrow (bi,\, n)\\\\
    n + 1 < \elength{\Omega.\emem\, bi}\\\\
    \tau = \tau_0 \vee \tau = \tau_1
  }{
    \vdash \&\Omega(p.0 : \tau) \readarrow (bi,\, n + 1)
  }

  \inferrule[PL-Read]{
    \vdash \&\Omega(p : \tau) \readarrow (bi,\, n)\\\\
    (\Omega.\emem\, bi)[n;\, \esizeof{\tau}[ = [\vec{v}]
  }{
    \vdash \Omega(p : \tau) \readarrow [\vec{v}]
  }
}

\begin{figure}
  \centering
  \smaller
  \begin{mathpar}
    \PlReadRules*
  \end{mathpar}
  \caption{Read Judgments for PL}
  \flabel{pl-read}
\end{figure}

\NewDocumentCommand{\PlWriteRules}{s}{
   \inferrule[PL-Write]{
     \vdash \&\Omega(p : \tau) \readarrow (bi, n)\\
     \Omega' = \left((\Omega.\emem\, bi)[n;\, \esizeof{\tau}[ := [\vec{v}]\right)
   }{
     \vdash \Omega(p) \updtoperator [\vec{v}] \updtarrow \Omega'
   }
}

\begin{figure}
  \centering
  \smaller
  \begin{mathpar}
    \PlWriteRules*
  \end{mathpar}
  \caption{Write Judgment for PL}
  \flabel{pl-write}
\end{figure}

We define the rules to evaluate expressions in \Figure{pl-expr-rules}
and the rules to evaluate statements in \Figure{pl-statement-rules}.
The judgment $\Omega \vdash op \exprarrow [\vec{v} : \tau]$ has to be read as:
evaluating $op$ in $\Omega$ results in the sequence of values $[\vec{v}]$ of
length $\esizeof{\tau}$ (the type is only used for the length of the sequence).
One has to note that we encode sum values as tagged unions, and the first cell
is reserved for the tag (\Rule{PL-E-Constructor-Left}, \Rule{PL-E-Constructor-Right}).
The judgment $\Omega \vdash s \rightsquigarrow (r,\, \Omega')$ has to be read
as: evaluating statement $s$ in state $\Omega$ leads to result $r$ in state $\Omega'$.

\NewDocumentCommand{\PlExprRules}{s}{
   \inferrule[PL-E-Pointer]{
     \vdash \&\Omega(p : \tau) \readarrow addr
   }{
     \Omega(p) \vdash \{ \&,\, \&\mathsf{mut},\,\&\mathsf{reserved} \}\, p
     \exprarrow addr : *\tau
   }

   \inferrule[PL-E-Move]{
     \vdash \Omega(p : \tau) \readarrow [\vec{v}]
   }{
     \Omega(p) \vdash \emove p
     \exprarrow [\vec{v}] : \tau
   }

   \inferrule[PL-E-Copy]{
     \vdash \Omega(p : \tau) \readarrow [\vec{v}]
   }{
     \Omega(p) \vdash \ecopy p
     \exprarrow [\vec{v}] : \tau
   }

   \inferrule[PL-E-Constructor-Pair]{
     \Omega \vdash op_0 \exprarrow [\vec{v_0}] : \tau_0 \\
     \Omega \vdash op_1 \exprarrow [\vec{v_1}] : \tau_1
   }{
     \Omega \vdash (op_0,\, op_1) \exprarrow
     ([\vec{v_0}] \econcat [\vec{v_1}]) : (\tau_0,\, \tau_1)
   }

   \inferrule[PL-E-Constructor-Left]{
     \Omega \vdash op \exprarrow [\vec{v}] : \tau_0\\
     [\vec{v'}] = [0] \econcat [\vec{v}] \econcat [\vec{\bot}] \\
     \elength{[\vec{v'}]} = \esizeof{(\tau_0 + \tau_1)}
   }{
     \Omega \vdash \eleft op \rightsquigarrow \vec{v'} : \tau_0 + \tau_1
   }

   \inferrule[PL-E-Constructor-Right]{
     \Omega \vdash op \exprarrow [\vec{v}] : \tau_1\\
     [\vec{v'}] = [1] \econcat [\vec{v}] \econcat [\vec{\bot}] \\
     \elength{[\vec{v'}]} = \esizeof{(\tau_0 + \tau_1)}
   }{
     \Omega \vdash \eright op \rightsquigarrow \vec{v'} : \tau_0 + \tau_1
   }
}

\begin{figure}
  \centering
  \smaller
  \begin{mathpar}
    \PlExprRules*
  \end{mathpar}
  \caption{Evaluating Expressions in PL}
  \flabel{pl-expr-rules}
\end{figure}

\NewDocumentCommand{\PlStatementRules}{s}{
   \inferrule[PL-E-Assign]{
     \Omega \vdash op \rightsquigarrow [\vec{v}] : \tau\\
     \updtstateplace{\Omega}{p}{[\vec{v}]} \updtarrow \Omega'
   }{
     \Omega \vdash \eassign p op \stmtres{()}{\Omega'}
   }

   \inferrule[PL-E-IfThenElse-T]{
     \Omega \vdash op \rightsquigarrow [\kw{true}]\\
     \Omega \vdash s_0 \stmtres{r}{\Omega'}
   }{
     \Omega \vdash \eite{op}{s_0}{s_1} \stmtres{r}{\Omega'}
   }

   \inferrule[PL-E-IfThenElse-F]{
     \Omega \vdash op \rightsquigarrow [\kw{false}]\\
     \Omega \vdash s_1 \stmtres{r}{\Omega'}
   }{
     \Omega \vdash \eite{op}{s_0}{s_1} \stmtres{r}{\Omega'}
   }

   \inferrule[PL-E-Match-Left]{
     \vdash \Omega(p) \readarrow [0] \econcat [v] \\
     \Omega \vdash s_0 \stmtres{r}{\Omega'}
   }{
     \Omega \vdash \ematchsum{p}{s_0}{s_1} \stmtres{r}{\Omega'}
   }

   \inferrule[PL-E-Match-Right]{
     \vdash \Omega(p) \readarrow [1] \econcat [v] \\
     \Omega \vdash s_1 \stmtres{r}{\Omega'}
   }{
     \Omega \vdash \ematchsum{p}{s_0}{s_1} \stmtres{r}{\Omega'}
   }

   \inferrule[PL-E-Return]{
   }{
     \Omega \vdash \ereturn \stmtres{\ereturn}{\Omega}
   }

   \inferrule[PL-E-Panic]{
   }{
     \Omega \vdash \ereturn \stmtres{\epanic}{\Omega}
   }

   \inferrule[PL-E-Break]{
   }{
     \Omega \vdash \ebreak i \stmtres{\ebreak i}{\Omega}
   }

   \inferrule[PL-E-Continue]{
   }{
     \Omega \vdash \econtinue i \stmtres{\econtinue i}{\Omega}
   }

   \inferrule[PL-E-New]{
     bi \text{ fresh}\\
     \Omega \vdash op \rightsquigarrow [\vec{v}] : \tau\\\\
     \Omega' = \{\, \Omega\; \kw{with}\; \emem := \Omega.\emem \cup
       \{\, bi \rightarrow [\vec{v}] \,\} \,\}\\
   }{
     \Omega \vdash \enew{op} \stmtres{[(bi,\, 0)] : \ebox{\tau}}{\Omega'}
   }

   \inferrule[PL-E-Free]{
     \vdash \Omega(p : \ebox{\tau}) \readarrow [(bi,\, 0)]\\\\
     \Omega' = \{\, \Omega\; \kw{with}\; \emem := \Omega.\emem
       \setminus \{\, bi \,\}\,\}
   }{
     \Omega \vdash \efree{p} \stmtres{()}{\Omega'}
   }

  \inferrule[PL-E-Loop-Break-Inner]{
    \Omega \vdash s \stmtres{\ebreak{0}}{\Omega}
  }{
    \Omega \vdash \eloop{s} \stmtres{()}{\Omega}
  }

  \inferrule[PL-E-Loop-Break-Outer]{
    \Omega \vdash s \stmtres{\ebreak{i + 1}}{\Omega}
  }{
    \Omega \vdash \eloop{s} \stmtres{\ebreak{i}}{\Omega}
  }

  \inferrule[PL-E-Loop-Continue-Inner]{
    \Omega \vdash s \stmtres{\econtinue{0}}{\Omega'}\\\\
    \Omega' \vdash \eloop{s} \stmtres{r}{\Omega''}
  }{
    \Omega \vdash \eloop{s} \stmtres{r}{\Omega''}
  }

  \inferrule[PL-E-Loop-Continue-Outer]{
    \Omega \vdash s \stmtres{\econtinue{i + 1}}{\Omega}
  }{
    \Omega \vdash \eloop{s} \stmtres{\econtinue{i}}{\Omega}
  }

  \inferrule[PL-E-Loop-Panic]{
    \Omega \vdash s \stmtres{\epanic}{\Omega}
  }{
    \Omega \vdash \eloop{s} \stmtres{\epanic}{\Omega}
  }

  \inferrule[PL-E-Loop-Return]{
    \Omega \vdash s \stmtres{\ereturn}{\Omega}
  }{
    \Omega \vdash \eloop{s} \stmtres{\ereturn}{\Omega}
  }

  \inferrule[PL-E-PushStack]{
    \vec{bi} \text{ fresh}\\
    \Omega' = \{\; \Omega\; \kw{with}\;
    \kw{env} = [\overrightarrow{x \rightarrow (bi,\, 0)}] \econcat \Omega.\kw{env},\;
    \kw{mem} = [\overrightarrow{bi \rightarrow [\vec{v}]}] \econcat \Omega.\kw{mem},\;
    \kw{stack} = [\vec{x}] :: \Omega.\kw{stack}\;
    \}
  }{
    \plpushstack [\overrightarrow{x \rightarrow \vec{v}}]\; \Omega\; =\; \Omega'
  }

  \inferrule[PL-E-PopStack]{
    \Omega.\kw{stack} = ([x_{\text{ret}}] \econcat [\overrightarrow{x_i}]) :: \kw{stack'}\\
    \Omega.env = [x_{\text{ret}} \rightarrow bi_{\text{ret}}]
      \econcat [\overrightarrow{x_i \rightarrow [\overrightarrow{bi_i}]}] \econcat
      \kw{env'}\\
    \Omega.\kw{mem} = [\overrightarrow{v_{\text{ret}}}]\\
    \kw{env'} = \Omega.\kw{env} \setminus \{ bi_{\text{ret}},\, \overrightarrow{bi_i} \}\\
    \Omega' = \{\; \Omega\; \kw{with}\;
    \kw{stack}\; =\; \kw{stack'},\;
    \kw{env}\;=\;\kw{env'},\;
    \kw{mem}\;=\;\kw{mem'} \;\}
  }{
    \plpopstack \Omega\; =\; ([\overrightarrow{v_{\text{ret}}}],\; \Omega')
  }

  \inferrule[PL-E-Call]{
    \feqsig{\vec{\_}}\\
    \forall\; j, \Omega_j \vdash op_j \exprres{[\overrightarrow{v_j}]}{\Omega_{j+1}}\\
    \plpushstack
      \left([x_{\text{ret}} \rightarrow [\vec{\bot}]] \econcat
      [\overrightarrow{x_j \rightarrow [\overrightarrow{v_j}]}] \econcat
      [\overrightarrow{y_k \rightarrow [\vec{\bot}]}]\right) \Omega_m = \Omega_{\text{begin}}\\
    \Omega_{\text{begin}} \vdash body \stmtres{r}{\Omega_{\text{end}}}\\
    (r',\, \Omega_1) = \begin{cases}
      (\epanic,\, \Omega_{\text{end}}) & \text { if } r = \epanic \cr
      ((),\; \Omega''_{\text{end}}) &
        \text { if } r = \ereturn \;\wedge
        \plpopstack \Omega_{\text{end}}\;=\; (\overrightarrow{[v_{\text{ret}}}],\; \Omega'_{\text{end}})
        \;\wedge
        \updtstateplace{\Omega'_{\text{end}}}{p}{[\overrightarrow{v_{\text{ret}}}]} \ueqmut \Omega''_{\text{end}}
    \end{cases}
  }{
    \Omega_0 \vdash \eassign{p}{\fcall} \stmtres{r'}{\Omega_1}
  }
}

\begin{figure}
  \centering
  \smaller
  \begin{mathpar}
    \PlStatementRules*
  \end{mathpar}
  \caption{Evaluating Statements in PL}
  \flabel{pl-statement-rules}
\end{figure}

In order to relate the PL states and the HLPL states we introduce a concretization
function which turns HLPL states into PL states. The concretization
function is parameterized by several auxiliary functions:

\begin{itemize}
\item $\kw{blockof} : x + \ell^b \underset {\text{partial, inj}} \longrightarrow bi : \tau$:
  partial, injective function from variable or box identifier to typed block
  (a pair of a block identifier and a type).
\item $\kw{addrof} : \ell \underset {\text{partial}} \longrightarrow addr$:
  partial function from a loan or a box identifier to an address.
\end{itemize}

We define the concretization functions $\ecstate{\Omega}$ and $\ecval{v}$, for functions and
states respectively. We define the concretization function for states below, and the
concretization for values in \Figure{hlpl-concretize}.
\begin{align*}
\ecstate\Omegahlpl &:= \;\{\\
&\kw{env} := \lambda x\{x\in \Omegahlpl.\kw{env}\}.\;
  \eblockof{x}\\
&\kw{mem := \lambda bi\{(bi,\,\tau) \in \kw{image(\kw{blockof})}\}},\\
  &\qquad\ecval{(\Omegahlpl.\kw{mem}\;(\eblockofinv{bi}) : \tau)},\\
&\kw{stack} := \Omegahlpl.\kw{stack}\; \}
\end{align*}

\begin{figure}
  \centering
  \smaller
  \begin{mathpar}
  
   \inferrule[Concrete-Lit]{
     $v$ \text{ is a literal value}
   }{
     \ecval{(v : \tau)} = [i]
   }

   \inferrule[Concrete-Bot]{
     s = [\overrightarrow\eundef]\\
     \elength s = \esizeof\tau
   }{
     \ecval{(\bot : \tau)} = s
   }

   \inferrule[Concrete-Pair]{
     \ecval{v_0 : \tau_0} = [\vec{v}]\\\\
     \ecval{v_1 : \tau_1} = [\vec{w}]
   }{
     \ecval{((v_0,\, v_1) : (\tau_0,\, \tau_1))} = [\vec{v}] \econcat [\vec{w}]
   }

  \inferrule[Concrete-Sum-Left]{
     \ecval{v : \tau_0} = [\vec{v}]\\\\
     s = [0] \econcat [\vec{v}] \econcat [\overrightarrow{\eundef}]\\\\
     \elength{s} = \esizeof{(\tau_0 + \tau_1)}
   }{
     \ecval{(\eleft{v} : \tau_0 + \tau_1)} = s
   }

  \inferrule[Concrete-Sum-Right]{
     \ecval{v : \tau_1} = [\vec{v}]\\\\
     s = [0] \econcat [\vec{v}] \econcat [\overrightarrow{\eundef}]\\\\
     \elength{s} = \esizeof{(\tau_0 + \tau_1)}
   }{
     \ecval{(\eright{v} : \tau_0 + \tau_1)} = s
   }
  
   \inferrule[Concrete-Loc]{
     \ecval{(v : \tau)} = [\vec{v}] : \tau
   }{
     \ecval{\eloc{\ell}{(v : \tau)}} = [\vec{v}] : \tau
   }

  \inferrule[Concrete-Ptr-Location]{
     \eaddrof{\ell^p} = addr
   }{
     \ecval{(\eright{\eptr{\ell^p}})} = [addr] : *\tau
   }

  \inferrule[Concrete-Ptr-Box]{
     \eblockof{\ell^b} = bi
   }{
     \ecval{(\eright{\eptr{\ell^b}})} = [(bi,\, 0)] : *\tau
   }

  \end{mathpar}
  \caption{Concretizing HLPL Values}
  \flabel{hlpl-concretize}
\end{figure}

In order to properly relate the structured values in the HLPL states to a lower-level
view with addresses and memory cells,
we introduce the judgment $\readaddress\Omegahlpl {addr} (v : \tau)$
in \Figure{hlpl-read-address} which, given some mappings $\eblockofname$ and $\eaddrofname$,
defines what it means to read a value \emph{of a given type} at an \emph{address}
in the HLPL state \Omegahlpl.
In order to define this judgment, we also introduce an auxiliary judgment
$\readvaladdress{(v : \tau)}{n}{v' :  \tau'}$
which states that reading the sub-value of $v : \tau$ at offset $n$
yields $v' : \tau'$.

\begin{figure}
  \centering
  \smaller
  \begin{mathpar}
  
   \inferrule[HLPL-Read-Address]{
     \Omegahlpl.\kw{env}\;(\eblockofinv{((bi,\, n))}) = v : \tau\\\\
     \readvaladdress{(v : \tau)}{n}{v' : \tau}
   }{
     \readaddress{\Omegahlpl}{(bi, n)}{v' : \tau}
   }

   \inferrule[HLPL-Read-Offset-0]{
   }{
     \readvaladdress{(v : \tau)}{0}{v :  \tau}
   }
  
   \inferrule[HLPL-Read-Offset-Pair-Left]{
     n \le \esizeof{\tau_0}\\\\
     \readvaladdress{(v_0 : \tau_0)}{n}{v' :  \tau'}
   }{
     \readvaladdress{((v_0,\, v_1) : (\tau_0,\, \tau_1))}{n}{v' :  \tau'}
   }

   \inferrule[HLPL-Read-Offset-Pair-Right]{
     n \ge \esizeof{\tau_0}\\\\
     \readvaladdress{(v_1 : \tau_1)}{n - \esizeof{\tau_0}}{v' :  \tau'}
   }{
     \readvaladdress{((v_0,\, v_1) : (\tau_0,\, \tau_1))}{n}{v' :  \tau'}
   }

   \inferrule[HLPL-Read-Offset-Sum-Left]{
     1 \le n \le 1 + \esizeof{\tau_0}\\\\
     \readvaladdress{(v_0 : \tau_0)}{n - 1}{v' :  \tau'}
   }{
     \readvaladdress{(\eleft{v} : \tau_0 + \tau_1))}{n}{v' :  \tau'}
   }

   \inferrule[HLPL-Read-Offset-Sum-Right]{
     1 \le n \le 1 + \esizeof{\tau_1}\\\\
     \readvaladdress{(v_1 : \tau_1)}{n - 1}{v' :  \tau'}
   }{
     \readvaladdress{(\eright{v} : \tau_0 + \tau_1))}{n}{v' :  \tau'}
   }

   \inferrule[HLPL-Read-Offset-Loc]{
     \readvaladdress{(v : \tau)}{n}{v' :  \tau'}
   }{
     \readvaladdress{\eloc{\ell}{(v : \tau)}}{n}{v' :  \tau'}
   }
  \end{mathpar}
  \caption{Reading Addresses in HLPL}
  \flabel{hlpl-read-address}
\end{figure}

We define a compatibility predicate between an HLPL state $\Omegahlpl$ and the auxiliary
concretization functions $\eblockofname$ and $\eaddrofname$ in \Figure{pl-compat}.
The predicate $\ecompatible{\Omegahlpl}$ states
that $\eblockofname$ must be defined
for all the variables and box identifiers in $\Omegahlpl$, and that
$\eaddrofname$ must be consistent with the low-level view of $\Omegahlpl$
with addresses defined in \Figure{hlpl-read-address}.

\begin{figure}
  \centering
  \smaller
  \begin{align*}
  &\kw{compatible}\; \eblockofname\; \eaddrofname\; \Omegahlpl :=\\
  &\quad\kw{domain}\;\Omegahlpl.\kw{env} \subset \kw{domain}\;\eblockofname \eand\\
  &\quad(\forall\; \ell^b \in \Omegahlpl,\; \ell^b \in \kw{domain}\; \eblockofname) \eand\\
  &\quad(\forall\; addr\;\ell\;v,\; (\readaddress{\Omegahlpl}{addr} \eloc{\ell}{v})
   \Rightarrow \eaddrof{\ell} = addr) \eand\\
  &\quad(\forall\; \ell \in \Omegahlpl,\; \exists\;addr\;v,\;
    \readaddress{\Omegahlpl}{addr} \eloc{\ell}{v})
  \end{align*}
  \caption{Compatibility Predicate for the Auxiliary Concretization Functions}
  \flabel{pl-compat}
\end{figure}

We define a relation $\le$ between PL states in \Figure{pl-le}.
We need this relation because of some semantic discrepancies between PL and HLPL.
When concretizing a sum value, we might need to pad it with a sequence of $\eundef$
so that it has the proper size. As a result, we might get discrepancies between
a concretized PL state in which we update the variant of a sum and the concretization
of the HLPL state after doing the same update.
Also, we get a discrepancy when evaluating the move operator: a move invalidates
the value in the HLPL state but not in its concretized PL state.
The $\le$ relation states that $\Omega_0 \le \Omega_1$
if $\Omega_0$ and $\Omega_1$ are equal everywhere but at the cells where we have
$\eundef$ in $\Omega_1$.

\begin{figure}
  \centering
  \smaller
  \begin{array}[t]{llll}
  &\Omega \le \Omega' &:=& \text{States} \\
  &&\Omega.\kw{env} = \Omega'.\kw{env} \eand&\\
  &&\Omega.\kw{stack} = \Omega'.\kw{stack} \eand&\\
  &&\Omega.\kw{mem} \le \Omega'.\kw{mem}\eand&\\[1ex]
  
  &m \le m' &:=& \text{Heaps}\\
  &&\kw{domain}\; m = \kw{domain}\; m' \eand &\\
  &&\forall\; bi \in \kw{domain}\; m,\; m\; bi \le m'\; bi &\\[1ex]
  
  &[\vec{v}] \le [\vec{w}] &:= &\text{Sequences of cells}\\
  && \elength{[\overrightarrow{v_i}]} = \elength{[\overrightarrow{w_i}]} \eand&\\
  &&\forall\; i,\; v_i \le w_i &
  \end{array}
  \caption{The $\le$ Relation For PL States}
  \flabel{pl-le}
\end{figure}

We now define a relation $\le$ between PL states and HLPL states.
Contrary to the other relations in this paper and because the states of PL and HLPL are
quite difference, $\le$ is not defined as a series of transformations between states but
in a more standard manner.

\begin{definition}[Refinement Between PL and HLPL]
\deflabel{pl-hlpl-rel}
For $\Omegapl$ a PL state and $\Omegahlpl$ an HLPL state, we state
that $\Omegapl$ and $\Omegahlpl$ are in relation, noted $\Omegapl \le \Omegahlpl$,
if there exists $\eblockofname$, $\eaddrofname$ such that:
\begin{align*}
&\ecompatible{\Omegahlpl} \eand \Omegapl \le \ecstate{\Omegahlpl}
\end{align*}
\end{definition}

We now turn to the proof of the forward simulation between PL and HLPL.
We want to prove the theorem \thref{pl-hlpl-rel}.

\begin{theorem}[Forward Simulation From HLPL to PL]
For all $\Omegapl$ PL state and $\Omegahlpl$ HLPL state we have:
  \begin{align*}
  &\Omegapl \le \Omegahlpl \Rightarrow\\
  &\forall\, s\, r\, \Omegahlpl_1,\,
    \Omegahlpl \vdashHlpl\, s \stmtres{r}{\Omegahlpl_1} \Rightarrow\\
  &\exists\, \Omegapl_1,\,
    \Omegapl \vdashPl\, s \stmtres{r}{\Omegapl_1} \eand \Omegapl_1 \le \Omegahlpl_1
  \end{align*}
\thlabel{pl-hlpl-rel}
\thmvspace
\end{theorem}

We need a series of auxiliary lemmas.

The following lemma is straightforward to prove, but provides a crucial property
to make the proofs work for the expressions and assignments, which are the difficult
cases. In particular, it gives us that assignments don't move or duplicate locations,
meaning the view of locations as addresses remains consistent between the HLPL
state and the PL state.

\begin{lemma}[HLPL-Rvalue-NoLoc]
For all $\Omegahlpl$ HLPL state, $rv$, $v$ and $\Omegahlpl_1$ such that
$\Omegahlpl \vdashHlpl\, rv \exprres{v}{\Omegahlpl_1}$,
there are no $\kw{loc}$ values in $v$.
\thlabel{hlpl-rvalue-noloc}
\end{lemma}

\noindent\textbf{Proof}\\
By induction on $\Omegahlpl \vdashHlpl\, rv \exprres{v}{\Omegahlpl_1}$.
\begin{itemize}
\item Case $\ecopy p$. Trivial induction on the value we read.
  We use the fact that copying a value in HLPL returns
  the same value but where the locations have been removed (see \Rule{Copy-Loc}
  in particular).
\item Case $\emove p$. Trivial by the premises of \Rule{HLPL-E-Move}.
\item Case \Rule{E-Pointer} ($\ebrw p$ $\erbrw p$, $\embrw p$).
  Trivial by the fact that the value is a pointer.
\item Case $\enew op$. Trivial by the fact that the value is a pointer.
\item Case constant. Trivial.
\item Case Adt constructor. Trivial.
\item Case unary/binary operations (not, neg, +, -, etc.). Trivial.
\end{itemize}

\medskip

\begin{lemma}[HLPL-PL-Read]
For all $\Omegapl$ PL state and $\Omegahlpl$ HLPL state, $\eblockofname$,
$\eaddrofname$ we have:
  \begin{align*}
  &\ecompatible{\Omegahlpl}\Rightarrow\\
  &\Omegapl \le \ecstatefull{\eblockofname}{\eaddrofname}{\Omegahlpl} \Rightarrow\\
  &\forall\, p\, k\, v\, \tau,\, (\vdash \Omegahlpl(p) \eqk v : \tau) \Rightarrow\\
  &\exists\, \vec{v},\, \vdash \Omegapl(p : \tau) \eqk [\vec{v}] \eand
   [\vec{v}] \le \ecvalfull{\eblockofname}{\eaddrofname}{v}
  \end{align*}
\thlabel{hlpl-pl-read}
\end{lemma}
\noindent\textbf{Proof}\\
Straightforward by induction on $p$.

\begin{lemma}[Rvalue-Preserves-PL-HLPL-Rel]
For all $\Omegapl$ PL state and $\Omegahlpl$ HLPL state, $\eblockofname$,
$\eaddrofname$ we have:
  \begin{align*}
  &\ecompatible{\Omegahlpl}\Rightarrow\\
  &\Omegapl \le \ecstatefull{\eblockofname}{\eaddrofname}{\Omegahlpl} \Rightarrow\\
  &\forall\; rv\, v\, \Omegahlpl_1,\,
    \Omegahlpl \vdashHlpl\, rv \exprres{v}{\Omegahlpl_1} \Rightarrow\\
  &\exists\; \eblockofname_1\; \eaddrofname_1\; \vec{v},\,\\
    &\quad\Omegapl \vdashPl\, rv \exprarrow \vec{v}\eand\\
    &\quad\ecompatiblefull{\eblockofname_1}{\eaddrofname_1}{\Omegahlpl_1}\eand\\
    &\quad\Omegapl_1 \le \ecstatefull{\eblockofname_1}{\eaddrofname_1}{\Omegahlpl_1}\eand\\
    &\quad\vec{v} \le \ecvalfull{\eblockofname_1}{\eaddrofname_1}{v}
  \end{align*}
\thlabel{rvalue-pl-hlpl-rel}
\end{lemma}

\noindent\textbf{Proof}\\
By induction on $\Omegahlpl \vdashHlpl\, rv \exprres{v}{\Omegahlpl_1}$.

\begin{itemize}
\item Case $\ecopy p$. We use Lemma \thref{hlpl-pl-read} and the fact
  \Rule{E-Copy} leaves the state unchanged.
\item Case $\emove p$.
  We use Lemma \thref{hlpl-pl-read} to relate the values we evaluate to.
  Relating the updated HLPL state to the PL state is slightly more technical.
  The goal is implied by the following auxiliary lemma (the proof is straightforward by
  induction on the path $P$):
  \begin{align*}
  &\forall\, P\, v_0\, v_1\, \tau\, \overrightarrow{v_0}\, \overrightarrow{v_1}\,
   \Omegahlpl_1,\\
  &\ecompatible{\Omegahlpl}\Rightarrow\\
  &\Omegahlpl \vdash P(v_0) \eqmove v_1 : \tau \Rightarrow\\
  &\text{no location} \in v_1 \Rightarrow\\
  &\Omegapl \vdash P([\overrightarrow{v_0}]) \readarrow [\overrightarrow{v_1}] : \tau
   \Rightarrow\\
  &[\overrightarrow{v_1}] \le \ecvalfull{\eblockofname}{\eaddrofname}{v_1} \Rightarrow\\
  &\Omegahlpl \vdash P(v_0) \updtk (\bot : \tau) \wmove{v'_1}{\Omegahlpl_1}\Rightarrow\\
  &\Omegapl \le \ecstatefull{\eblockofname}{\eaddrofname}{\Omegahlpl_1} \eand
   [\overrightarrow{v_1}] \le \ecvalfull{\eblockofname}{\eaddrofname}{v'_1}
  \end{align*}
\item Case \Rule{E-Pointer} ($\ebrw p$ $\erbrw p$, $\embrw p$).
  Similar to the move case, but if we insert a fresh location, we have to update the
  $\eaddrofname$ function to insert a mapping for the fresh location identifier.
\item Case $\enew op$. We use the induction hypothesis, and need to insert a binding
  in $\eblockofname$ for the fresh box.
\item Case constant. Trivial.
\item Case Adt constructor. Straightforward by using the induction hypotheses
  and the rules to evaluate and concretize Adts.
\item Case unary/binary operations ($\neg$, +, -, etc.). Trivial.
\end{itemize}

\begin{lemma}[Assign-Preserves-PL-HLPL-Rel]
For all $\Omegapl$ PL state and $\Omegahlpl$ HLPL state we have:
  \begin{align*}
  &\Omegapl \le \Omegahlpl \Rightarrow\\
  &\forall\; p\, rv\, \Omegahlpl_1,\,
    \Omegahlpl \vdashHlpl\, \eassign p rv \stmtres{()}{\Omegahlpl_1} \Rightarrow\\
  &\exists\; \Omegapl_1,\,
    \Omegapl \vdashPl\, \eassign p rv \stmtres{()}{\Omegapl_1} \eand
    \Omegapl_1 \le \Omegahlpl_1
  \end{align*}
\thlabel{assign-pl-hlpl-rel}
\end{lemma}

\noindent\textbf{Proof}\\
We do the proof by induction on $p$. It is very similar to the move case
of Lemma \thref{rvalue-pl-hlpl-rel}. In particular, we use the fact that
there are no locations in
the value we move
(by Lemma \thref{hlpl-rvalue-noloc}), 
and in the value we overwrite and gets saved to an anonymous value (by the
premises of \Rule{HLPL-E-Assign})
\footnote{The fact that we need the overwritten value in HLPL to not contain
locations to be able to this proof
(otherwise we break the relation between the locations in the HLPL state and
the addresses in the PL state) is the reason why we ultimately need to forbid
overwriting values which contain outer loans in LLBC (see \Rule{E-Move}).}.

\medskip

\begin{lemma}[Reorg-Preserves-PL-HLPL-Rel]
For all $\Omegapl$ PL state and $\Omegahlpl$ HLPL state we have:
  \begin{align*}
  &\Omegapl \le \Omegahlpl \Rightarrow
  \forall\; \Omegahlpl_1,\,
    \Omegapl \hookrightarrow \Omegahlpl \Rightarrow
  \Omegahlpl \le \Omegahlpl_1
  \end{align*}
\thlabel{reorg-pl-hlpl-rel}
\end{lemma}

\noindent\textbf{Proof}\\
By induction on $\Omegapl \hookrightarrow \Omegahlpl$.
By $\Omegapl \le \Omegahlpl$ there exists $\eblockofname$, $\eaddrofname$ such
that:
\begin{align*}
&\ecompatible{\Omegahlpl} \eand \Omegapl \le \ecstate{\Omegahlpl}
\end{align*}

\begin{itemize}
\item Case \Rule{Reorg-None}. Trivial.
\item Case \Rule{Reorg-Seq}. Trivial by the induction hypotheses.
\item Case \Rule{Reorg-End-Pointer}.
  There exists $\Omega[.]$, $\ell$ such that:
  \begin{align*}
  &\Omegahlpl = \Omega[\eptr{\ell}]\\
  &\Omegahlpl_1 = \Omega[\bot]
  \end{align*}
  We have to show that for all $bi$ box identifier we have:
  \begin{align*}
  &\Omegapl.\kw{mem}\; bi \le \ecval{(\Omegahlpl_1.\kw{mem}\;(\eblockofinv{bi}))}
  \end{align*}
  Let's pose:
  \begin{align*}
  &[\vec{v}] := \Omegapl.\kw{mem}\; bi\\
  &v_0 = \Omegahlpl_0.\kw{mem}\;(\eblockofinv{bi})\\
  &v_1 = \Omegahlpl_1.\kw{mem}\;(\eblockofinv{bi})
  \end{align*}
  We have $[\vec{v}] \le \ecval{v_0}$ and we want to show $[\vec{v}] \le \ecval{v_1}$.
  
  There are two cases, depending on whether the hole is in $v_1$ or not.
  More formally, either we have $v_1 = v_0$, in which case the proof is trivial,
  or there exists $V[.]$ such that $v_0 = V[\eptr{\ell}]$ and
  $v_1 = V[\bot]$.
  
  The end of the proof is implied by the following auxiliary theorem, which is straightforward to
  prove by induction on $V[.]$:
  \begin{align*}
  &[\vec{v}] \le \ecval{v_0} \Rightarrow\\
  &v_0 = V[\eptr\ell] \Rightarrow v_1 = V[\bot] \Rightarrow\\
  &[\vec{v}] \le \ecval{v_1}
  \end{align*}
\item Case \Rule{Reorg-End-Loc}.
  Similar to above, but this time we have to update $\eaddrofname$
  to remove the location identifier that we eliminate.

  There exists $\Omega[.]$, $\ell$, $v_s$ such that:
  \begin{align*}
  &\Omegahlpl = \Omega[\esloan{\ell}{v_s}]\\
  &\Omegahlpl_1 = \Omega[v_s]
  \end{align*}
  
  Let's pose $\kw{addrof}_1 := \kw{addrof}_{\vert(\kw{domain}\;\kw{addrof})\setminus\{\ell\}}$
  the restriction of $\kw{addrof}$ to its domain from which we removed $\ell$.
  We show that:
  \begin{flalign}
  &\Omegapl \le \ecstatefull{\eblockofname}{\eaddrofname_1}{\Omegahlpl_1} \eand\\
  &\ecompatiblefull\eblockofname{\eaddrofname_1}{\Omegahlpl_1}
  \end{flalign}

  We show (1) the same way as in the \Rule{Reorg-End-Pointer} case.
  For (2), the difficult part is the last two conjuncts:
  \begin{flalign}
  &(\forall\; addr\;\ell\;v,\; (\readaddress{\Omegahlpl_1}{addr} \eloc{\ell}{v})
   \Rightarrow \eaddrof{\ell} = addr) \eand\\
  &(\forall\; \ell \in \Omegahlpl_1,\; \exists\;addr\;v,\;
    \readaddress{\Omegahlpl_1}{addr} \eloc{\ell}{v})
  \end{flalign}
  For (1), we pose $(bi,\, n)$ an address and do the proof by induction
  on $\Omegahlpl_1.\kw{mem}\; bi$ (straightforward).
  For (2), the compatibility assumption for $\Omegahlpl$ gives us a candidate
  address $(bi,\,n)$, and we conclude the proof also by induction on
  $\Omegahlpl_1.\kw{mem}\; bi$.
  
\end{itemize}

\medskip

We can finally turn to the proof of the target theorem (\thref{pl-hlpl-rel}).

\begin{itemize}
\item Case empty statement. Trivial.
\item Case \Rule{E-Reorg}. By Lemma \thref{reorg-pl-hlpl-rel}.
\item Case $s_0;\; s_1$. Trivial by the induction hypotheses.
\item Case $\eassign p rv$. By Lemma \thref{assign-pl-hlpl-rel}.
\item Case \li+if then else+. Trivial by the induction hypotheses.
\item Case \li+match+. Trivial by the induction hypotheses.
\item Case $\efree p$. We remove one block from the map and end the corresponding
  pointer: straightforward.
\item Case $\ereturn$. Trivial.
\item Cases $\epanic$, $\ebreak i$, $\econtinue i$. Trivial.
\item Case loop. Trivial by the induction hypotheses.
\item Case function call.
  Similar to the same case in the proof of \Thm{Eval-Preserves-HLPL+-Rel}.
  We leverage the fact that \Rule{PL-E-Call} and \Rule{E-Call} have a very
  similar structure.
  We use the Lemma \thref{rvalue-pl-hlpl-rel} to relate the states resulting
  from evaluating the operands given as inputs to the function (trivial induction
  on the number of inputs).
  We easily get that the states resulting from $\pushstack$ are
  related (again, induction on the number of inputs then on the number of
  local variables which are not used as inputs; we have to take care to
  extend the map $\kw{blockof}$ to account for the blocks freshly allocated
  for the local variables).
  The induction hypothesis gives us that the states are related after evaluation
  of $body$. If the tag is $\epanic$ we are done.
  If it is $\ereturn$, we prove that the states resulting from $\popstack$
  are related (induction on the number of local variables, and we use
  \thref{assign-pl-hlpl-rel} for the assignments which ``drop'' the local variables;
  this time we have to remove the blocks which were allocated for the local variables
  from the map $\kw{blockof}$).
  Finally, we do a reasoning similar to the one we did in
  \thref{assign-pl-hlpl-rel} to show that the states are still related after
  assigning to the destination.
\end{itemize}
\section{Forward Simulation Between LLBC and \llbcs}
\appendixlabel{llbc+}

We show the additional rules for \llbcs in \Figure{llbcs-full}.
We show the full rules for the $\le$ relation about \llbcp states in
\Figure{llbcp-rel-prim}.
\Figure{to-region-rules} lists the rules we use to transform values into
fresh region abstractions (this judgment is used by \Rule{Le-ToAbs}),
while \Figure{merge-rules} describes how to merge two region abstractions
into one (used by \Rule{Le-MergeAbs}).
Some of the rules we show in \Figure{llbcp-rel-selected},
like \Rule{Le-Reborrow-MutBorrow-Abs}, are actually not part of the definition
of $\le$, but derived from more primitive rules. We list those derived rules in
\Figure{le-rules-derived}.

\begin{figure}
  \smaller
  \arraycolsep=1pt %
  \centering

  \begin{array}[t]{llll}
    v & ::= & & \text{value} \\
    && \ktrue \mid \kfalse \mid n_{\mathsf{i32}} \mid n_{\mathsf{u32}} \mid \ldots &
      \text{literal constants} \\
    && \eleft{v} \mid \eright{v} & \text{sum value} \\
    && (v_0,\, v_1) & \text{pairs}\\
    && \bot & \text{bottom (invalid) value}\\
    && \emloan{\ell} & \text{mutable loan}\\
    && \emborrow{\ell}{v} & \text{mutable borrow}\\
    && \esloan{\ell}{v} & \text{shared loan}\\
    && \esborrow{\ell} & \text{shared borrow}\\
    && \erborrow{\ell} & \text{reserved borrow}\\
    && \sigma & \text{symbolic value}\\
    && \_ & \text{ignored value (only used in region abstractions)}
    \\[1ex]

    id & ::= & & \text{environment binding identifier}\\
    && x & \text{variable identifier}\\
    && \_x & \text{anonymous (ghost) variable identifier}\\
    \\[1ex]

    A & ::= & \Abs{A_{id}}{[v]} & \text{region abstraction}\\[1ex]

    \Omega^{\text{LLBC}} & ::= & \{\;
      \kw{env} : id \underset {\text{partial}} \longrightarrow v,\;
      \kw{abs} : A_{id} \underset {\text{partial}} \longrightarrow A,\;
      \kw{stack} : [[x]]\; \} \; & \text{state}\\[1ex]
  \end{array}
  \flabel{llbcs-grammar}
  \caption{Grammar of \llbcs States and Values}
\end{figure}

\begin{figure}
  \centering
  \smaller
  \begin{mathpar}
    \LlbcsRules*
  \end{mathpar}
  \caption{Additional Rules for \llbcs}
  \flabel{llbcs-full}
\end{figure}

\begin{figure}
  \centering
  \smaller
  \begin{mathpar}
    \LlbcLeRules*
  \end{mathpar}
  \caption{The Relation $\le$ about \llbcp States}
  \flabel{llbcp-rel-prim}
\end{figure}

\begin{figure}
  \centering
  \smaller
  \begin{mathpar}
    \ToAbsRules*
  \end{mathpar}
  \caption{Rules to Transform Values to Region Abstractions.}
  \flabel{to-region-rules}
\end{figure}

\begin{figure}
  \centering
  \smaller
  \begin{mathpar}
    \MergeAbsRules*
  \end{mathpar}
  \caption{Rules to Merge Region Abstractions.}
  \flabel{merge-rules}
\end{figure}

\begin{figure}
  \centering
  \smaller
  \begin{mathpar}
    \LlbcLeDerivedRules*
  \end{mathpar}
  \caption{Derived Rules for the $\le$ Relation About \llbcp States}
  \flabel{le-rules-derived}
\end{figure}

\subsection{Forward Simulation for \llbcp}
We now turn to the proof that evaluation for \llbcp preserves the relation
$\le$ over \llbcp states. We need several auxiliary lemmas.

\begin{lemma}[Rvalue-Preserves-LLBC+-Rel]
\thlabel{llbc+-rvalue-rel}
For all $\Omega_l$ and $\Omega_r$ \llbcp states and $rv$ right-value we have:
\begin{align*}
&\Omega_l \le \Omega_r \Rightarrow
\forall\; v_r\, \Omega'_r, \ \Omega_r \vdashLlbcp rv \exprres{v_r}{\Omega'_r} \Rightarrow\\
&\exists\; v_l\, \Omega'_l,\ \Omega_l \vdashLlbcp rv \exprres{v_l}{\Omega'_l} \wedge
(v_l,\, \Omega'_l) \le (v_r,\, \Omega'_r)
\end{align*}

where we define $(v_l,\, \Omega'_l) \le (v_r,\, \Omega'_r)$ as:
\begin{align*}
(v_l,\, \Omega'_l) \le (v_r,\, \Omega'_r) := (\Omega'_l,\, \_ \rightarrow v_l) \le (\Omega'_r,\, \_ \rightarrow v_r)
\end{align*}
\end{lemma}

\noindent\textbf{Proof}\\
By induction on $\Omega'_r, \ \Omega_r \vdashLlbcp rv \exprres{v_r}{\Omega'_r}$.

\begin{itemize}
\item Case $\ecopy p$. By induction on $\Omega_l \le \Omega_r$.
\begin{itemize}
\item Case \Rule{Le-ToSymbolic}. The interesting case happens if in the right state we
  copy the symbolic value we just introduced. If this happens, we use the fact that
  copying a symbolic value introduces a fresh symbolic value (\Rule{Copy-Symbolic})
  and apply \Rule{Le-ToSymbolic} twice to relate the states (once for the original
  value, once for the copied value).
\item Case \Rule{Le-MoveValue}. We note that moving a value can only
  constrain the places we can directly access in the right state.
  It also doesn't have any effect on the values we indirectly access through
  shared borrows (thanks to the premise that the hole is not inside a shared
  loan).
  Following this intuition, we prove by induction on $p$ that $\ecopy p$ reduces
  to the same value in the left state and the right state, and leaves the states
  unchanged; we conclude by
  using \Rule{Le-MoveValue}.
\item Case \Rule{Le-Fresh-MutLoan}. Similar to \Rule{Le-MoveValue}.
  We prove by induction on $p$ that $\ecopy p$ reduces to the same value in the left state
  and the right state, and leaves the states unchanged; we conclude by \Rule{Le-Fresh-MutLoan}.
\item Case \Rule{Le-Fresh-SharedLoan}. Similar to above. Here, the copy actually succeeds
  in the right state if and only if it succeeds in the left state, and it yields
  the same value.
\item Case \Rule{Le-Reborrow-MutBorrow}. Similar to case \Rule{Le-Fresh-SharedLoan}.
\item Case \Rule{Le-Fresh-SharedBorrow}. Similar to case \Rule{Le-Fresh-SharedLoan}.
\item Case \Rule{Le-Reborrow-SharedLoan}. We might use a shared borrow to copy
  the fresh symbolic value in the right state and the original value in the
  left state. Similarly to the case \Rule{Le-ToSymbolic}, we use the fact that
  \Rule{Copy-Symbolic} introduces a fresh symbolic value and conclude by using
  \Rule{Le-ToSymbolic} twice if it happens.
\item Case \Rule{Le-Abs-End-SharedLoan}. The shared loan in the region abstraction
  is not accessible from the outside because there are no remanining borrows pointing
  to this value. The copied value is the same in both states, and we conclude
  by \Rule{Le-Abs-End-SharedLoan}.
\item Case \Rule{Le-Abs-End-DupSharedBorrow}. Ending a shared borrow in a region
  abstraction doesn't have any effect on the $\vdash \ecopy p$ judgement.
  We conclude by \Rule{Le-Abs-End-DupSharedBorrow}.
\item Case \Rule{Le-Reborrow-SharedBorrow}. Similar to the case \Rule{Le-Reborrow-SharedLoan}.
\item Case \Rule{Le-ToAbs}. We note that the shared loans, which are accessible from
  the non-anonymous values (and so through the copy) are preserved by the $\toabs$
  rules (in particular, \Rule{ToAbs-SharedLoan}).
  We prove by induction on the value that:
  $\forall\; \esloan{\ell}{v} \in \Omega_l, \esloan{\ell}{v} \in \Omega_r$.
  This allows us to prove that the copy operation reduces to the same values
  in the left and right states, and we conclude by \Rule{Le-ToAbs}.
\item Case \Rule{Le-MergeAbs}. Similar to \Rule{Le-Abs}: the shared loans
  are preserved (by induction on the $\MergeAbs$ derivation).
\item Cases \Rule{Le-ClearAbs}, \Rule{Le-Abs-ClearValue},
      \Rule{Le-Abs-DeconstructPair}, \Rule{Le-Abs-DeconstructSum}.
      Similar to \Rule{Le-ToAbs}: the shared loans are preserved.
\item Case \Rule{Le-AnonValue}. Similar to above.
\end{itemize}

\item Case $\emove p$. By induction on $\Omega_l \le \Omega_r$.
\begin{itemize}
\item Case \Rule{Le-ToSymbolic}. Case disjunction on wether we move the symbolic
  value or not. We conclude by \Rule{Le-ToSymbolic}.
 \item Case \Rule{Le-MoveValue}. By induction on $p$, we get that $\emove$
   reduces to the same value in both states, and yields two states related by \Rule{Le-MoveValue}.
 \item Case \Rule{Le-Fresh-MutLoan}. Similar to \Rule{Le-MoveValue}.
 \item Case \Rule{Le-Fresh-SharedLoan}. Similar to \Rule{Le-MoveValue}.
 \item Case \Rule{Le-Reborrow-MutBorrow}. Similar to \Rule{Le-MoveValue}.
 \item Case \Rule{Le-Fresh-SharedBorrow}. Similar to \Rule{Le-MoveValue}.
 \item Case \Rule{Le-Reborrow-SharedLoan}. Similar to \Rule{Le-MoveValue}.
   We use the fact that we can't move a loaned value (\Rule{R-SharedLoan}, \Rule{W-SharedLoan})
   to show that $\emove p$ reduces to the same value in both states.
\item Case \Rule{Le-Abs-End-SharedLoan}. Similar to the case \Rule{Le-Reborrow-SharedLoan}.
\item Case \Rule{Le-Abs-End-DupSharedBorrow}. Similar to \Rule{Le-MoveValue} (and simpler).
\item Case \Rule{Le-Reborrow-SharedBorrow}.
  Similar to \Rule{Le-Reborrow-SharedLoan}, but here we use the fact that we can't
  dereference shared borrows when moving values.
\item Case \Rule{Le-ToAbs}.
  We note that we can not dereference shared borrows when moving values; this
  forbids us from jumping to anonymous values or region abstractions.
\item Case \Rule{Le-MergeAbs}.
  Same as \Rule{Le-ToAbs}.
\item Cases \Rule{Le-ClearAbs}, \Rule{Le-Abs-ClearValue},
      \Rule{Le-Abs-DeconstructPair}, \Rule{Le-Abs-DeconstructSum}.
      Same as \Rule{Le-ToAbs}.
\item Case \Rule{Le-AnonValue}. Similar to above.
\end{itemize}

\item Cases $\ebrw p$, $\erbrw$. By induction on $\Omega_l \le \Omega_r$.
\begin{itemize}
\item Case \Rule{Le-ToSymbolic}. Case disjunction on wether we borrow the symbolic
  value or not (we might insert a shared borrow). We conclude by \Rule{Le-ToSymbolic}.
 \item Case \Rule{Le-MoveValue}.
   By induction on $p$, we get that $\emove$
   reduces to the same value in both states,
   and yields two states related by \Rule{Le-MoveValue} (we might insert a shared
   loan in the moved value).
 \item Case \Rule{Le-Fresh-MutLoan}. Similar to \Rule{Le-MoveValue}.
 \item Case \Rule{Le-Fresh-SharedLoan}. Similar to \Rule{Le-MoveValue}.
 \item Case \Rule{Le-Reborrow-MutBorrow}. Similar to \Rule{Le-MoveValue}.
 \item Case \Rule{Le-Fresh-SharedBorrow}. Similar to \Rule{Le-MoveValue}.
 \item Case \Rule{Le-Reborrow-SharedLoan}.
   Similar to \Rule{Le-MoveValue} and
   \Rule{Le-ToSymbolic} (we might borrow the fresh symbolic value or one of the
   shared values which were moved to the region abstraction; in the first
   case we have to convert one more shared borrow).
\item Case \Rule{Le-Abs-End-SharedLoan}. The shared loan inside the region abstraction
  is not accessible as there are no remaining borrows to this value.
\item Case \Rule{Le-Abs-End-DupSharedBorrow}. This has no impact on the evaluation.
\item Case \Rule{Le-Reborrow-SharedBorrow}.
  Similar to \Rule{Le-Reborrow-SharedLoan}.
\item Case \Rule{Le-ToAbs}.
  Similar to the $\ecopy p$ case: we use the fact that all the shared loans are preserved
  in the region abstraction.
\item Case \Rule{Le-MergeAbs}.
  Same as \Rule{Le-ToAbs}.
\item Cases \Rule{Le-ClearAbs}, \Rule{Le-Abs-ClearValue},
      \Rule{Le-Abs-DeconstructPair}, \Rule{Le-Abs-DeconstructSum}.
      Same as \Rule{Le-ToAbs}.
\item Case \Rule{Le-AnonValue}. Similar to above.
\end{itemize}

\item Case $\embrw p$. By induction on $\Omega_l \le \Omega_r$.
  This is similar to the $\emove p$ case, in particular because we can't mutably
  borrow shared values.

\begin{itemize}
\item Case \Rule{Le-ToSymbolic}. Case disjunction on wether we mutably borrow the symbolic
  value or not. We conclude by \Rule{Le-ToSymbolic}.
 \item Case \Rule{Le-MoveValue}. By induction on $p$, we get that $\ebrw p$
   reduces to the same value in both states, and yields two states related by \Rule{Le-MoveValue}.
 \item Case \Rule{Le-Fresh-MutLoan}. Similar to \Rule{Le-MoveValue}.
 \item Case \Rule{Le-Fresh-SharedLoan}. Similar to \Rule{Le-MoveValue}.
 \item Case \Rule{Le-Reborrow-MutBorrow}. Similar to \Rule{Le-MoveValue}.
 \item Case \Rule{Le-Fresh-SharedBorrow}. Similar to the move case.
 \item Case \Rule{Le-Reborrow-SharedLoan}. Similar to the move case.
 \item Case \Rule{Le-Abs-End-SharedLoan}. Similar to the move case.
 \item Case \Rule{Le-Abs-End-DupSharedBorrow}. Similar to the move case.
 \item Case \Rule{Le-Reborrow-SharedBorrow}. Similar to the move case.
 \item Case \Rule{Le-ToAbs}. Similar to the move case.
 \item Case \Rule{Le-MergeAbs}. Similar to the move case.
 \item Cases \Rule{Le-ClearAbs}, \Rule{Le-Abs-ClearValue},
      \Rule{Le-Abs-DeconstructPair}, \Rule{Le-Abs-DeconstructSum}.
      Similar to the move case.
 \item Case \Rule{Le-AnonValue}. Similar to above.
\end{itemize}
\item Case $\enew op$. We use the induction hypothesis.
\item Case constant. Trivial: the states have no impact on the reduction
  of the constants, and are left unchanged.
\item Case Adt constructor. We use the induction hypothesis (the proof is
  similar to the $\enew op$) case.
\item Case unary/binary operations ($\neg$, neg, +, -, etc.). Trivial by
  the induction hypotheses.
\end{itemize}

We now prove the following lemma about assignments.

\begin{lemma}[Assign-Preserves-LLBC+-Rel]
\thlabel{llbc+-assign-rel}
For all $\Omega_l$ and $\Omega_r$ \llbcp states, $rv$ right-value and $p$ place we have:
\begin{align*}
&\Omega_l \le \Omega_r \Rightarrow
\forall\, \Omega'_r, \ \Omega_r \vdashLlbcp \eassign p rv \stmtres{()}{\Omega'_r} \Rightarrow\\
&\exists\, \Omega'_l\,\ \Omega_l \vdashLlbcp \eassign p rv \stmtres{()}{\Omega'_l} \wedge
\Omega'_l \le \Omega'_r
\end{align*}
\end{lemma}

\noindent\textbf{Proof}\\
Lemma \thref{llbc+-rvalue-rel} gives us that there exists $v_r$, $\Omega''_r$,
$v_l$, $\Omega''_l$ such that:
\begin{align*}
&\Omega_r \vdashLlbcp rv \exprres{v_r}{\Omega''_r} \eand\\
&\Omega_l \vdashLlbcp rv \exprres{v_l}{\Omega''_l} \eand\\
&(v_l,\, \Omega''_l) \le (v_r,\, \Omega''_r)
\end{align*}

We do the proof by induction on $(v_l,\, \Omega''_l) \le (v_r,\, \Omega''_r)$, then
on the path $p$. The reasoning is very similar to what we saw in the proof of lemma
\thref{llbc+-rvalue-rel}. Something important to note is that the value that gets
overwritten is always saved in a fresh (ghost) anonymous value; this allows us
to conclude in most situations.

\begin{itemize}
\item Case \Rule{Le-ToSymbolic}. We have to consider several possibilities:
  1. the symbolic value is in the value we move; 2. the symbolic value gets moved to an anonymous value because
  of the assignment. We conclude by \Rule{Le-ToSymbolic}.
\item Case \Rule{Le-MoveValue}. Similar to above.
\item Case \Rule{Le-Fresh-MutLoan}. Similar to above.
\item Case \Rule{Le-Fresh-SharedLoan}. Similar to above.
\item Case \Rule{Le-Reborrow-MutBorrow}. Similar to above.
\item Case \Rule{Le-Fresh-SharedBorrow}. Similar to above.
\item Case \Rule{Le-Reborrow-SharedLoan}. Similar to above.
\item Case \Rule{Le-Abs-End-SharedLoan}. Modifications in region abstractions
  have no impact on writes with the move capability.
\item Case \Rule{Le-Abs-End-DupSharedBorrow}. Similar to above.
\item Case \Rule{Le-Reborrow-SharedBorrow}. Similar to above.
\item Case \Rule{Le-ToAbs}. We can't use a move capability to write to an anonymous value,
  so we can conclude by \Rule{Le-ToAbs}.
\item Case \Rule{Le-MergeAbs}. Similar to above.
\item Cases \Rule{Le-ClearAbs}, \Rule{Le-Abs-ClearValue},
      \Rule{Le-Abs-DeconstructPair}, \Rule{Le-Abs-DeconstructSum}. Similar to above.
\item Case \Rule{Le-AnonValue}. Similar to above.
\end{itemize}

We now prove that reorganizations preserve the relation between \hlplp states.
\begin{lemma}[Reorg-Preserves-LLBC+-Rel]
\thlabel{llbc+-reorg-rel}
For all $\Omega_l$ and $\Omega_r$ \hlplp states we have:
\begin{align*}
&\Omega_l \le \Omega_r \Rightarrow
\forall\; \Omega'_r, \ \Omega_r \hookrightarrow \Omega'_r \Rightarrow\\
&\exists\; \Omega'_l,\ \Omega_l \hookrightarrow \Omega'_l \wedge
\Omega'_l \le \Omega'_r
\end{align*}
\end{lemma}
\noindent\textbf{Proof}\\
By induction on $\Omega_r \hookrightarrow \Omega'_r$.

\begin{itemize}
\item Case \Rule{Reorg-None}. Trivial.
\item Case \Rule{Reorg-Seq}. Trivial by the induction hypotheses.
\item Case \Rule{Reorg-End-MutBorrow}. By induction on $\Omega_l \le \Omega_r$.
  \begin{itemize}
  \item Case \Rule{Le-ToSymbolic}. Case disjunction on whether the symbolic
    value is inside the mutably borrowed value or not. We conclude by
    \Rule{Le-ToSymbolic}.
  \item Case \Rule{Le-MoveValue}. Similar. An important point is that the moved
    value can't come from a region abstraction (this is important because
    we can't end borrows which are inside region abstractions: we have to
    end the abstraction first).
  \item Case \Rule{Le-Fresh-MutLoan}. Similar.
  \item Case \Rule{Le-Fresh-SharedLoan}. Similar.
  \item Case \Rule{Le-Reborrow-MutBorrow}.
    We have to make a case disjunction on whether we end the fresh borrow $\ell_1$
    or not. If no, we conclude by \Rule{Le-Reborrow-MutBorrow}.
    If yes, the premises of \Rule{Reorg-End-MutBorrow} give us that
    the hole is not inside a shared loan, and there are no loans in v.
    This allows us to conclude by using \Rule{Le-MoveValue}.
  \item Case \Rule{Le-Fresh-SharedBorrow}. We conclude by \Rule{Le-Fresh-SharedBorrow}.
  \item Case \Rule{Le-Reborrow-SharedLoan}. We conclude by \Rule{Le-Reborrow-SharedLoan}.
  \item Case \Rule{Le-Abs-End-SharedLoan}. We conclude by \Rule{Le-Abs-End-SharedLoan}.
  \item Case \Rule{Le-Abs-End-DupSharedBorrow}. We conclude by \Rule{Le-Abs-End-DupSharedBorrow}.
  \item Case \Rule{Le-Reborrow-SharedBorrow}. We conclude by \Rule{Le-Abs-End-DupSharedBorrow}.
  \item Case \Rule{Le-ToAbs}. We can not end borrows in region abstractions with
    \Rule{Reorg-End-MutBorrow} (we have to use \Rule{Reorg-End-Abstraction}).
    However, we can end a mutable loan inside a region abstraction.
    If the mutable loan is not inside the region abstraction we just introduced,
    we conclude by \Rule{Le-ToAbs}.
    Otherwise, we prove that ending the same mutable loan on the left (it must
    be in the anonymous variable we convert to an abstraction) then
    converting the anonymous variables to a region abstraction by \Rule{Le-ToAbs}
    yields the same state as on the right (the proof is by induction on
    the value we convert).
  \item Case \Rule{Le-MergeAbs}. Similar to \Rule{Le-ToAbs}. The difficult
    case is \Rule{MergeAbs-Mut}; the important point to notice is that it makes
    borrows and loans disappear in the \emph{right} state, meaning there are
    more borrows and loans in the left state (this is important, because we
    don't have a well-formedness assumption). In particular, if the mutable
    loan we end is in the merged abstraction, then it is also present in
    one of the two initial abstractions in the left state, and the corresponding
    borrow is not inside a region abstraction.
  \item Cases \Rule{Le-ClearAbs}, \Rule{Le-Abs-ClearValue},
        \Rule{Le-Abs-DeconstructPair}, \Rule{Le-Abs-DeconstructSum}. Trivial.
  \item Case \Rule{Le-AnonValue}. Trivial.
  \end{itemize}

\item Case \Rule{Reorg-End-SharedReservedBorrow}. By induction on $\Omega_l \le \Omega_r$.
  \begin{itemize}
  \item Case \Rule{Le-ToSymbolic}. Trivial.
  \item Case \Rule{Le-MoveValue}. Case disjunction on wether the borrow
    is moved or not; we conclude by \Rule{Le-MoveValue}.
    Similarly to the \Rule{Reorg-End-MutBorrow} case, an important point is that the moved
    value can't come from a region abstraction (this is important because
    we can't end borrows which are inside region abstractions: we have to
    end the abstraction first).
  \item Case \Rule{Le-Fresh-MutLoan}. The ended borrows is necessarily
    independent from the fresh mutable loan; we conclude by \Rule{Le-Fresh-MutLoan}.
  \item Case \Rule{Le-Fresh-SharedLoan}. We conclude by \Rule{Le-Fresh-SharedLoan}.
  \item Case \Rule{Le-Reborrow-MutBorrow}. Similar to above.
  \item Case \Rule{Le-Fresh-SharedBorrow}. On the right we might terminate the fresh
    shared borrow in which case we conclude by \Rule{Le-AnonValue}; otherwise we
    conclude with \Rule{Le-Fresh-SharedBorrow}.
  \item Case \Rule{Le-Reborrow-SharedLoan}. We can not end any of the borrows
    inside the fresh region abstraction, meaning we can not end the fresh borrow
    nor the borrows inside the shared value. We conclude by \Rule{Le-Reborrow-SharedLoan}.
  \item Case \Rule{Le-Abs-End-SharedLoan}. We conclude by \Rule{Le-Abs-End-SharedLoan}.
  \item Case \Rule{Le-Abs-End-DupSharedBorrow}. We can not directly end
    a borrow inside a region abstraction; we conclude by \Rule{Le-Abs-End-DupSharedBorrow}.
  \item Case \Rule{Le-Reborrow-SharedBorrow}.
    This one is slightly technical.
    There exists $\onehole\Omega .$, $\ell_0$, $\ell_1$, $v$, $\sigma$, $A_0$ such that:
    \begin{align*}
    &\bot, \emloanv, \eborrowv \notin v\\
    &\esloan{\ell_0}{v} \in \Omega_l\\
    &\Omega_l = \onehole{\Omega}{\esborrow{\ell_0}}\\
    &\Omega_r = \onehole{\Omega}{\esborrow{\ell_1}},\, \AbsI{0}{\esborrow{\ell_0},\, \esloan{\ell_1}{\sigma}}
    \end{align*}
    The difficult case happens when we end the borrow $\ell_1$, so this is the
    case we focus on (in the other case, we conclude by \Rule{Le-Reborrow-SharedBorrow}).
    We do no reorganization on the left. We have to prove:
    \begin{align*}
    &\onehole{\Omega}{\esborrow{\ell_0}}\le
     \onehole{\Omega}{\bot},\, \AbsI{0}{\esborrow{\ell_0},\, \esloan{\ell_1}{\sigma}}
    \end{align*}
    Because we could end $\ell_1$ on the right, necessarily the hole on the left is
    not inside a shared loan. This means we can move it to an anonymous value
    by using \Rule{Le-MoveValue}.
    We get:
    \begin{align*}
    &\onehole{\Omega}{\ell_0}\le\onehole{\Omega}{\bot},\, \_ \rightarrow \esborrow{\ell_0}
    \end{align*}
    We now apply \Rule{Le-Reborrow-SharedBorrow} to $\esborrow{\ell_0}$.
    We get:
    \begin{align*}
    &\onehole{\Omega}{\ell_0}\le\onehole{\Omega}{\bot},\, \_ \rightarrow
      \esborrow{\ell_1},\,
      \AbsI{1}{\esborrow{\ell_0},\, \esloan{\ell_1}{\sigma}}
    \end{align*}
    We apply \Rule{Le-ToAbs} to $\esborrow{\ell_0}$, then merge the two region
    abstractions with \Rule{Le-MergeAbs}, taking care of using \Rule{MergeAbs-Shared}
    to get rid of $\esborrow{\ell_1}$. QED.
    
  \item Case \Rule{Le-ToAbs}.
    We can not end a borrow which was moved to a region abstraction; the borrow
    we end on the right state was thus not in the anonymous value we converted to
    an abstraction; we can end it in the left state and conclude by \Rule{Le-ToAbs}.
  \item Case \Rule{Le-MergeAbs}. Similar to \Rule{Le-ToAbs}.
  \item Cases \Rule{Le-ClearAbs}, \Rule{Le-Abs-ClearValue},
        \Rule{Le-Abs-DeconstructPair}, \Rule{Le-Abs-DeconstructSum}. Trivial.
  \item Case \Rule{Le-AnonValue}. Trivial.
  \end{itemize}

\item Case \Rule{Reorg-End-SharedLoan}. By induction on $\Omega_l \le \Omega_r$.
  \begin{itemize}
  \item Case \Rule{Le-ToSymbolic}. Trivial.
  \item Case \Rule{Le-MoveValue}. Trivial.
  \item Case \Rule{Le-Fresh-MutLoan}. Trivial.
  \item Case \Rule{Le-Fresh-SharedLoan}. We might end the fresh shared loan,
    in which case the right state become equal to the left state; we conclude
    by reflexivity of $\le$.
  \item Case \Rule{Le-Reborrow-MutBorrow}. Trivial.
  \item Case \Rule{Le-Fresh-SharedBorrow}. Trivial (we can not end the shared loan
    from which we created a new shared borrow).
  \item Case \Rule{Le-Reborrow-SharedLoan}.
    Because there is $\esborrow{\ell_1}$ in the region abstraction, we can not
    end $\esloan{\ell_1}{\sigma}$ on the right. We can end the shared loan $\ell_0$,
    if there are no shared borrows for $\ell_0$. If it is the case, we do no
    reorganization on the left, then use \Rule{Le-Reborrow-SharedLoan} in combination
    with \Rule{Le-Abs-End-SharedLoan}. In the other situations, we can conclude
    by \Rule{Le-Reborrow-SharedLoan}.
  \item Case \Rule{Le-Abs-End-SharedLoan}. Trivial.
  \item Case \Rule{Le-Abs-End-DupSharedBorrow}. Trivial.
  \item Case \Rule{Le-Reborrow-SharedBorrow}. We can not end any of the shared loans
    for $\ell_0$ or $\ell_1$ on the right; we easily conclude by \Rule{Le-Reborrow-SharedBorrow}.
  \item Case \Rule{Le-ToAbs}. If we end one of the shared loans which were moved
    to the fresh region abstraction, we use \Rule{Le-ToAbs} then
    \Rule{Le-Abs-End-SharedLoan}.
    Otherwise we end the corresponding shared loan on the left then conclude
    by \Rule{Le-ToAbs}.
  \item Case \Rule{Le-MergeAbs}. If the shared loan we end doesn't appear inside
    the merged region abstraction, we end the corresponding loan on the left and conclude
    by \Rule{Le-MergeAbs}.
    Otherwise, we simply use \Rule{Le-MergeAbs} then \Rule{Le-Abs-EndSharedLoan}.
  \item Cases \Rule{Le-ClearAbs}, \Rule{Le-Abs-ClearValue},
        \Rule{Le-Abs-DeconstructPair}, \Rule{Le-Abs-DeconstructSum}. Trivial.
  \item Case \Rule{Le-AnonValue}. Trivial.
  \end{itemize}
\item Case \Rule{Reorg-Activate-Reserved}. By induction on $\Omega_l \le \Omega_r$.
  \begin{itemize}
  \item Case \Rule{Le-ToSymbolic}. Trivial.
  \item Case \Rule{Le-MoveValue}. Trivial.
  \item Case \Rule{Le-Fresh-MutLoan}. Trivial.
  \item Case \Rule{Le-Fresh-SharedLoan}. Trivial.
  \item Case \Rule{Le-Reborrow-MutBorrow}. Trivial.
  \item Case \Rule{Le-Fresh-SharedBorrow}. The borrow we activate can't have
    the same identifier as the fresh shared borrow; we can thus activate the
    same borrow on the left and conclude by \Rule{Le-Fresh-SharedBorrow}.
  \item Case \Rule{Le-Reborrow-SharedLoan}. The borrow we activate can not be $\ell_1$ because
    there is a corresponding shared borrow, nor $\ell_0$, because the shared
    loan is in a region abstraction. We can thus activate the same borrow on
    the left and conclude by \Rule{Le-Reborrow-SharedLoan}.
  \item Case \Rule{Le-Abs-End-SharedLoan}. Trivial.
  \item Case \Rule{Le-Abs-End-DupSharedBorrow}. Trivial.
  \item Case \Rule{Le-Reborrow-SharedBorrow}. Similar to \Rule{Le-Reborrow-SharedLoan}.
  \item Case \Rule{Le-ToAbs}. We can not move reserved borrows to region abstractions,
    and can not activate borrows associated to loans in region abstractions.
    We can thus activate the same borrow on the left and conclude by \Rule{Le-ToAbs}.
  \item Case \Rule{Le-MergeAbs}. Similar to \Rule{Le-ToAbs}.
  \item Cases \Rule{Le-ClearAbs}, \Rule{Le-Abs-ClearValue},
        \Rule{Le-Abs-DeconstructPair}, \Rule{Le-Abs-DeconstructSum}. Trivial.
  \item Case \Rule{Le-AnonValue}. Trivial.
  \end{itemize}

\item Case \Rule{Reorg-End-Abstraction}. By induction on $\Omega_l \le \Omega_r$.
  \begin{itemize}
  \item Case \Rule{Le-ToSymbolic}. Trivial.
  \item Case \Rule{Le-MoveValue}. Trivial (the moved value can't be inside a region
    abstraction; we can thus end the
    region abstraction on the left and conclude by \Rule{Le-MoveValue}).
  \item Case \Rule{Le-Fresh-MutLoan}. Trivial (we can't end a region abstraction
    in which there is a loan).
  \item Case \Rule{Le-Fresh-SharedLoan}. Trivial (similar to the case
    \Rule{Le-Fresh-MutLoan}).
  \item Case \Rule{Le-Reborrow-MutBorrow}. The mutable borrow might be inside
    the region abstraction we end, but we can still conclude \Rule{Le-Reborrow-MutBorrow}.
    We also conclude by \Rule{Le-Reborrow-MutBorrow} in the other cases.
  \item Case \Rule{Le-Fresh-SharedBorrow}. Trivial (note that we can't end a region
    abstraction in which there is a loan).
  \item Case \Rule{Le-Reborrow-SharedLoan}. Trivial (note that we can't end a region
    abstraction in which there is a loan).
  \item Case \Rule{Le-Abs-End-SharedLoan}. We might end the region abstraction
    in which we just ended the loan. In this case, we also end the loan and
    the region abstraction on the left, and conclude by reflexivity of $\le$.
    Otherwise, we end the corresponding region on the left, and conclude
    by \Rule{Le-Abs-End-SharedLoan}.
  \item Case \Rule{Le-Abs-End-DupSharedBorrow}.
    If we end the region abstraction from which we removed the duplicated borrow,
    we can end the same region abstraction on the left, and end one of the
    borrows which were reintroduced into the context.
    We then need to remove the anonymous variable containing $\bot$ that is
    left in place of the borrow (there is no primitive rule for $\le$ to do
    that, but we can actually achieve the same result by using \Rule{Le-ToAbs}
    on this value).
  \item Case \Rule{Le-Reborrow-SharedBorrow}. Trivial.
  \item Case \Rule{Le-ToAbs}. If the abstraction we end is not the one we just
    introduced, we can end the corresponding abstraction on the left an conclude
    by \Rule{Le-ToAbs}. If it is the same region abstraction, then this abstraction
    doesn't contain any loans. This means that the anonymous value we converted
    to a region abstraction only contains borrows (we get this by a simple
    induction on the value). Also, it doesn't contain nested borrows
    (because it doesn't contain shared loans, and because of the premises of
    \Rule{ToAbs-MutBorrow}).
    Finally, the mutably borrowed values don't contain $\bot$ (again, because
    of the premises of \Rule{ToAbs-MutBorrow}).
    We can thus apply \Rule{Le-ToSymbolic} on all the mutably borrowed value
    of the anonymous value
    (note that when ending a region abstraction, we reintroduce the mutable
    borrows with fresh symbolic values in the context; see \Rule{Reorg-End-Abstraction}).
    We can then move all the borrows from the anonymous to their own anonymous
    values (by \Rule{Le-MoveValue}; we showed before that those borrows
    don't appear in shared loans) and finally get rid of the original anonymous
    value by using the same technique as in the case \Rule{Le-Abs-End-DupSharedBorrow}.
  \item Case \Rule{Le-MergeAbs}. If the abstraction we end is not the merged abstraction,
    we end the corresponding abstraction on the left and conclude by \Rule{Le-MergeAbs}.
    If it is the same abstraction, we can end the first merged abstraction on the left,
    the borrows it reintroduced in the context and whose corresponding loans
    are in the second abstraction, then the second abstraction.
    In particular, we note that we could end the merged abstraction only if,
    for all pairs borrow/loan such that the borrow was in the first abstraction
    and the loan in the second, we correctly used \Rule{Merge-AbsMut}.
    The borrows consumed during the merge by using \Rule{Merge-AbsMut} are the ones
    we need to end on the left, after ending the first abstraction, so that we
    can end the second. Finally, we conclude by the reflexivity of $\le$.
  \item Cases \Rule{Le-ClearAbs}, \Rule{Le-Abs-ClearValue},
        \Rule{Le-Abs-DeconstructPair}, \Rule{Le-Abs-DeconstructSum}. Trivial.
  \item Case \Rule{Le-AnonValue}. Trivial.
\end{itemize}
\end{itemize}

We now turn to the proof of the target theorem.

\begin{theorem}[Eval-Preserves-LLBC+-Rel]
\thlabel{llbc+-statement-rel}
For all $\Omega_l$ and $\Omega_r$ \llbcp states we have:
\begin{align*}
&\Omega_l \le \Omega_r \Rightarrow
\forall\; s\, v_r\, \Omega'_r, \ \Omega_r \vdashLlbcp s \stmtres{r}{\Omega'_r} \Rightarrow\\
&\exists\; \Omega'_l,\ \Omega_l \vdashLlbcp s \exprres{r}{\Omega'_l} \wedge
\Omega'_l \le \Omega'_r
\end{align*}
\thmvspace
\end{theorem}

\noindent\textbf{Proof}\\
By induction on $\Omega_r \vdashLlbcp\, s \rightsquigarrow \stmtres{r}{\Omega_r'}$.

\begin{itemize}
\item Case empty statement. Trivial.
\item Case \Rule{E-Reorg}. By Lemma \thref{llbc+-reorg-rel}.
\item Case $s_0;\; s_1$. We use the induction hypotheses, then have to consider
  all the states resulting from evaluating $s_0$; this is straightforward.
\item Case $\eassign p rv$. By Lemma \thref{llbc+-assign-rel}.
\item Case \li+if then else+. Trivial by the induction hypotheses.
\item Case \li+match+. Trivial by the induction hypotheses.
\item Case $\efree p$. The reasoning is similar to the $\emove p$ and $\eassign p$
  cases in Lemmas \thref{llbc+-rvalue-rel} and \thref{llbc+-assign-rel}.
  The proof is straightforward by induction on $\Omega_l \le \Omega_r$.
\item Case $\ereturn$. Trivial.
\item Cases $\epanic$, $\ebreak i$, $\econtinue i$. Trivial.
\item Case loop. Trivial by the induction hypotheses.
\item Case function call.
  Similar to the same case in the proof of \Thm{Eval-Preserves-HLPL+-Rel}.
  Note that we study \llbcp here, not \llbcs, meaning the rule we need to consider
  is \Rule{E-Call}, not \Rule{E-Call-Symbolic}.
  We use the Lemma \thref{llbc+-rvalue-rel} to relate the states resulting
  from evaluating the operands given as inputs to the function (trivial induction
  on the number of inputs).
  We easily get that the states resulting from $\pushstack$ are
  related (again, induction on the number of inputs then on the number of
  local variables which are not used as inputs).
  The induction hypothesis gives us that the states are related after evaluation
  of $body$. If the tag is $\epanic$ we are done.
  If it is $\ereturn$, we get that the states resulting from $\popstack$
  are related (induction on the number of local variables, and we use
  \thref{llbc+-assign-rel} for the assignments which ``drop'' the local variables).
  Finally, we do a reasoning similar to what we did in
  \thref{llbc+-assign-rel} to show that the states are still related after
  assigning to the destination.
\end{itemize}

\section{Forward Simulation for \llbcp and \llbcs}
\appendixlabel{llbcs-proof}

We present the definitions of $\kw{init}$, $\kw{final}$, $\kw{inst-sig}$
in figures \fref{proj-input}, \fref{proj-output}, \fref{init-extern} and
\fref{inst-sig}.

\begin{figure}
  \centering
  \smaller
  \begin{mathpar}
  \inferrule[ProjInput-Ignore]{
    \rho\notin\tau
  }{
    \projinput{\rho}{(v : \tau)}{\_}
  }

  \inferrule[ProjInput-Pair]{
    \projinput{\rho}{v_0}{v'_0}\\
    \projinput{\rho}{v_1}{v'_1}
  }{
    \projinput{\rho}{(v_0,\, v_1)}{(v'_0,\, v'_1)}
  }


  \inferrule[ProjInput-Shared]{
    \text{no borrows} \in\tau
  }{
    \projinput{\rho}{(\esborrow{\ell} : \&\rho\,\tau)}{\esborrow\ell}
  }

  \inferrule[ProjInput-Mut]{
    \text{no borrows} \in v
  }{
    \projinput{\rho}{(\emborrow{\ell}{v} : \&\rho\,\kw{mut}\,\tau)}{\emborrow{\ell}{\_}}
  }
  \end{mathpar}

  \caption{Projecting Input Borrows for Region Abstractions}
  \flabel{proj-input}
\end{figure}

\begin{figure}
  \centering
  \smaller
  \begin{mathpar}
  \inferrule[ProjOutput-Symbolic]{
     \sigma \text{ fresh}\\
     \text{no borrows}\in\tau
  }{
    \projoutput{\vec{\rho}}{\tau}{\sigma}{\vecabsrho}
  }

  \inferrule[ProjOutput-Pair]{
    \projoutput{\vec{\rho}}{\tau_0}{v_0}{\vecabsrhoi 1}\\\\
    \projoutput{\vec{\rho}}{\tau_1}{v_1}{\vecabsrhoi 2}\\\\
    \forall\;\rho,\; A(\rho) = A_1(\rho)\cup A_2(\rho)
  }{
    \projoutput{\vec{\rho}}{(\tau_0,\, \tau_1)}{(v_0,\,v_1)}{\vecabsrho}
  }

  \inferrule[ProjOutput-Shared]{
    \text{no borrows}\in\tau\\
    \sigma,\,\ell \text{ fresh}\\\\
    A(\rho) = \{\; \esloan{\ell}{\sigma} \;\}\\
    \forall\;\rho'\neq\rho,\;A(\rho') = \{\}
  }{
    \projoutput{\vec{\rho}}{(\&\rho\,\tau)}{\esborrow{\ell}}{\vecabsrho}
  }

  \inferrule[ProjOutput-Mut]{
    \text{no borrows}\in\tau\\
    \sigma,\,\ell \text{ fresh}\\\\
    A(\rho) = \{\; \emloan{\ell} \;\}\\
    \forall\;\rho'\neq\rho,\;A(\rho') = \{\}
  }{
    \projoutput{\vec{\rho}}{(\&\rho\,\kw{mut}\,\tau)}{\emborrow{\ell}{\sigma}}{\vecabsrho}
  }
  \end{mathpar}
  \caption{Generating Output Values for Function Calls}
  \flabel{proj-output}
\end{figure}

\newcommand\initextern[3]{\kw{init\_extern}\;{#1}\;{#2}={#3}}
\begin{figure}
  \centering
  \smaller
  \begin{mathpar}
  \inferrule[InitExtern-Symbolic]{
     \text{no borrows}\in\tau\\
     \forall\;\rho,\;A_{\rho} = \{\}
  }{
    \initextern{\vec{\rho}}{\tau}{\vecabsrho}
  }

  \inferrule[InitExtern-Pair]{
    \initextern{\vec{\rho}}{\tau_0}{\vecabsrhoi 0}\\\\
    \initextern{\vec{\rho}}{\tau_1}{\vecabsrhoi 1}\\\\
    \forall\;\rho,\; A(\rho) = A_0(\rho)\cup A_1(\rho)
  }{
    \initextern{\vec{\rho}}{(\tau_0,\;\tau_1)}{\vecabsrho}
  }

  \inferrule[InitExtern-Shared]{
    \text{no borrows}\in\tau\\
    \ell \text{ fresh}\\\\
    A(\rho) = \{\; \esborrow{\ell} \;\}\\
    \forall\;\rho'\neq\rho,\;A(\rho') = \{\}
  }{
    \initextern{\vec{\rho}}{(\&\rho\,\tau)}{\vecabsrho}
  }

  \inferrule[InitExtern-Mut]{
    \text{no borrows}\in\tau\\
    \ell \text{ fresh}\\\\
    A(\rho) = \{\; \emborrow{\ell}{\_} \;\}\\
    \forall\;\rho'\neq\rho,\;A(\rho') = \{\}
  }{
    \initextern{\vec{\rho}}{(\&\rho\,\kw{mut}\,\tau)}{\vecabsrho}
  }
  \end{mathpar}
  \caption{Generating the External Borrows for State Initialization}
  \flabel{init-extern}
\end{figure}

\begin{figure}
  \centering
  \smaller
  \begin{mathpar}
  \inferrule[Init]{
    \forall\;i,\;\initextern{\vec\rho}{\tau_i}{\overrightarrow{A_i^{\text{ext}}(\rho)}}\\\\
    \forall\;i,\;\projoutput{\vec\rho}{\tau_i}{v_i}{\overrightarrow{A_i^{\text{out}}(\rho)}}\\\\
    \forall\;\rho,\;A(\rho) = \underset i \cup \left(A_i^{\text{ext}}(\rho) \cup A_i^{\text{out}}(\rho) \right)
  }{
    \kw{init}\;(\vec\rho,\,\vec\tau) = (\overrightarrow{v_i},\, \vecabsrho)
  }

  \inferrule[Final]{
    \forall\;i,\;\initextern{\vec\rho}{\tau_i}{\overrightarrow{A_i^{\text{ext}}(\rho)}}\\\\
    \projoutput{\vec\rho}{\tau}{v}{\overrightarrow{A^{\text{out}}(\rho)}}\\\\
    \forall\;\rho,\;A(\rho)=A^{\text{out}}(\rho)\cup \left( \underset i \cup A_i^{\text{ext}}(\rho) \right)
  }{
    \kw{final}\;(\vec\rho,\;\vec\tau,\;\tau) = (v,\, \vecabsrho)
  }

  \inferrule[InstSig]{
    \forall\;\rho,\; A^{\text{in}}(\rho) = \{ \kw{proj\_input}\;\rho\;v_i \}\\\\
    \projoutput{\vec\rho}{\tau}{\vout}{\overrightarrow{A^{\text{out}}(\rho)}}\\\\
    \forall\,\esborrow{\ell}\in\vec v,\;\exists\;v,\;\esloan{\ell}{v}\in\Omega\\\\
    \forall\;\rho,\; \abssigrho = A^{\text{in}}(\rho) \cup A^{\text{out}}(\rho)
  }{
    \kw{inst\_sig}\;(\Omega,\;\vec\rho,\;\vec v,\;\tau) = \vecabssigrho,\; \vout
  }
  \end{mathpar}
  \caption{Init and Final \llbcs States for Borrow Checking and Signature Instantiation}
  \flabel{inst-sig}
\end{figure}

The following substitution lemmas are trivial but crucial in several situations.

\begin{lemma}[Le-Subst]
For all $\Omega_l$, $\Omega_r$ \llbcp states
and $\kw{subst}$ identifier substitution we have (where $\kw{subst}\;\Omega$
is a pointwise substitution):
\begin{align*}
&\Omega_l \le \Omega_r \Rightarrow \kw{subst}\;\Omega_l \le \kw{subst}\;\Omega_r
\end{align*}
\thlabel{le-subst}
\end{lemma}
\textbf{Proof.} The proof is straightforward by induction on
$\Omega \le \Omega'$.
\begin{itemize}
\item Reflexive case. Trivial.
\item Transitive case. Trivial by the induction hypothesis.
\item Case \Rule{Le-ToSymbolic}.
  There exists $\Omega[.]$, $v$ and $\sigma$ fresh such that:
  \[
  \text{borrows, loans, $\bot$} \not\in v \eand \Omega_l = \Omega[v] \eand \Omega_r = \Omega[\sigma]
  \]
  A trivial induction on $v$ gives us that, as there are no borrows, loans
  and $\bot$ in $v$ (premise of the rule) then there are no borrows, loans
  and $\bot$ in $\kw{subst}\;v$.
  Moreover, as $\kw{subst}$ is injective we have that $\kw{subst}\;\sigma$
  if fresh for $\Omega_l$. We pose:
  \begin{align*}
  &\Omega'[.] := (\kw{subst}\;\Omega)[.]\\
  &v' := \kw{subst}\;v'\\
  &\sigma' := \kw{subst}\;\sigma
  \end{align*}
  We have (trivial induction on $\Omega[.]$, comes from the fact that
  $\kw{subst}$ is applied pointwise):
  \begin{align*}
  \kw{subst}\;\Omega_l &= \kw{subst}\;(\Omega[v])\\
  &= (\kw{subst}\;\Omega)[\kw{subst}\;v])\\
  &= \Omega'[v']
  \end{align*}
  Similarly, we have: $\kw{subst}\;\Omega_r = \Omega'[\sigma']$.
  Hence: $\kw{subst}\;\Omega_l \le \kw{subst}\;\Omega_r$.
\item Case \Rule{Le-MoveValue}. Trivial. We use the fact that the substitution
  is pointwise (like in the \Rule{Le-ToSymbolic} case).
\item Case \Rule{Le-Fresh-MutLoan}. Similar to the \Rule{Le-ToSymbolic} case.
\item Case \Rule{Le-Fresh-SharedLoan}. Similar to the \Rule{Le-ToSymbolic} case.
\item Case \Rule{Le-Reborrow-MutBorrow}. Similar to the \Rule{Le-ToSymbolic} case.
\item Case \Rule{Le-Fresh-SharedBorrow}. Similar to the \Rule{Le-ToSymbolic} case.
\item Case \Rule{Le-Reborrow-SharedLoan}. Similar to the \Rule{Le-ToSymbolic} case.
\item Case \Rule{Le-Abs-End-SharedLoan}.  Similar to the \Rule{Le-ToSymbolic} case.
\item Case \Rule{Le-Abs-End-DupSharedBorrow}.Similar to the \Rule{Le-ToSymbolic} case.
\item Case \Rule{Le-Reborrow-SharedBorrow}. Similar to the \Rule{Le-ToSymbolic} case.
\item Case \Rule{Le-ToAbs}.
  We need the auxiliary lemma:
  \[
  \forall v \vec{A},\; v \toabs \vec{A} \Rightarrow \kw{subst}\;v\toabs\kw{subst}\;(\toabs \vec{A})
  \]
  The proof is straightforward by induction on $v \toabs \vec{A}$.
\item Case \Rule{Le-MergeAbs}.
  We need the auxiliary lemma:
  \[
  \forall A_0\;A_1\;A,\; \vdash A_0 \MergeAbs A_1 \merge A \Rightarrow
  \vdash \kw{subst}\;A_0 \MergeAbs \kw{subst}\;A_1 \merge \kw{subst}\;A
  \]
  The proof is straightforward by induction on $\vdash A_0 \MergeAbs A_1 \merge A$.
\item Cases \Rule{Le-ClearAbs}, \Rule{Le-Abs-ClearValue},
      \Rule{Le-Abs-DeconstructPair}, \Rule{Le-Abs-DeconstructSum}, \Rule{Le-AnonValue}.
      Similar to the \Rule{Le-MoveValue} case.
\end{itemize}

\medskip

\begin{lemma}[Eval-Subst-\llbcp]
For all $\Omega$, $\Omega'$ \llbcp states, $s$ statement, $r$ control-flow
tag and $\kw{subst}$ identifier substitution we have:
\begin{align*}
&\Omega \le \Omega' \Rightarrow\\
&\Omega \vdash s \stmtres{r}{\Omega'}\Rightarrow\\
&\kw{subst}\;\Omega \vdash s \stmtres{r}{\kw{subst}\;\Omega'}
\end{align*}
\thlabel{eval-subst}
\end{lemma}
\textbf{Proof.}
Like with the proof of the simulation for \llbcp, we have to introduce auxiliary lemmas
for evaluating rvalues and applying reorganizations.
All the proofs are straightforward by induction on the evaluation or reorganization
derivations, and are actually very similar to the proof of Lemma \thref{le-subst}.
\medskip

We have a lemma similar to the above one for \llbcs evaluations (omitted).\\

We need the following framing lemma for $\le$. Note that as the transformation rules for
$\le$ are quite local we don't need very strong disjointness conditions.
The disjointness condition is given by $\kw{le\_framable}$ and states
that: 1. the fresh identifiers introduced in $\Omega'$ are not present in the
frame (this allows us to preserve freshness);
2. the initial state $\Omega$ and the frame $\Omega_f$ don't use the same identifiers, put
aside the fact that we allow some borrows in the partial state $\Omega$ to be dangling, by
pointing to loans in the frame $\Omega_f$ (we will need this condition when handling function calls).

\newcommand\leframable[3]{\kw{framable}\;{#1}\;{#2}\;{#3}}
\newcommand\framable[3]{\kw{framable}\;{#1}\;{#2}\;{#3}}
\begin{lemma}[Frame Rule for $\le$ in \llbcp]
For all $\Omega$, $\Omega'$ \llbcp states, we have:
\begin{align*}
\Omega \le \Omega'\Rightarrow
\forall\; \Omega_f,\;
\leframable{\Omega}{\Omega'}{\Omega_f}\Rightarrow
\quad\Omega \cup \Omega_f \le \Omega' \cup \Omega_f
\end{align*}
where:
\setcounter{equation}{0}
\begin{align}
  &\leframable{\Omega}{\Omega'}{\Omega_f} :=\nonumber\\
  &\quad
    (\forall\;\ell\in\Omega',\;\ell\notin\Omega\Rightarrow\ell\notin\Omega_f)\eand\\
  &\quad
    (\forall\;\sigma,\;\sigma\in\Omega\eor\sigma\in\Omega'\Rightarrow\sigma\notin\Omega_f)\eand\\
  &\quad
    (\forall\;A,\;A\in\Omega\eor A\in\Omega'\Rightarrow A\notin\Omega_f)\eand\\
&\quad(\forall\;\ell\in\Omega,\;\ell\in\Omega_f\Rightarrow\nonumber\\
&\qquad(\emborrow{\ell}{v}\in\Omega \eand \emloan\ell\notin\Omega) \eor \nonumber\\
&\qquad(\esborrow{\ell}\in\Omega
  \eand\esborrow{\ell}\in\Omega'
  \eand\esloan{\ell}{v}\notin\Omega'))
\end{align}
\thlabel{le-frame}
\end{lemma}

\textbf{Proof.}
We do the proof by induction on $\Omega \le \Omega'$.

\begin{itemize}
\item Reflexive case. Trivial.
\item Transitive case.
  This is the difficult case: if we want to use the induction hypotheses we have
  to pay attention to the $\kw{framable}$ condition. We take care of this
  mostly by using the substitution lemma \thref{le-subst}.
  
  More precisely, the facts $\leframable{\Omega}{\Omega''}{\Omega_f}$
  and $\leframable{\Omega''}{\Omega'}{\Omega_f}$ don't necessarily hold.
  However, we can build $\kw{subst}$ s.t.:
  \begin{align*}
  &\leframable{\Omega}{\kw{subst}\;\Omega''}{\Omega_f}\eand\\
  &\leframable{\kw{subst}\;\Omega''}{\Omega'}{\Omega_f}\eand\\
  &\Omega \le \kw{subst}\;\Omega'' \le \Omega'
  \end{align*}
  
  For instance, if there exists $\sigma \in \Omega''$, we have to prove
  that $\sigma\notin\Omega_f$. If $\sigma \in \Omega$ or $\sigma\notin\Omega'$
  we simply use (2). However, it may happen that $\sigma\notin\Omega$ and
  $\sigma\notin\Omega'$, if a transformation (e.g., \Rule{Le-ToSymbolic}) introduces a fresh $\sigma$
  when transforming $\Omega$ to $\Omega''$, then another transformation (e.g.,
  \Rule{Le-MoveValue} then \Rule{Le-ToAbs}) removes it when transforming
  $\Omega''$ to $\Omega'$. In this situation, we just have to pick
  $\sigma'$ such that $\sigma' \notin \Omega,\;\Omega'\;\Omega'',\;\Omega_f$
  and pose $\kw{subst}$ such that $\kw{subst}$ is the identity but on
  $\sigma$ where we define it as: $\kw{subst}\;\sigma=\sigma'$. We then
  use \thref{le-subst} to prove: $\Omega \le \kw{subst}\;\Omega'' \le \Omega'$.
  By using this technique, we can make sure (2) and (3) are satisfied for
  $\Omega$ and $\Omega''$, and for $\Omega''$ and $\Omega'$; that is:
  \begin{align*}
  &(\forall\;\sigma,\;\sigma\in\Omega\eor\sigma\in\Omega''\Rightarrow\sigma\notin\Omega_f)\eand\\
  &(\forall\;A,\;A\in\Omega\eor A\in\Omega''\Rightarrow A\notin\Omega_f)\\
  \end{align*}
  and:
  \begin{align*}
  &(\forall\;\sigma,\;\sigma\in\Omega''\eor\sigma\in\Omega'\Rightarrow\sigma\notin\Omega_f)\eand\\
  &(\forall\;A,\;A\in\Omega''\eor A\in\Omega'\Rightarrow A\notin\Omega_f)\\
  \end{align*}
  
  (1) and (4) are more difficult.
  We can make sure that (1) is satisfied for $\Omega$ and $\Omega''$ (same technique).
  Then, given $\ell\in\Omega'$ such that $\ell\notin\Omega''$, we want to
  show that $\ell\notin\Omega_f$. If $\ell\notin\Omega$ it is trivial by (1).
  However, it may happen that $\ell\in\Omega$: it means that $\ell$ was originally
  in $\Omega$, that we removed it (because of \Rule{Le-Abs-End-SharedLoan} or \Rule{MergeAbs-Mut}),
  then reused it in a rule which introduces a fresh loan identifier.
  But this is actually forbidden by (4), and we use this fact.
  
  By induction on $\Omega \le \Omega'$ we prove that:
  \begin{align*}
  &(\forall\;\ell\in\Omega'',\;\ell\in\Omega_f\Rightarrow\\
  &\quad(\emborrow{\ell}{v}\in\Omega'' \eand \emloan\ell\notin\Omega'') \eor\\
  &\quad(\esborrow{\ell}\in\Omega''
  \eand\esborrow{\ell}\in\Omega'
  \eand\esloan{\ell}{v}\notin\Omega'))
  \end{align*}
  We then reuse this to prove (by contradiction and) by induction on $\Omega'' \le \Omega'$ that:
  $\forall\;\ell\in\Omega',\;\ell\notin\Omega''\Rightarrow\ell\notin\Omega_f$.
  
  This allows us to use the induction hypotheses, and conclude the proof.
    
\item Case \Rule{Le-ToSymbolic}.
  There exists $\Omega_h[.]$, $v$ and $\sigma$ fresh for $\Omega_l$ such that:
  \[
  \text{borrows, loans, $\bot$} \not\in v \eand \Omega = \Omega_h[v] \eand \Omega' = \Omega_h[\sigma]
  \]
  Because of (2) in the $\leframable{\Omega}{\Omega'}{\Omega_f}$ assumption we have
  that $\sigma$ is fresh for $\Omega \cup \Omega_f$.
  Posing $\Omega'_h[.] := \Omega_h[.] \cup \Omega_f$ we get:
  \[
  \Omega \cup \Omega_f = \Omega'_h[v]\\
  \Omega'\cup\Omega_f = \Omega'_h[\sigma]
  \]
  
  Hence: $\Omega \cup \Omega_f \le \Omega'\cup\Omega_f$.
  
\item Case \Rule{Le-MoveValue}. Trivial (we don't introduce fresh identifiers).
\item Case \Rule{Le-Fresh-MutLoan}. Similar to the \Rule{Le-ToSymbolic} case.
\item Case \Rule{Le-Fresh-SharedLoan}. Similar to the \Rule{Le-ToSymbolic} case.
\item Case \Rule{Le-Reborrow-MutBorrow}. Similar to the \Rule{Le-ToSymbolic} case.
\item Case \Rule{Le-Fresh-SharedBorrow}. Similar to the \Rule{Le-ToSymbolic} case.
\item Case \Rule{Le-Reborrow-SharedLoan}. Similar to the \Rule{Le-ToSymbolic} case.
\item Case \Rule{Le-Abs-End-SharedLoan}. We use (4) to prove that there is no
  $\esrborrow{\ell}\in\Omega_f$.
\item Case \Rule{Le-Abs-End-DupSharedBorrow}.  Similar to the \Rule{Le-ToSymbolic} case.
\item Case \Rule{Le-Reborrow-SharedBorrow}. Similar to the \Rule{Le-ToSymbolic} case.
\item Case \Rule{Le-ToAbs}. Similar to the \Rule{Le-ToSymbolic} case.
\item Case \Rule{Le-MergeAbs}. Trivial.
\item Cases \Rule{Le-ClearAbs}, \Rule{Le-Abs-ClearValue},
      \Rule{Le-Abs-DeconstructPair}, \Rule{Le-Abs-DeconstructSum}, \Rule{Le-AnonValue}. Trivial.
\end{itemize}

\bigskip

We need the framing lemma below for evaluation.
We have the same disjointness conditions as for the $\le$ framing lemma
(\thref{le-frame}).

\begin{lemma}[Frame Rule for Evaluation in \llbcp]
For all $\Omega$, $\Omega'$ \llbcp states, we have:
\begin{align*}
\Omega \vdash s \stmtres{r}{\Omega'}\Rightarrow
\forall\; \Omega_f,\; \framable{\Omega}{\Omega'}{\Omega_f} \Rightarrow
\quad\Omega \cup \Omega_f \vdash s \stmtres{r}{\Omega' \cup \Omega_f}
\end{align*}
\end{lemma}
\textbf{Proof.} The proof is similar to the proof of \thref{le-frame}.
Like with the simulation proofs, we have to split it into several auxiliarly
lemmas for the evaluation of rvalues and for reorganizations.
Exactly like with \thref{le-frame}, the difficult cases are:
1. the transitive case of reorganization, because it allows us to remove
borrows from the context; 2. the cases for statement evaluationwhere
we need to use the induction hypotheses (typically, \Rule{E-Seq-Unit}
and \Rule{E-Reorg}). We handle those by using the same strategy as in \thref{le-frame}.

\bigskip

We now move on to the proof of Theorem \thref{borrow-checks}. We do the proof by
induction on the step index. In order to do so we have to generalize the theorem
statement a bit.

\begin{lemma}
  For all $\Omega$ and $\Omegas$ \llbcs states, step $n$, statement $s$, and $\Ss$ set of
  states with outcomes, we have:
  \begin{align*}
  &(\forall\; f \in \mathcal{P},\; \borrowchecks{f}) \Rightarrow
  \Omega \le \Omegas \Rightarrow
  \Omegas \vdashLlbcs\, s \rightsquigarrow \Ss \Rightarrow\\
  &(\Omega \vdashLlbc s \steparrow{n} \infty)
   \eor (\exists\; \Omega_1,\, \Omega \vdashLlbc s \stmtstepres{n}{\epanic}{\Omega_1}) \eor\\
  &(\exists\; r\, \Omega_1\, \Omegas_1,\,
    r \in \{(),\,\ereturn,\,\ebreak i,\,\econtinue i \} \eand\\
  &\quad\Omega \vdashLlbc\, s \stmtres{r}{\Omega_1} \eand \Omega_1 \le \Omegas_1
    \eand (r,\; \Omegas_1) \in \Ss)
  \end{align*}
  \thlabel{borrow-checks-n}
\end{lemma}

\noindent\textbf{Proof.} We do the proof by induction on the step index.\\
\noindent Case $n = 0$: trivial.\\
Case $n = n' + 1$:\\
We do it by induction on the statement $s$. Note that because we do not dive into
function calls, we can prove the statement by induction on the syntax rather
than on the derivation rules.

When the statement does not contain a function call, we can reuse the proofs
that there is a forward simulation from \llbcs to \llbcs (also note that only function
calls decrease the step index).

There remains the interesting case, that is a call to some function
$\fsig{\overrightarrow{\rho}}$).
We use the fact that the rule \Rule{E-Call-Symbolic} and the predicate $\kw{borrow\_checks}$ were
written so that they "match".\\

Let us illustrate how we do on an example.
We start in the state $\Omega_0$ below and evaluate \li+let z = choose(true, move px, move py)+:
\begin{align*}
\smaller
\Omega_0 &=\\
&x0 \mapstoemloan{\ell_0}\\
&y0 \mapstoemloan{\ell_1}\\
&px \mapstoemborrow{\ell_0}{0}\\
&py \mapstoemborrow{\ell_1}{1}
\end{align*}

After pushing the stack (\Rule{E-PushStack}) we get $\Omega_1$ below (the field
\kw{stack} of the state, which contains the list of all the pushed stack variables, is implicit):
\begin{align*}
\smaller
\Omega_1 &=\\
&x0 \mapstoemloan{0}\\
&y0 \mapstoemloan{1}\\
&px \mapsto\bot\\
&py \mapsto\bot\\
&\xret \mapsto \bot\\
&x \mapstoemborrow{0}{0}\\
&y \mapstoemborrow{1}{1}
\end{align*}

The borrow checking assumption gives us that there \emph{exists} some loan identifiers,
symbolic values, states $\Omega_{\kw{init}}$ and $\Omega_{\kw{final}}$, and a set of
states with outcomes $\Ss$ such that:
\begin{align*}
&\Omega_{\kw{init}}\vdash \kw{choose}.\kw{body} \stmtarrow \Ss \eand\\
&\forall\; res \in \Ss,\;\exists\; \Omegas,\;
    res = (\epanic,\; \Omegas) \eor (res = (\ereturn,\; \Omegas) \eand
    \Omegas \le \Omegas_{\kw{final}})
\end{align*}

where:
\begin{align*}
\Omega_{\kw{init}} &=\\
&\Abs{A}{ \emborrow{\ell^0_x}{\_},\; \emborrow{\ell^0_y}{\_},\; \emloan{\ell_x},\; \emloan{\ell_y} },\\
&\xret\mapsto\bot\\
&x \mapstoemborrow{x}{\sigma_x},\;\\
&y \mapstoemborrow{y}{\sigma_y}),\;
\end{align*}

and:
\begin{align*}
\Omega_{\kw{final}} &=\\
&\Abs{A}{ \emborrow{\ell^0_x}{\_},\; \emborrow{\ell^0_y}{\_},\; \emloan{\ell_z} },\\
&\xret \mapstoemborrow{z}{\sigma_z}\\
&x \mapsto\bot,\;\\
&y \mapsto\bot,\;\\
\end{align*}

We want to transform $\Omega_1$ (by using $\le$ rules) in order to isolate
a local state which is equal to $\Omega_{\kw{init}}$, then apply the framing
rules.
One issue is that the borrow checking predicate fixes a choice
of loans, symbolic values and region abstractions identifiers (the $\ell_x$, $\ell_y$,
etc. above), which is not
necessarily the choice which suits us (the borrow checking predicate gives us
a ``there exists'', while we want a ``for all'').
Fortunately, we can apply the substitution lemmas to get the identifiers we want
(\thref{le-subst} and \thref{eval-subst}).

Another issue is that the ``dangling'' borrows in environment $\Omega_{\kw{init}}$
are pairwise disjoint. This is not necessarily the case in environment $\Omega_0$
for two reasons: 1. we don't enforce a well-formedness condition on the
state in the assumptions of our theorem, meaning there can be duplicated mutable
borrows (though in practice it won't happen of course); 2. it is perfectly valid
to use several times the same shared borrows.
We can take care of this problem by introducing reborrowing abstractions
($A_0$ and $A_1$ below, and $\vecabsreborrowrho$ in the proof of the general case
afterwards)
with rules \Rule{Le-Reborrow-MutBorrow} and \Rule{Le-Reborrow-SharedBorrow-Abs}: as those rules introduce fresh loan identifiers,
we make sure the external ``dangling'' borrows are pairwise distinct.

Coming back to the example with \li+choose+, by repeatedly applying
\Rule{Le-Reborrow-MutBorrow-Abs}, \Rule{Le-Reborrow-SharedBorrow}, \Rule{Le-MergeAbs} and
\Rule{Le-ToSymbolic}, we get that $\Omega_1 \le \Omega_2$, where:
\begin{align*}
\smaller
\Omega_2 &=\\
&x0 \mapstoemloan{0},\\
&y0 \mapstoemloan{1},\\
&px \mapsto\bot,\\
&py \mapsto\bot,\\
&\AbsI{0}{ \emborrow{\ell_0}{\_},\; \emloan{\ell^0_x} } ,\\
&\AbsI{1}{ \emborrow{\ell_1}{\_},\; \emloan{\ell^0_y} } ,\\
&\texttt{// Frame is above, local state is below}\\
&\Abs{A}{ \emborrow{\ell^0_x}{\_},\; \emborrow{\ell^0_y}{\_},\; \emloan{\ell_x},\; \emloan{\ell_y} },\\
&x \mapstoemborrow{x}{\sigma_x},\\
&y \mapstoemborrow{y}{\sigma_y},\\
&\xret \mapsto \bot\\
\end{align*}

We now apply the frame rule and get that:
\begin{align*}
&\Omega_{2}\vdash \kw{choose}.\kw{body} \stmtarrow \Ss \eand\\
&\forall\; res \in \Ss,\;\exists\; \Omegas,\;
    res = (\epanic,\; \Omegas) \eor (res = (\ereturn,\; \Omegas) \eand
    \Omegas \le \Omegas_3)
\end{align*}

where:
\begin{align*}
\smaller
\Omega_3 &=\\
&x0 \mapstoemloan{0},\\
&y0 \mapstoemloan{1},\\
&px \mapsto\bot,\\
&py \mapsto\bot,\\
&\AbsI{0}{ \emborrow{\ell_0}{\_},\; \emloan{\ell^0_x} } ,\\
&\AbsI{1}{ \emborrow{\ell_1}{\_},\; \emloan{\ell^0_y} } ,\\
&\texttt{// Frame is above, local state is below}\\
&\AbsI{\kw{final}}{ \emborrow{\ell^0_x}{\_},\; \emborrow{\ell^0_y}{\_},\; \emloan{\ell_z} },\\
&x \mapsto\bot,\;\\
&y \mapsto\bot,\;\\
&\xret \mapstoemborrow{z}{\sigma_z}\\
\end{align*}

We pop the stack of $\Omega_3$ (rule \Rule{E-PopStack}) and apply the assignemnt to \li+z+, and get $\Omega_4$
(once again, we make the \kw{stack} field explicit):
\begin{align*}
\smaller
\Omega_4 &=\\
&x0 \mapstoemloan{0},\\
&y0 \mapstoemloan{1},\\
&px \mapsto\bot,\\
&py \mapsto\bot,\\
&\AbsI{0}{ \emborrow{\ell_0}{\_},\; \emloan{\ell^0_x} } ,\\
&\AbsI{1}{ \emborrow{\ell_1}{\_},\; \emloan{\ell^0_y} } ,\\
&\AbsI{\kw{final}}{ \emborrow{\ell^0_x}{\_},\; \emborrow{\ell^0_y}{\_},\; \emloan{\ell_z} },\\
&z \mapstoemborrow{z}{\sigma_z}
\end{align*}

We finally merge the reborrowing abstractions ($A_0$ and $A_1$) into $A_{\kw{final}}$ and
get (this is exactly the state resulting from applying the rule \Rule{E-Call-Symbolic}):
\begin{align*}
\smaller
\Omega_5 &=\\
&x0 \mapstoemloan{0},\\
&y0 \mapstoemloan{1},\\
&px \mapsto\bot,\\
&py \mapsto\bot,\\
&\AbsI{\kw{final}}{ \emborrow{\ell_0}{\_},\; \emborrow{\ell_1}{\_},\; \emloan{\ell_z} },\\
&z \mapstoemborrow{z}{\sigma_z}
\end{align*}

\medskip
We illustrated the proof on the example of \li+choose+; we now come back to the general
case.

We need some auxiliary lemmas.

\begin{lemma}[Auxiliary Lemma: Frame Body - Value]
\begin{align*}\smaller
&\initextern{\vec\rho}{\tau}{\overrightarrow{A^{\kw{ext}}(\rho)}}\Rightarrow
\projoutput{\vec\rho}{\tau}{v'}{\overrightarrow{A^{\kw{ext\_in}}(\rho)}}\Rightarrow\\
&\left(\forall\;\rho,\;\absinitrho = \cup \left(A^{\kw{ext}}(\rho) \cup
  A^{\kw{ext\_in}}(\rho) \right) \right)\Rightarrow
\projoutput{\vec\rho}{\tauout}{\vout}{\overrightarrow{A^{\kw{out}}(\rho)}}\Rightarrow\\
&\left(\forall\;\rho,\;\absfinalrho=A^{\kw{out}}(\rho)\cup
  A^{\kw{ext}}(\rho)\right)\Rightarrow
\left(\forall\;\rho,\; A^{\kw{in}}(\rho) = \{\; \kw{proj\_input}\;\rho\;v \;\} \right)\Rightarrow\\
&\left(\forall\,\esborrow{\ell}\in v,\;\exists\;v,\;\esloan{\ell}{v}\in\Omegas_0\right)\Rightarrow
\left(\forall\;\rho,\; \abssigrho = A^{\kw{in}}(\rho) \cup
  A^{\kw{out}}(\rho)\right)\Rightarrow\\
&\exists\;\vecabsreborrowrho,\\
&\kw{let}\ \Omegas_{\kw{beg}} = \Omegas_0,\; x \rightarrow V[v : \tau],\;
    \xret\rightarrow\bot \\
&\kw{let}\ \Omegas_f = \Omegas_0,\; \vecabsreborrowrho\\
&\kw{let}\ \Omegas_{\kw{init\_local}} =
    \vecabsinitrho,\; x \rightarrow V[v' : \tau],\;
    \xret\rightarrow\bot\\
&\kw{let}\ \Omegas_{\kw{init}} = \Omegas_{\kw{init\_local}} \cup \Omegas_f\\
&\kw{let}\ \Omegas_{\kw{final\_local}} =
  \vecabsfinalrho,\; x\rightarrow V[\bot],\;
  \overrightarrow{y\rightarrow\bot},\;\xret\rightarrow\vout \\
&\kw{let}\ \Omegas_{\kw{final}} = \Omegas_{\kw{final\_local}} \cup \Omegas_f\\
&\kw{let}\ \Omegas_{\kw{end}} = \Omegas_0,\; \overrightarrow{\abssigrho},\; x\rightarrow V[\bot],\;
   \xret\rightarrow\vout \\
&\Omegas_{\kw{beg}} \le \Omegas_{\kw{init}} \eand
\Omega_{\kw{final}} \le \Omegas_{\kw{end}} \eand
\framable{\Omegas_{\kw{init\_local}}}{\Omegas_{\kw{final\_local}}}{\Omegas_f}
\end{align*}
\thlabel{frame-body-value}
\end{lemma}
\textbf{Proof.} By induction on $\tau$.
\begin{itemize}
\item Pair. By induction hypotheses and \Rule{Le-MergeAbs}.
\item $\&\; \tau$. We apply \Rule{Le-Reborrow-SharedBorrow} twice: once for the
  $\absinit$ abstraction and once for the $\absreborrow$ abstraction.
  We note that only \Rule{InitExtern-Shared}, \Rule{ProjOutput-Shared} and \Rule{ProjInput-Shared} apply
  for the projections (we do case disjunctions on $v$,
  $\initextern{\vec\rho}{\tau}{\overrightarrow{A^{\kw{ext}}(\rho)}}$,
  $\projoutput{\vec\rho}{\tau}{v'}{\overrightarrow{A^{\kw{ext\_in}}(\rho)}}$
  and $\kw{proj\_input}\;\rho\;v$). For $\kw{framable}$ we leverage the freshness of the
  identifiers introduced by the rules and the fact that the $\absinitrho$ and
  $\absfinalrho$
  abstractions don't have identifiers which are also used elsewhere, at the exception
  of dangling borrows pointing to loans in the $\absreborrow$ abstractions.
\item $\& \kw{mut}\; \tau$. Similar to the $\&\; \tau$ case, but this time with \Rule{Le-Reborrow-MutBorrow-Abs}.
\item Sum. We apply \Rule{Le-ToSymbolic} (there can't be borrows in sums).
\item Literal. We apply \Rule{Le-ToSymbolic}.
\end{itemize}

\begin{lemma}[Auxiliary Lemma: Frame Body]
\begin{align*}\smaller
&\left(\forall\;i,\;\initextern{\vec\rho}{\tau_i}{\overrightarrow{A_i^{\kw{ext}}(\rho)}}\right)\Rightarrow
\left(\forall\;i,\;\projoutput{\vec\rho}{\tau_i}{v'_i}{\overrightarrow{A_i^{\kw{ext\_in}}(\rho)}}\right)\Rightarrow\\
&\left(\forall\;\rho,\;\absinitrho = \underset i \cup \left(A_i^{\kw{ext}}(\rho) \cup
  A_i^{\kw{ext\_in}}(\rho) \right) \right)\Rightarrow
\projoutput{\vec\rho}{\tauout}{\vout}{\overrightarrow{A^{\kw{out}}(\rho)}}\Rightarrow\\
&\left(\forall\;\rho,\;\absfinalrho=A^{\kw{out}}(\rho)\cup \left( \underset i \cup
  A_i^{\kw{ext}}(\rho) \right)\right)\Rightarrow
\left(\forall\;\rho,\; A^{\kw{in}}(\rho) = \{\; \kw{proj\_input}\;\rho\;v_i \;\} \right)\Rightarrow\\
&\left(\forall\,\esborrow{\ell}\in\overrightarrow{v_i},\;\exists\;v,\;\esloan{\ell}{v}\in\Omegas_0\right)\Rightarrow
\left(\forall\;\rho,\; \abssigrho = A^{\kw{in}}(\rho) \cup
  A^{\kw{out}}(\rho)\right)\Rightarrow\\
&\exists\;\vecabsreborrowrho,\\
&\kw{let}\ \Omegas_{\kw{beg}} = \Omegas_0,\; \overrightarrow{x_i \rightarrow v_i},\;
    \overrightarrow{y\rightarrow\bot},\;\xret\rightarrow\bot \\
&\kw{let}\ \Omegas_f = \Omegas_0,\; \vecabsreborrowrho\\
&\kw{let}\ \Omegas_{\kw{init\_local}} =
    \vecabsinitrho,\; \overrightarrow{x_i \rightarrow v'_i},\;
    \overrightarrow{y\rightarrow\bot},\;\xret\rightarrow\bot\\
&\kw{let}\ \Omegas_{\kw{init}} = \Omegas_{\kw{init\_local}} \cup \Omegas_f\\
&\kw{let}\ \Omegas_{\kw{final\_local}} =
  \vecabsfinalrho,\; \overrightarrow{x_i\rightarrow\bot},\;
  \overrightarrow{y\rightarrow\bot},\;\xret\rightarrow\vout \\
&\kw{let}\ \Omegas_{\kw{final}} = \Omegas_{\kw{final\_local}} \cup \Omegas_f\\
&\kw{let}\ \Omegas_{\kw{end}} = \Omegas_0,\; \overrightarrow{\abssigrho},\; \overrightarrow{x_i\rightarrow\bot},\;
  \overrightarrow{y\rightarrow\bot},\;\xret\rightarrow\vout \\
&\Omegas_{\kw{beg}} \le \Omegas_{\kw{init}} \eand
\Omega_{\kw{final}} \le \Omegas_{\kw{end}} \eand
\framable{\Omegas_{\kw{init\_local}}}{\Omegas_{\kw{final\_local}}}{\Omegas_f}
\end{align*}
\thlabel{frame-body}
\end{lemma}
\textbf{Proof.} By induction on $\kw{length}\;\overrightarrow{x_i}$.
The base case is trivial: we note that $\vecabsinitrho = \emptyset$ and
$\vecabsfinalrho = \vecabssigrho$, and pose
$\vecabsreborrowrho = \emptyset$.
In the recursive case, the list of local input variables is: $x_n,\; \overrightarrow{x_i}$.
We use the induction hypothesis with $\Omegas_0 := \Omegas_0,\; x_n \rightarrow v_n$,
then apply \thref{frame-body-value} to introduce reborrowing abstractions for $v_n$.

\bigskip

Given \thref{frame-body}, the rest of the proof of the recursive case of
\thref{borrow-checks-n} is straightforward.

\bigskip

Theorem \thref{borrow-checks-n} implies the target theorem \thref{borrow-checks}.

\section{Proof of Join and Collapse}
\appendixlabel{join-collapse}

We show the full rules for the join and collapse operations in \Figure{join-rules-values},
\Figure{join-rules-states}, \Figure{join-rules-values-precise}, and \Figure{collapse-rules}.

\begin{figure}
  \centering
  \smaller
  \begin{mathpar}
    \JoinValueRules*
  \end{mathpar}
  \caption{Rules to Join Values}
  \flabel{join-rules-values}
\end{figure}

\begin{figure}
  \centering
  \smaller
  \begin{mathpar}
    \JoinStateRules*
  \end{mathpar}
  \caption{Rules to Join States}
  \flabel{join-rules-states}
\end{figure}

\begin{figure}
  \centering
  \smaller
  \begin{mathpar}
    \JoinValueRulesPrecise
  \end{mathpar}
  \caption{More Precise Rules to Join Borrows}
  \flabel{join-rules-values-precise}
\end{figure}

\begin{figure}
  \centering
  \smaller
  \begin{mathpar}
    \CollapseRules*
  \end{mathpar}
  \caption{Rules to Collapse States}
  \flabel{collapse-rules}
\end{figure}

\son{TODO: We need the rule \Rule{Le-DeconstructSharedLoans} to deconstruct the nested
shared loans inside region abstractions before doing a join.
This is useful
to simplify the context by losing ``structural'' information if we perform
different operations about shared values in the two different branches.
In particular, it will be extremely useful when considering the loops:
for instance, in one branch we might dive into a recursive data structure
by reborrowing sub-values of this structure - think of lists and the nil
and cons branches.
This rule derives from the combination of \Rule{Le-Reborrow-SharedLoan}
to move the inner shared value into a different region abstraction,
\Rule{Le-MergeAbs} to merge the two region abstractions together
(we use \Rule{MergeAbs-Shared} to get rid of the fresh borrow we introduced
in the new abstraction), and finally \Rule{Le-Abs-End-SharedLoan} to get
rid of the now orphan inner loan.
}

We prove Theorem \thref{join-collapse-le}.

We introduce an auxiliary function $\projmarkedname$ to formalize what it means to project
values and states to keep only the values coming from the left state or the
values coming from the right state. The term $\projmarked{l}{v}$ (respectively,
$\projmarked{r}{v}$) is the value $v$ where we discard the values marked as coming from
the right (respectively, left) state. We naturally extend this definition to region
abstractions and states.

\begin{align*}
&\projmarked{l}{v} = v\\
&\projmarked{l}{\lbox{v}} = \projmarked{r}{\rbox{v}} = v\\
&\projmarked{l}{\rbox{v}} = \projmarked{r}{\lbox{v}} = \emptyset\\
&\ldots \text{ (Omitting tuples, etc.)}\\
&\forall\; m \in \{l,\,r\},\, \projmarked{m}{\Abs{A}{\overrightarrow{v_i}}} =
 \Abs{A}{\overrightarrow{\projmarked{m}{v_i}}}\\
&\ldots \text{ (Omitting the rules for states)}
\end{align*}

We introduce the auxiliary predicate $\sloansinclname$ for the proofs.
We need it because of the premises of \Rule{Join-SharedBorrows}
and \Rule{Join-SharedBorrows-Precise} (which in turn come from the premises of
\Rule{Le-Reborrow-SharedBorrow}).
The environments $\Omega_0$ and $\Omega_1$ on the left of
$\Omega_0,\, \Omega_1 \vdash \join_v\;{v_0}\;{v_1}$ are actually needed only so that they can be
mentioned by those rules.

\begin{align*}
\sloansincl{\Omega}{\Omega'} :=
\forall\; \ell\, v,\, \esloan{\ell}{v} \in \Omega \Rightarrow
\exists\; v',\, \esloan{\ell}{v'} \in \Omega'
\end{align*}

The proof requires several lemmas.

\begin{lemma}[Join-Values-Le]
  For all $\Omega^{0}_l$, $\Omega^{0}_r$, $\Omega_l[.]$, $\Omega_r[.]$,
  $v_l$, $v_r$, $v_j$, $\overrightarrow{A}$ we have:
  \begin{align*}
  &(\forall\; m \in \{l,\,r\},\, \Omega^{0}_m \le \Omega_m[v_m] \wedge
    \sloansincl{\Omega^{0}_m}{\Omega_m} \wedge
    \wfjoinhole{\Omega_m[.]}{v_m}) \Rightarrow\\
  &\Omega^{0}_l,\, \Omega^{0}_r \vdash
   \join{v_l}{v_r} \rightsquigarrow v_j \joinres \overrightarrow{A} \Rightarrow\\
  &\forall\; m \in \{l,\,r\},\, \Omega^{0}_m \le \Omega_m[v_j], \projmarked{m}{\overrightarrow{A}} \wedge
   \sloansincl{\Omega^{0}_m}{(\Omega_m[v_j], \projmarked{m}{\overrightarrow{A}})}
  \end{align*}
  
  where:
  
  \begin{align*}
  &\wfjoinhole{\Omega[.]}{v} :=\\
  &\text{hole of $\Omega[.]$ inside a shared loan} \Rightarrow
    \bot \notin v
  \end{align*}

\end{lemma}

\noindent\textbf{Proof}:\\
By induction on
$\Omega^{0}_l,\, \Omega^{0}_r \vdash \join{v_l}{v_r} \rightsquigarrow v_j
 \joinres \overrightarrow{A}$.
\begin{itemize}
\item Case \Rule{Join-Same}: trivial by reflexivity of $\le$.
\item Case \Rule{Join-Symbolic}: we use \Rule{Le-ToSymbolic}.
\item Case \Rule{Join-Bottom}: we use \Rule{Le-MoveValue}. We need the assumption
  $\wfjoinhole{\Omega_m[.]}{v_m}$ for the premise that the hole of $\Omega_l[.]$
  or $\Omega_r[.]$ (depending on whether $v_l = \bot$ or $v_r = \bot$) is not inside a shared loan.
\item Case \Rule{Join-MutBorrows}: we the the induction hypothesis for the inner
  value (with states $\Omega_l[\emborrow{\ell_0}{[.]}]$ and
  $\Omega_r[\emborrow{\ell_1}{[.]}]$) then \Rule{Le-Reborrow-MutBorrow-Abs}
  to introduce the fresh region abstraction with the reborrow.
\item Case \Rule{Join-SharedBorrows}: we use \Rule{Le-Reborrow-SharedBorrow}; the premises
  stating that there are corresponding shared loans in the context are proven by using the
  premise of \Rule{Join-SharedBorrows} in combination with the assumption
  $\sloansincl{\Omega^{0}_m}{\Omega_m}$.
\item Case \Rule{Join-MutLoans}: we use \Rule{Le-Reborrow-MutLoan-Abs} to insert
  a fresh mutable loan and move the current loan to a fresh region abstraction;
  the rules for $\toabs$ (\Rule{ToAbs-MutBorrow} then \Rule{ToAbs-MutLoan}) allow
  us to conclude that the fresh region abstraction has the proper shape.
\item Case \Rule{Join-SharedLoans}: we use \Rule{Le-Reborrow-SharedLoan}.
\item Cases \Rule{Join-MutLoan-Other}, \Rule{Join-Other-MutLoan}:
  we use \Rule{Le-Fresh-MutLoan-Abs} to introduce a fresh loan on the side
  which doesn't have one, then use the induction hypothesis.
\item Cases \Rule{Join-SharedLoan-Other}, \Rule{Join-Other-SharedLoan}:
  we use \Rule{Le-Fresh-SharedLoan} to introduce a fresh loan on the side
  which does't have one, then use the induction hypothesis.
\item Cases \Rule{Join-Tuple}, \Rule{Join-Sum}: trivial by the induction
  hypotheses.
\item Case \Rule{Join-Same-MutBorrow}: trivial by the induction hypothesis.
\item Case \Rule{Join-Same-SharedLoan}: trivial by the induction hypothesis;
  we have to use the premise that there are no $\bot$ in the inner values
  to prove that we can maintain the assumption $\sloansincl{\Omega^{0}_m}{\Omega_m}$.
\item Case \Rule{Join-MutBorrows-Precise}: same as for the case
  \Rule{Join-MutBorrows-Precise}, but we have to introduce
  additional region abstractions (with \Rule{Le-Reborrow-MutBorrow-Abs}).
\item Case \Rule{Join-SharedBorrows-Precise}: same as for the case
  \Rule{Join-SharedBorrows-Precise}, but we have to introduce
  additional region abstractions (with \Rule{Le-Reborrow-SharedBorrow-Abs}).
\end{itemize}

\begin{lemma}[Join-States-Le]
  \thlabel{join-states-le}
  For all $\Omega^{0}_l$, $\Omega^{0}_r$, $\Omega_{acc}$, $\Omega_l$, $\Omega_r$,
  $\Omega_j$ we have:
  \begin{align*}
  &(\forall\; m \in \{l,\,r\},\,
    \Omega^{0}_m \le ((\projmarked{m}{\Omega_{acc}}) \cup \Omega_m)
    \wedge\\
    &\sloansincl{\Omega^{0}_m}{((\projmarked{m}{\Omega_{acc}}) \cup \Omega_m}))
    \Rightarrow\\
  &\Omega^{0}_l,\, \Omega^{0}_r \vdash \join{\Omega_l}{\Omega_r} \rightsquigarrow \Omega_j
   \Rightarrow\\
  &\forall\; m \in \{l,\,r\},\,
   \Omega^{0}_m \le \projmarked{m}{(\Omega_{acc} \cup \Omega_j)}
  \end{align*}
\end{lemma}

\noindent\textbf{Proof}:\\
By induction on
$\Omega^{0}_l,\, \Omega^{0}_r \vdash \join{\Omega_l}{\Omega_r} \rightsquigarrow \Omega_j$.
\begin{itemize}
\item Case \Rule{Join-SameAbs}: we pose $\Omega'_{acc} := \Omega_{acc},\, A$ (we add
  the region abstraction $A$ to $\Omega_{acc}$) and use
  the induction hypothesis.
\item Case \Rule{Join-Same-Anon}: similar to case \Rule{Join-SameAbs}; we pose
  $\Omega'_{acc} := \Omega_{acc}, \_ \rightarrow v$ and use the induction hypothesis.
\item Case \Rule{Join-AbsLeft}: we pose $\Omega'_{acc} := \Omega_{acc},\, \lbox{A}$
  and use the induction hypothesis.
\item Case \Rule{Join-AbsRight}: we pose $\Omega'_{acc} := \Omega_{acc},\, \rbox{A}$
  and use the induction hypothesis.
\item Case \Rule{Join-Var}: we use \Thm{Join-Values-Le}, pose
  $\Omega'_{acc} := \Omega_{acc},\, \overrightarrow{A}$, use the fact that
  $\sloansinclname$ is transitive, and use the induction hypothesis.
\end{itemize}

\begin{lemma}[Collapse-Merge-Abs-Le]
  For all $\Omega_l$, $\Omega_r$, $\Omega_{acc}$, $\Omega_l$, $\Omega_r$,
  $\Omega_j$ we have:
  \begin{align*}
  \vdash A_0 \MergeAbs A_1 \merge A \Rightarrow
  \forall\; m \in \{l,\,r\},\,
  \projmarked{m}{A_0} \MergeAbs \projmarked{m}{A_1} \merge \projmarked{m}{A}
  \end{align*}
\end{lemma}

\noindent\textbf{Proof}:\\
By induction on $\vdash A_0 \MergeAbs A_1 \merge A$.
\begin{itemize}
\item Case \Rule{MergeAbs-Union}: by the induction hypothesis.
\item Case \Rule{MergeAbs-Mut}: by the induction hypothesis.
\item Case \Rule{MergeAbs-Shared}: by the induction hypothesis.
\item Case \Rule{MergeAbs-Mut-MarkedLeft}: we use \Rule{MergeAbs-Mut} then
  the induction hypothesis for the left projection, and directly use the
  induction hypothesis for the right projection.
\item Case \Rule{MergeAbs-Mut-MarkedRight}: symmetric of previous case.
\item Case \Rule{MergeAbs-Shared-MarkedLeft}: we use \Rule{MergeAbs-Shared}
  then the induction hypothesis for the left projection, and directly use
  the induction hypothesis for the right projection.
\item Case \Rule{MergeAbs-Shared-MarkedRight}: symmetric of previous case.
\end{itemize}

\begin{lemma}[Collapse-Le]
  \thlabel{collapse-le}
  For all $\Omega$, $\Omega'$ we have:
  \begin{align*}
  \vdash \Omega \collapse \Omega' \Rightarrow
  \forall\; m \in \{l,\,r\},\,
  \projmarked{m}{\Omega} \le \projmarked{m}{\Omega'}
  \end{align*}
\end{lemma}

\noindent\textbf{Proof}:\\
By induction on $\vdash \Omega \collapse \Omega'$.
\begin{itemize}
\item Case \Rule{Collapse-Merge-Abs}. By \Thm{Collapse-Merge-Abs} and
  \Rule{Le-MergeAbs}.
\item Case \Rule{Collapse-Dup-MutBorrow}.
  We note that, for $m \in \{l,\,r\}$, we have:
  \begin{align*}
    \projmarked{m}{\left(A \cup \{\;
      \lbox{\emborrow{\ell}{\_}},\;
      \rbox{\emborrow{\ell}{\_}}
    \;\}\right)} &=
     (\projmarked{m}{A}) \cup \{\;
      \emborrow{\ell}{\_}
    \;\}\\
    &=
     \projmarked{m}{(A \cup \{\;
      \emborrow{\ell}{\_}
    \;\})}
  \end{align*}
\item Case \Rule{Collapse-Dup-MutLoan}. Similar to previous case.
\item Case \Rule{Collapse-Dup-SharedBorrow}. Similar to previous case.
\item Case \Rule{Collapse-Dup-SharedLoan}. Similar to previous case,
  but we use \Rule{Le-ToSymbolic} if we need to introduce a fresh
  symbolic value.
\end{itemize}

\medskip
Finally, the combination of theorems \thref{join-states-le} and \thref{collapse-le} gives
us the target theorem (\thref{join-collapse-le}).

\section{Loops}
\appendixlabel{loop-list}

We describe here a more realistic example which uses a loop.
In this example, we recursively dive into a list so as to get a mutable borrow
to its last element. We could then use this borrow to append another list, for instance.
We define the type list as: $\elist \tau := \mu X.\, () + (\tau,\, \ebox X)$.
We also pose $\enil := \eleft{()}$ and $\econs{x}{t} := \eright{(x,\, \ebox t)}$ as
syntactic shortcuts.
We start in an environment where \li+l+ mutably borrows a list (symbolic) value from another
variable \li+l0+.

\begin{minted}[xleftmargin=1.5em,mathescape=true,escapeinside=||]{rust}
// $l_0 \mapstoemloan{0},\; l \mapstoemborrow{0}{\sigma_0}$
loop {
  match *l {
    Nil => { // $l_0 \mapstoemloan{0},\; l \mapstoemborrow{0}{\enil}$
      break; // No need to compute a join here
    }
    Cons => { // $l_0 \mapstoemloan{0},\; l \mapstoemborrow{0}{(\econs{\sigma_1}{\sigma_2})}$
      l = &mut (*(*l as Cons).1); // $l_0 \mapstoemloan{0},\; \_\mapstoemborrow{0}{(\econs{\sigma_1}{\emloan{\ell_1}})},\; l \mapstoemborrow{1}{\sigma_2}$
      continue; // We join with the environment at the entry of the loop
    }
  }
}
\end{minted}

At the continue, we convert the anonymous value to a region abstraction
(with \Rule{Le-ToAbs}, \Rule{ToAbs-MutBorrow}, \Rule{ToAbs-Sum}, \Rule{ToAbs-Box},
\Rule{ToAbs-MutLoan}).
We join the environment at the entry of the loop with the environment at the
\li+continue+. There is nothing to do for \li+l_0+ because its value is unchanged (we use .
\Rule{Join-Same}).
For $A_0$ we use \Rule{Join-AbsRight}. Finally, for \li+l+ we use
\Rule{Join-MutBorrows}. We get:

\begin{align*}
&l_0 \mapstoemloan{0},\\
&\AbsI{0}{\rbox{\emborrow{\ell_0}{\_}},\, \rbox{\emloan{\ell_1}} },\\
&l \mapstoemborrow{2}{\sigma_3}\\
&\AbsI{1}{ \lbox{\emborrow{\ell_0}{\_}},\, \rbox{\emborrow{\ell_1}{\_}},\, \emloan{\ell_2} }
\end{align*}

We collapse the environment by merging $A_0$ and $A_1$.
By \Rule{MergeAbs-Mut-MarkedLeft} we simplify $\rbox{\emloan{\ell_1}}$
and $\rbox{\emborrow{\ell_1}{\_}}$. We then use \Rule{Collapse-Dup-MutBorrow}
to simplify $\rbox{\emborrow{\ell_0}{\_}}$ and $\lbox{\emborrow{\ell_0}{\_}}$
into $\emborrow{\ell_0}{\_}$. We get the following environment, which is a fixed-point.

\begin{align*}
&l_0 \mapstoemloan{0},\\
&\AbsI{2}{ \emborrow{\ell_0}{\_},\, \emloan{\ell_2} }\\
&l \mapstoemborrow{2}{\sigma_3}\\
\end{align*}

\fi 

\end{document}